\documentclass[a4paper]{article}
\usepackage{localrus}
\usepackage{hyperref}

\begin{document}
\title{Выборы и статистика:\\ казус <<Единой России>> (2009--2020)}
\author{A.\,Шень}
\date{22 октября 2009,  декабрь 2011~--  апрель 2012, май 2018,\\[1ex] июль 2020, сентябрь 2020}
\maketitle

\begin{abstract}

\eng{This survey contains statistics on \emph{elections} in Russia published in different places and available online. This data is discussed from the viewpoint of statistical model selection. The current version is updated including the materials up to September, 2020.}

\medskip

В этом обзоре собраны графики, отражающие статистику российских \emph{выборов} и опубликованные в открытой печати и сети Интернет. Объясняется смысл этих графиков и на их примере иллюстрируется процесс выбора статистических гипотез.

Обзор дополнен материалами по голосованиям следующих лет (до 2020 года). Разобраны некоторые появившиеся за это время публикации.
\end{abstract}

\clearpage

\tableofcontents

\clearpage

\begin{figure}[!h]
\begin{center}
\includegraphics[width=0.8\hsize,height=0.57\hsize]{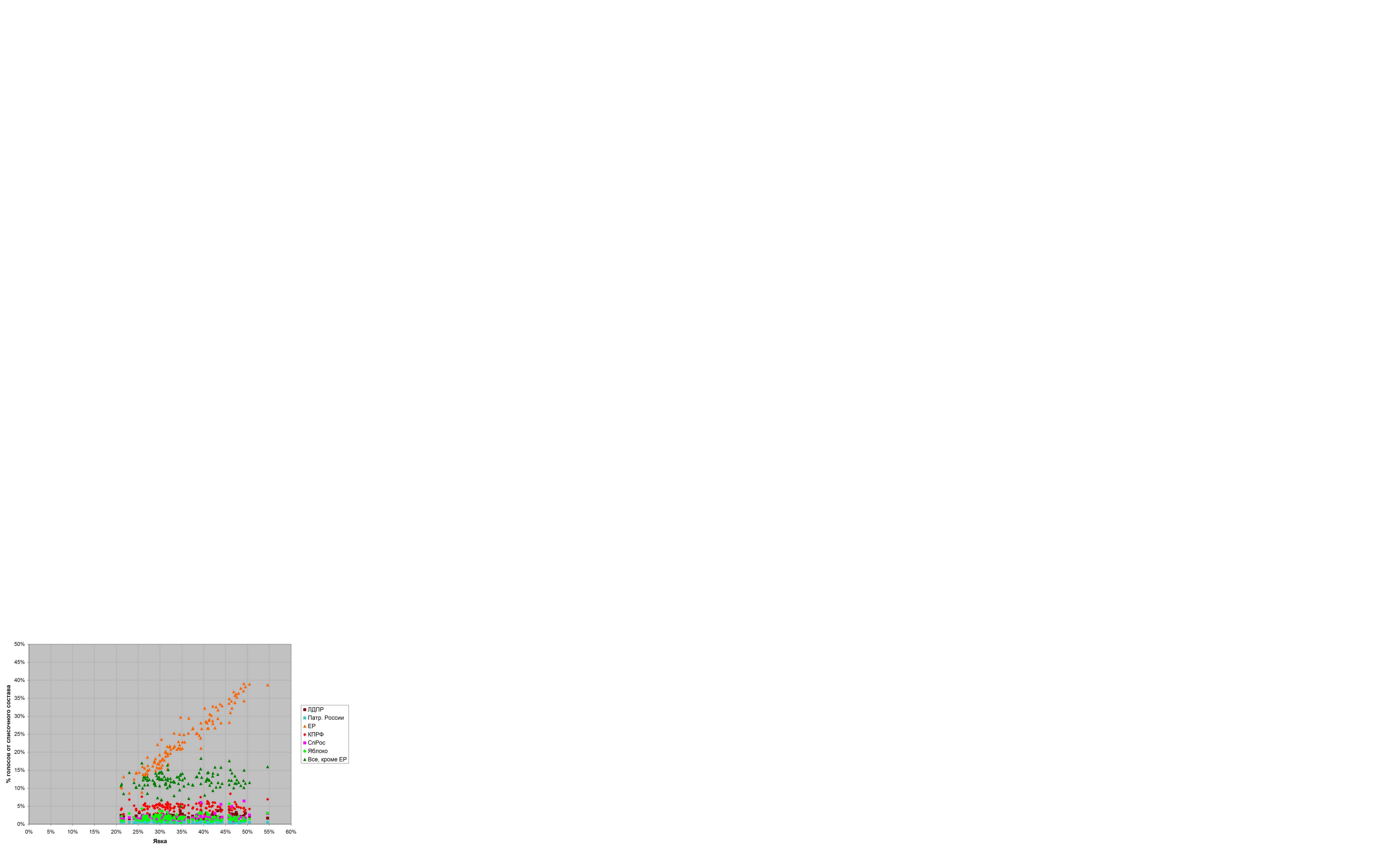}
\end{center}
\caption{Данные по районам~\cite{uborshizzza-674242}; добавлена сумма всех партий, кроме <<Единой России>>; рисунок растянут по вертикали, чтобы сделать масштабы по осям примерно равными.}
\label{raion-pdf}
\end{figure}

\section{Графики и обсуждение}

\begin{figure}[!h]
\begin{center}
\includegraphics[width=0.8\hsize,height=0.7\hsize]{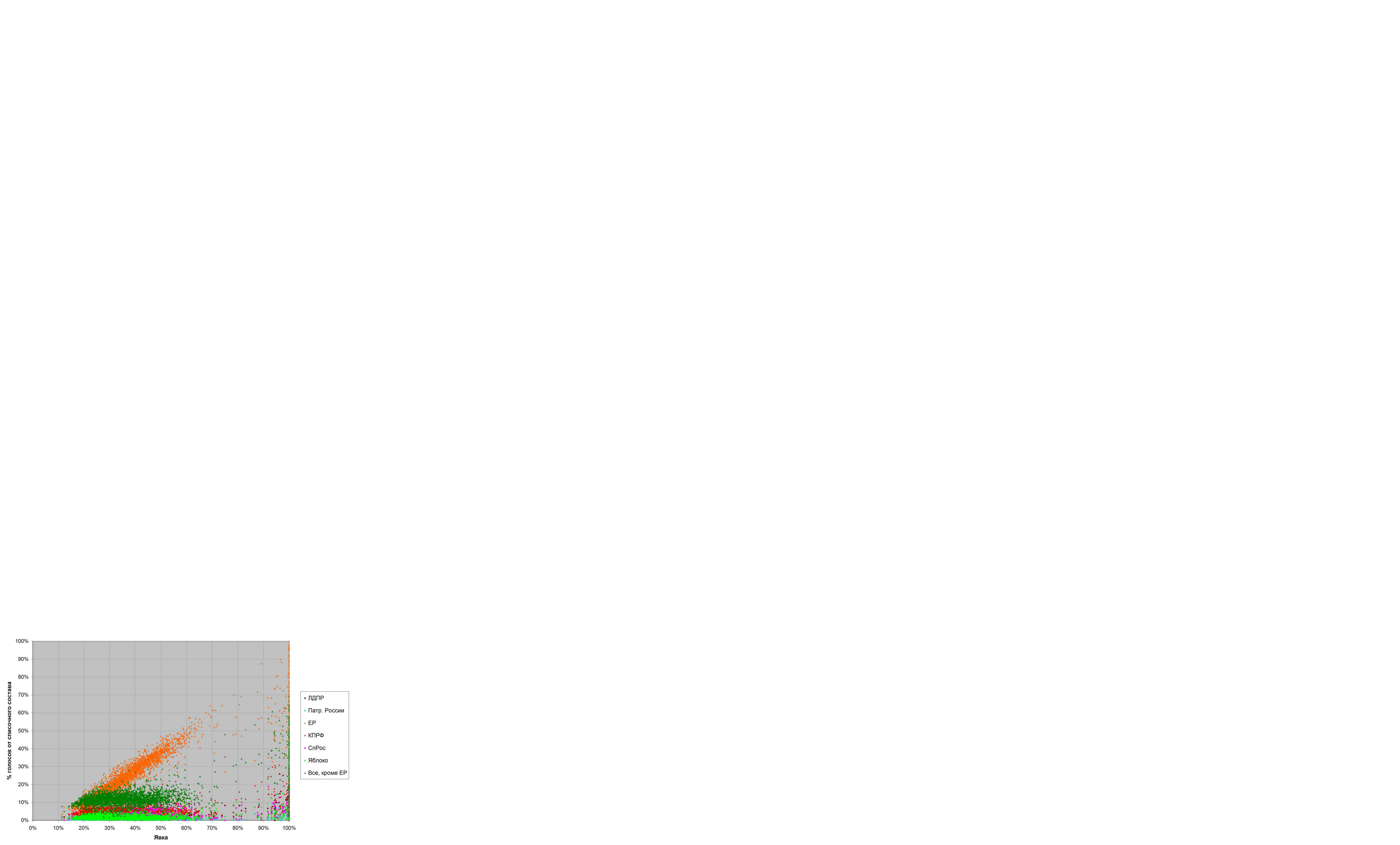}
\end{center}
\caption{Данные по участкам~\cite{podmoskovnik-56185}; добавлена сумма всех партий, кроме <<Единой России>>; рисунок растянут по вертикали, чтобы сделать масштабы по осям примерно равными.}
\label{uchastok-pdf}
\end{figure}

Данные голосования на выборах\footnote{Мы употребляем для краткости термин <<выборы>> наравне с безоценочным термином <<голосования>>.} в Мосгордуму 2009 года были опубликованы в интернете (на официальном сайте избиркома~\cite{izbirkom}), и всякий желающий мог их списать и анализировать. Авторы <<Живого Журнала>> (livejournal.com) построили на основе этих данных графики (рис.~\ref{raion-pdf},~\ref{uchastok-pdf}) и высказали предположение о том, что на этих выборах на этапе подсчёта голосов имели место подтасовки. (Другие этапы тоже любопытны, но заслуживают отдельного обсуждения.) Мне кажется, что этот случай (<<казус>>, как говорят юристы) интересен и с точки зрения знакомства с окружающей действительностью, и как иллюстрация некоторых простейших методов математической статистики.

Предположение о подтасовках (фальсификациях) обсуждалось в различных печатных и сетевых публикациях, в том числе критически (см., например, забавную паническую отповедь~\cite{anonim}).

\section{О статистических доказательствах}

Может ли вообще статистика что-то доказать? Нет, если мы понимаем доказательство <<математически>> (как сказал бы Достоевский)~--- как нечто, что не может быть опровергнуто никакими новыми данными и наблюдениями. Сразу по многим причинам:

\begin{itemize}
\item \textbf{Вероятность и достоверность}. Допустим, вы играете в казино в рулетку (будем для простоты считать выигрыш и проигрыш равновероятными) и проигрываете двадцать раз подряд. Можно ли считать доказанным, что казино жульничает? Большинство скажет, что да: первый проигрыш случается в половине случаев, два проигрыша~--- в четверти случаев (половина из этой половины), три проигрыша~--- в одной восьмой и так далее, так что двадцать проигрышей подряд~--- это примерно один шанс на миллион, пренебрежимо мало. С другой стороны, если в игре участвуют миллионы людей, скорее всего с кем-то такое случится и при честной игре со стороны казино. (См. обсуждение в~\cite{uss,shen-philosophy,borel}; о простейших понятиях теории вероятностей можно прочесть в~\cite{shen-popular}.)

\item \textbf{Априорные вероятности гипотез}. Пытаясь оценить (по наблюдениям) различные варианты объяснений (статистические гипотезы, как говорят), мы исходим из некоторых начальных представлений о их правдоподобии. В теории вероятностей это называют <<формулой Байеса>>, которую можно пояснить на таком примере. Пусть при болезни А симптом Х появляется в $90\%$ случаев, в при болезни Б~--- в $10\%$ случаев. Врач видит больного с симптомом X. Что более вероятно~--- болезнь А или Б? На первый взгляд~--- А, но на самом деле ответ зависит от <<априорных вероятностей>>: если болезнь А редкая (один случай на миллион), а Б~--- более частая (один случай из тысячи), то среди обладателей симптома X преобладают жертвы болезни Б (посчитайте в качестве упражнения, во сколько раз их больше: должен получиться ответ $111\frac{1}{9}$).

\item \textbf{Математические модели}. Наконец, выбор математической модели ещё ничего не означает. Скажем, в нашем первом примере можно обвинить владельцев казино в жульничестве. Но тот же самый результат может быть следствием, скажем, проседания грунта, в результате чего первоначально горизонтальная рулетка наклонилась. В этом случае злого умысла нет, а есть халатность~--- хотя математически ситуации неотличимы.

\end{itemize}

Всё это к тому, что от статистического анализа нельзя ждать окончательных ответов (тем более в <<социальных науках>>). Легенда про Колмогорова говорит, что после его первого научного доклада (на семинаре по русской истории) ему сказали, что его доказательство убедительно, но этого мало, хорошо бы иметь и независимые подтверждения~--- после чего он занялся наукой, где таких независимых подтверждений не требуется (математикой). Так и здесь~--- статистические наблюдения имеют смысл лишь в сочетании с другими обстоятельствами, и при этом желательно понимать самому, в чём именно состоит статистическая аргументация (а не полагаться на специалистов)

\section{Что изображено на графиках}

Представим себе, что на одном из участков явились все $100\%$ избирателей, и из них $60\%$ проголосовали за партию А, а $40\%$~--- за партию Б. Изобразим эти результаты на графике (рис.~\ref{elections-pdf}) в виде двух точек: по горизонтали отложим явку в диапазоне от нуля то $100\%$, а по вертикали~--- доли сторонников. (Партия А на цветном рисунке изображается синим, а Б~--- красным.) На другом участке, допустим, при явке в $50\%$ соотношение сил такое же: $60\%$ из пришедших проголосовали за А, остальные $40\%$ за Б. Это означает, что $30\%$ списочного состава проголосовали за А и $20\%$ списочного состава проголосовали за Б.

\begin{figure}[!h]
\begin{center}
\includegraphics[scale=1]{elections-1.mps}
\end{center}
\caption{Представление данных на графике}
\label{elections-pdf}
\end{figure}

На таком графике

\begin{itemize}
\item каждому участку соответствует набор точек (по одной каждого цвета~--- для каждой партии):
\item сумма вторых координат этих точек (суммарная доля проголосовавших за какую-либо из партий) не больше первой координаты (явки); равенство означает, что не было унесённых или недействительных бюллетеней.
\end{itemize}
Можно было бы составить другой график, откладывая по вертикальной оси процент сторонников от числа явившихся (как это обычно делается на выборах). Но наш график более наглядно отражает некоторые линейные закономерности, как мы увидим.

Именно так построены графики рис.~\ref{raion-pdf} и \ref{uchastok-pdf}.
Но перед тем, как внимательно смотреть на эти графики, полезно потренироваться и ответить на несколько контрольных вопросов.

\begin{itemize}
\item Как будет выглядеть график, если на всех участках явка одна и та же (скажем, $50\%$)? [Ответ: все точки будут на вертикальной прямой с абсциссой $50\%$.]
\item Как будет выглядеть график, если явка всегда составляет $100\%$ и все голосуют за одну партию? [Ответ: точки для этой партии будут в правом верхнем углу $(100\%, 100\%)$, точки для остальных~--- на оси абсцисс под ними.]
\item Как будет выглядеть график, если явка колеблется от $0$ до $100\%$, но все приходящие голосуют за одну партию? [Ответ: точки  этой партии лягут на прямую, идущую из начала координат под углом $45^\circ$, точки других партий будут на оси абсцисс.]
\item Как будет выглядеть график, если явка колеблется от $0$ до $100\%$, но все приходящие забирают бюллетени с собой? [Ответ: все точки будут на оси абсцисс.]
\end{itemize}

Теперь мы уже готовы посмотреть на график рисунка~\ref{raion-pdf}.
Вот что можно заметить, глядя на него:

\begin{itemize}

\item Явка колеблется примерно от $20\%$ до $55\%$.
\item Светло-коричневые точки (партия <<Единая Россия>>) образуют облако, близкое к наклонной прямой.
\item Тёмно-красные точки (КПРФ) образуют облако, близкое к горизонтальной прямой.
\item Во всех районах результат ЕР больше результата КПРФ (точки ЕР выше точек КПРФ с той же абсциссой);
\item Разрыв этот тем больше, чем больше явка: на результат КПРФ явка влияет мало (облако почти горизонтально), а на результат ЕР существенно: повышение явки с $20\%$ до $30\%$ приводит к увеличению проголосовавших за ЕР почти на те же $10\%$ списочного состава: от $10\%$ до $20\%$; дальнейшее повышение явки на $10\%$ до $40\%$ даёт рост ещё почти на $10\%$ до $30\%$, и так далее: прямая идёт под наклоном чуть меньше~$45^\circ$.

\end{itemize}

Аналогичный вид имеет и график по участкам (рис.~\ref{uchastok-pdf}), но есть и некоторые отличия:

\begin{itemize}

\item Есть небольшое количество точек с большой явкой и большим разбросом результатов. (Это не удивительно, так как есть участки с небольшим числом голосующих, скажем, в больницах, и там результаты сильно зависят от местных случайностей. На общих результатах это сказывается мало из-за небольшого числа участников.)

\item Заметно, что количество проголосовавших за другие партии (в частности, КПРФ и <<Яблоко>>) не растёт с ростом явки, а даже, похоже, немного убывает при увеличении явки от $40\%$ до $60\%$). Это не относится к участку небольшой явки ($15$--$20\%$).

\end{itemize}

\section{Варианты объяснений}

Для начала обсудим, чем может быть вызван такой заметный разброс по явке. Его трудно списать на индивидуальные случайности~--- участки в основном большие (тысячи человек), и если бы, скажем, каждый бросал монетку, чтобы решить, идти или не идти на выборы, то колебания явки были бы порядка $5\%$ (как учит теория вероятностей, размах колебаний убывает пропорционально $1/\sqrt{N}$ при $N$ независимых испытаниях). Но в принципе ничего удивительного в разбросе нет: если участок в неудобном месте, или на дверях подъездов не было объявлений, то можно ожидать существенно меньшего числа голосующих. Если же, скажем, избирательные комиссии, как в советское время, ходили по квартирам напоминать, голосующих могло быть больше. Бывают избирательные участки в общежитиях.

Удивительно другое: что этот фактор (или комбинация факторов) действует так избирательно: почти все дополнительно пришедшие, как обнаруживается из графика, голосуют за << Единую Россию>>. Как это можно объяснить?

Одно из возможных объяснений такое: реально пришло народу гораздо меньше, чем утверждается (около $20\%$, как в районе начала графика), а в результатах голосования были подтасовки: были вброшены дополнительные бюллетени за <<Единую Россию>> или просто изменены цифры в протоколах. (Если допустить саму возможность вброса, то не удивительно, что вбрасываются бюллетени за <<партию власти>>.) Небольшое уменьшение в правой части, правда, так не объясняется~--- тут надо дополнительно предположить, что в некоторых местах был не только вброс, но и перераспределение голосов.

Альтернативное объяснение состоит в поиске возможного фактора, который создаёт разброс в явке на участках и действует только на голосующих за <<Единую Россию>>. Например, можно считать, что сторонники других партий тверды в своём выборе (пойти голосовать за эти партии или бойкотировать выборы) и призывы принять участие в голосовании на них никак не действуют, но сильно влияют на участие в выборах голосующих за <<Единую Россию>>. Как указывалось в обсуждениях в Livejournal~\cite{rico}, <<подобный эффект должен получаться автоматически при наличии инертной массы, которая поддерживает едро, и очень активных сторонников других партий, которых, однако, очень мало. Тогда явка зависит в основном от того, насколько у едра получится отмобилизовать своих избирателей. Ну и процент за едро тоже будет расти от этого>>. (Но надо ещё предположить, что нет сторонников других партий, которые ленятся идти на выборы, но сходят, если про это им будут напоминать~--- ведь напоминают всем подряд, а не только любителям <<едра>>.  Важное обстоятельство, которое не всегда замечают: тот факт, что сторонники одних партий ходят на выборы чаще, чем сторонники других, сам по себе лишь меняет соотношение голосов, но не объясняет разброса явки по участкам.)

Другой комментатор~\cite{ivalnick} предлагает возможный конкретный механизм:  <<Имея опыт организации розничной торговли и общественного питания на избирательных участках на выборах в Санкт-Петербурге в 1997 году, могу сказать, что явка избирателей на участках, где все было организовано по высшему разряду (100 грамм + бутерброд за смешные деньги) была в три раза выше, чем на участках, где избирателям предлагали только чай и булочки>>. Чтобы это объяснение сработало, надо ещё предположить, что любители <<Единой России>> падки на халявную закуску и выпивку (что ещё можно себе представить) и что сторонники других партий (в том числе коммунистической!) к ней совершенно равнодушны (что выглядит менее правдоподобно).

Ещё одно возможное объяснение~--- принудительное голосование за <<Единую Россию>> больших групп с понуждением их к конкретному выбору. (Но это уже приближается к вбросам, особенно если одни и те же люди голосуют многократно или голосуют за других.)

Наконец, можно предположить, что избирательные участки Москвы отличаются по составу населения и их можно разбить на группы, для каждой из которых характерна своя явка и свои распределения предпочтений по кандидатам. (Сходная гипотеза была выдвинута в \cite{cik-arlazarov}). Но тогда остаётся без объяснения, почему все эти гипотетические группы хорошо укладываются на одну прямую (наклонную для <<Единой России>> и почти горизонтальные для остальных). См. ниже обсуждение статьи~\cite{cik-arlazarov}.

\section{Обсуждение гипотез}

Мы видим, что оценивать правдоподобность предлагаемых объяснений можно по-разному, и представляют интерес другие данные, которые могут поддержать ту или иную гипотезу. Среди них есть и статистические данные, и, что называется, \eng{anecdotal evidence}.

\subsection{Неожиданные пики}

\begin{figure}[!h]
\begin{center}
\includegraphics[width=0.9\textwidth]{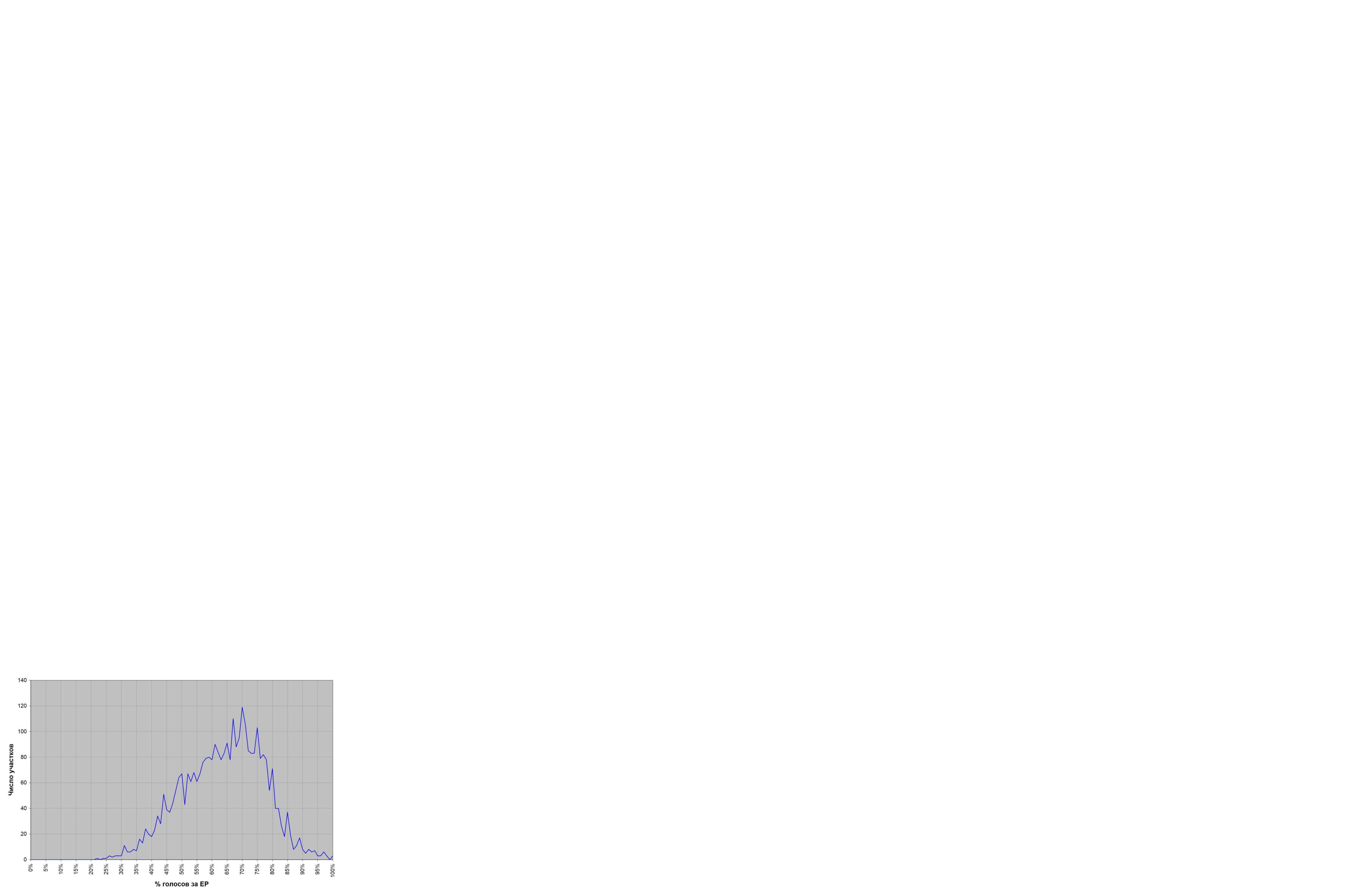}
\end{center}
\caption{Количество участков, где <<Единая Россия>>
получила данный процент голосов (гистограмма), выборы в Мосгордуму 2009~года~\cite{piki-resultat}}
\label{piki-resultat}
\end{figure}

На рис.~\ref{piki-resultat} показана так называемая <<гистограмма>>: для каждого числа от $1$ до $100$ по вертикали отложено количество участков, на которых был именно такой процент голосов за <<Единую Россию>>. Видны пики на $70\%$, $75\%$, $80\%$, $85\%$, которые можно объяснить наличием <<контрольных нормативов>>, которым старались следовать при подтасовках. В принципе, конечно, это можно считать и случайным совпадением.

\begin{figure}[!h]
\begin{center}
\includegraphics[width=0.9\textwidth]{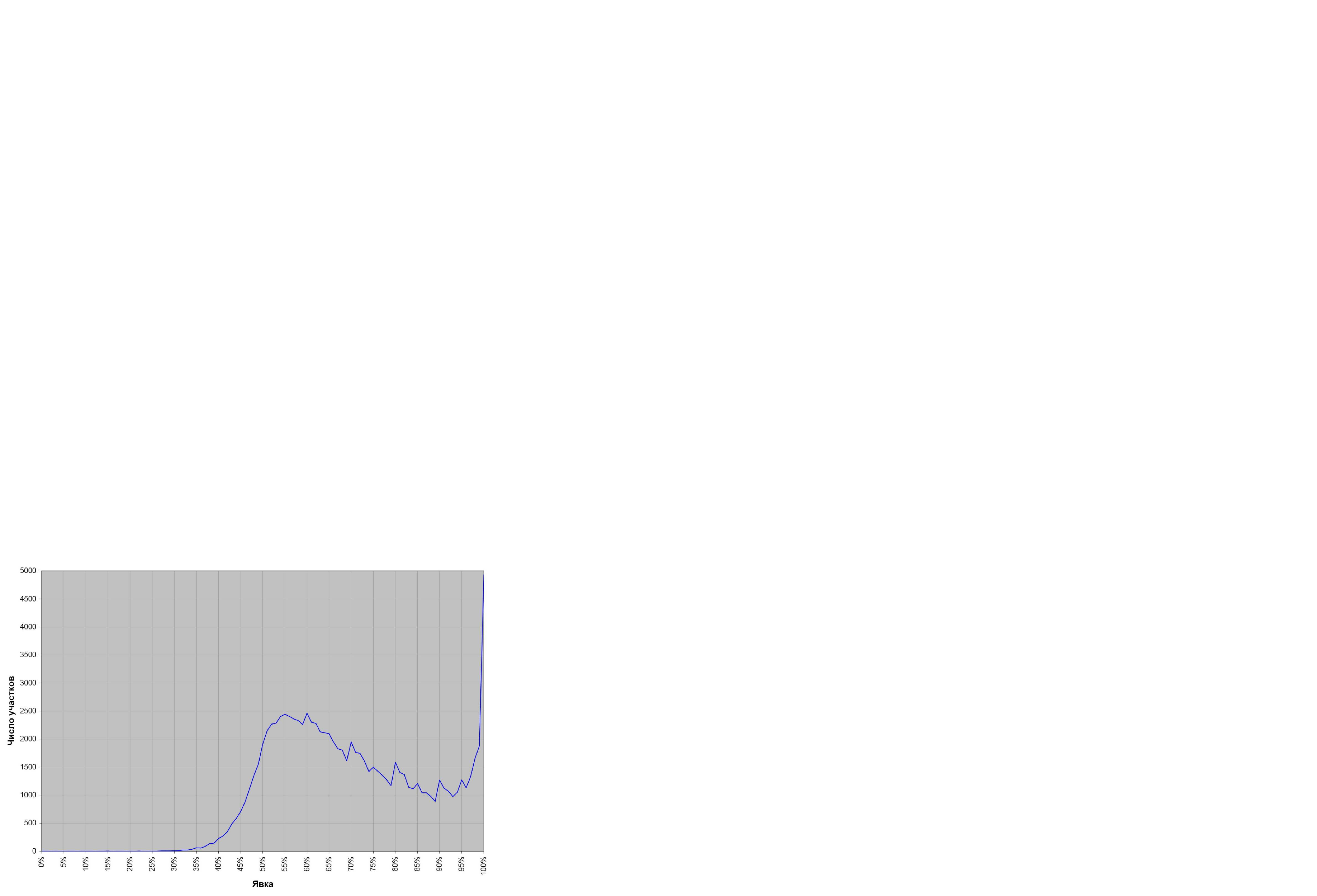}
\end{center}
\caption{Количество участков, где зарегистрирована данная явка
 (гистограмма), выборы в Госдуму 2007~года~\cite{piki-yavki}}
\label{piki-yavki}
\end{figure}

\begin{figure}[!h]
\begin{center}
\includegraphics[width=0.9\textwidth]{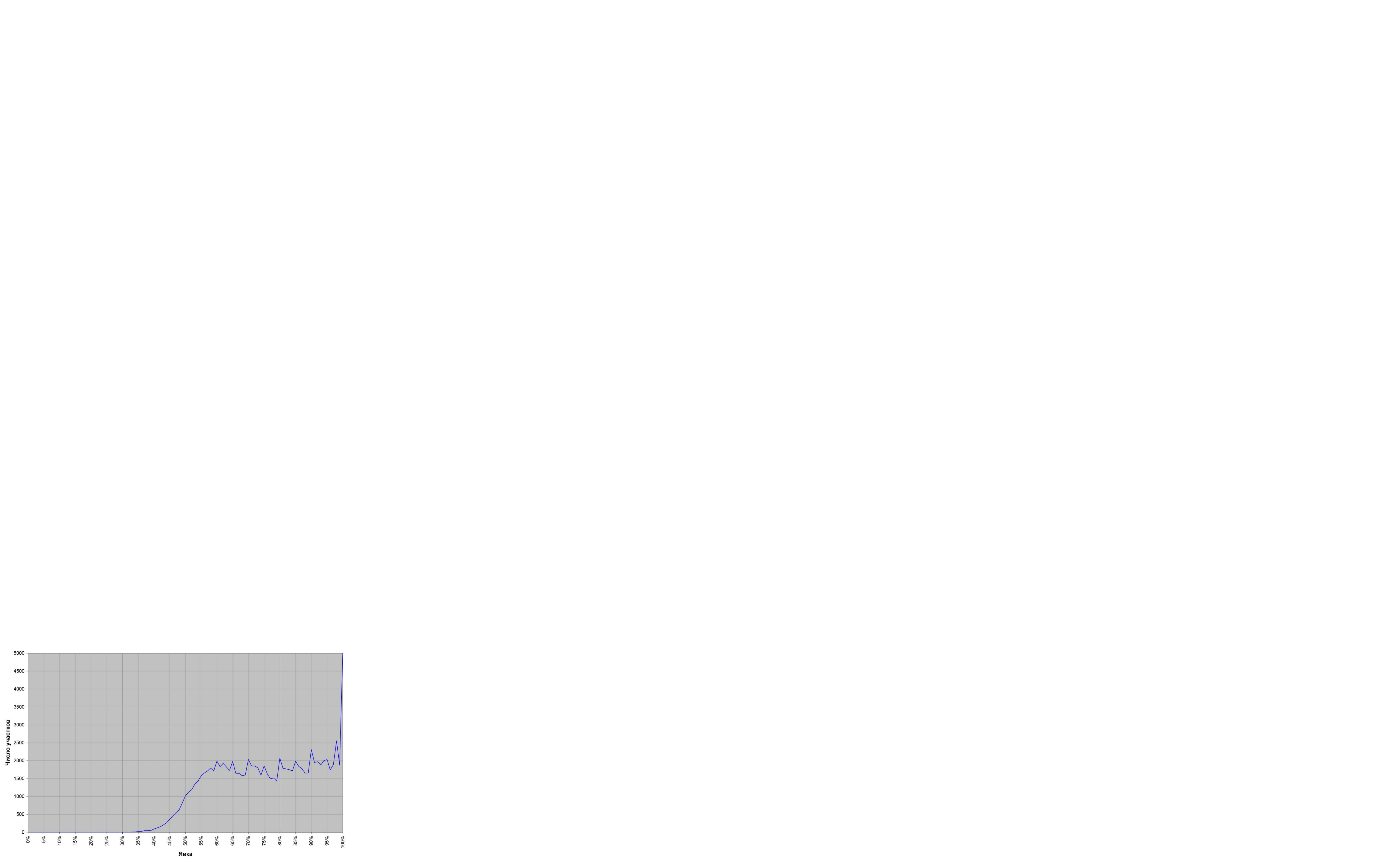}
\end{center}
\caption{Количество участков, где зарегистрирована данная явка
 (гистограмма), выборы президента 2008~года~\cite{piki-yavki-2}}
\label{piki-yavki-2}
\end{figure}

Любопытно сравнить это с аналогичной гистограммой для <<явки>> на выборах 2007~года в Госдуму (рис.~\ref{piki-yavki}) и для выборов президента в 2008~году (рис.~\ref{piki-yavki-2}). Там тоже заметны характерные пики на <<круглых>> числах, но для явки, а не для процента <<Единой России>>.

Интересно сравнить это с комментариями А.\,Шмелёва~\cite{shmelev}, где он говорит о недостатках последних выборов по сравнению с предыдущими (но не в Госдуму, а весной 2008~года):

\begin{quote}
Не знаю как насчет статистики, но то, что на прошлых московских выборах не было таких вбросов~--- знаю точно. Там перед комиссиями ставилась другая задача: явку повысить. Поэтому даже там, где вбросы делались, фиктивные бюллетени, как правило, заполнялись за всех кандидатов пропорционально. Ну, и число этих вброшенных бюллетеней было несоизмеримо меньше~--- в среднем по $50$ на участок\ldots

Вероятно, некоторые комиссии и тогда проявляли <<усердие не по разуму>>, но разнарядка тогда была исключительно на явку, а не на результаты. Уж поверьте мне~--- сведения совершенно надёжные.
\end{quote}

(Идея расчёта и пропорционального заполнения фиктивных бюллетеней на первый взгляд кажется слишком сложной для реализации; оценка в $50$ фиктивных бюллетеней на участок не очень согласуется со статистикой, см. ниже о КОИБ.)

\subsection{Данные  КОИБ}

\begin{figure}[!h]
\begin{center}
  \includegraphics[width=0.8\textwidth]{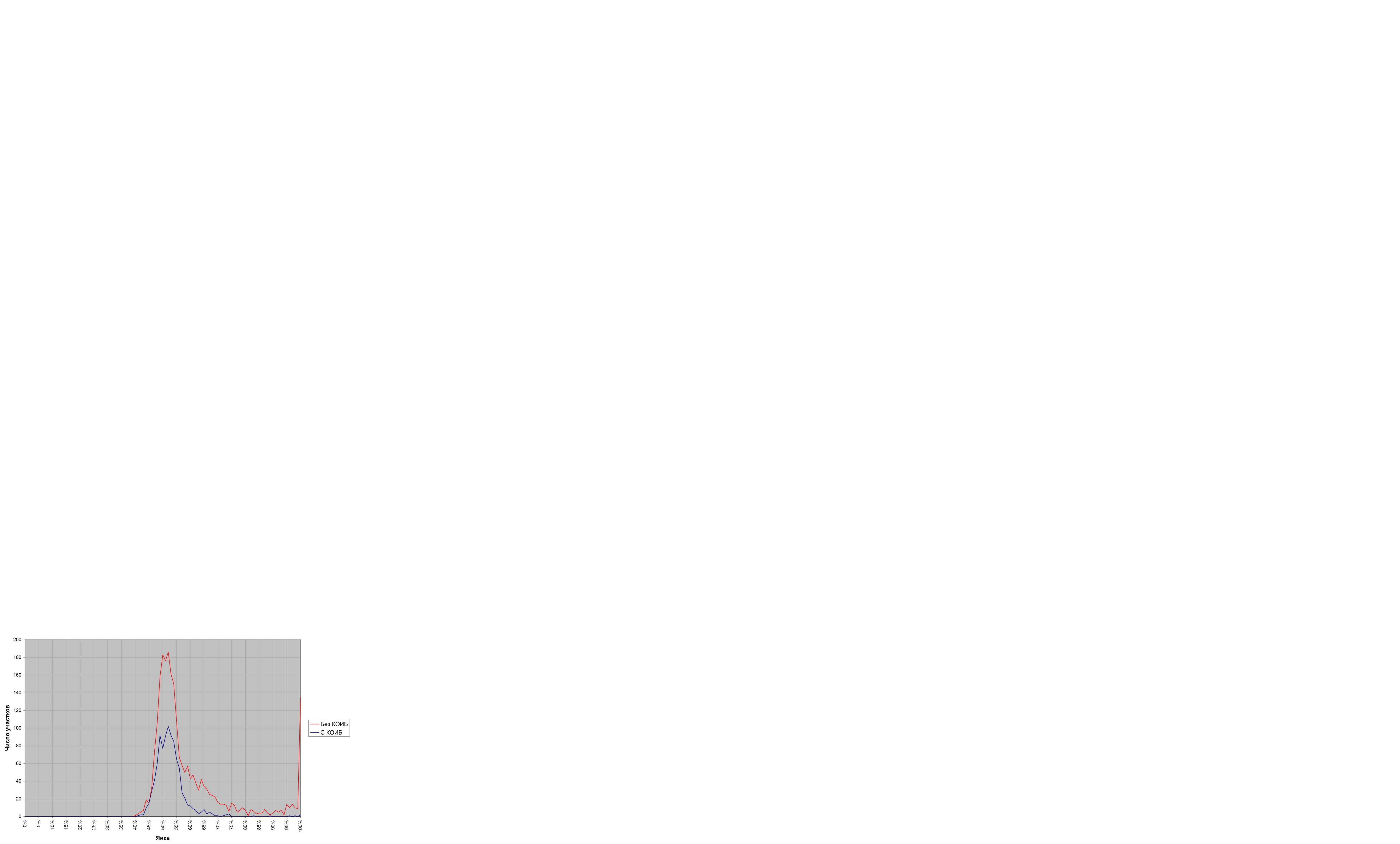}
\end{center}
\caption{Гистограмма явки \emph{по Москве} с КОИБ и без в 2007 г., выборы Госдумы}
\label{2007-KOIB}
\end{figure}

\begin{figure}[!h]
\begin{center}
  \includegraphics[width=0.8\textwidth]{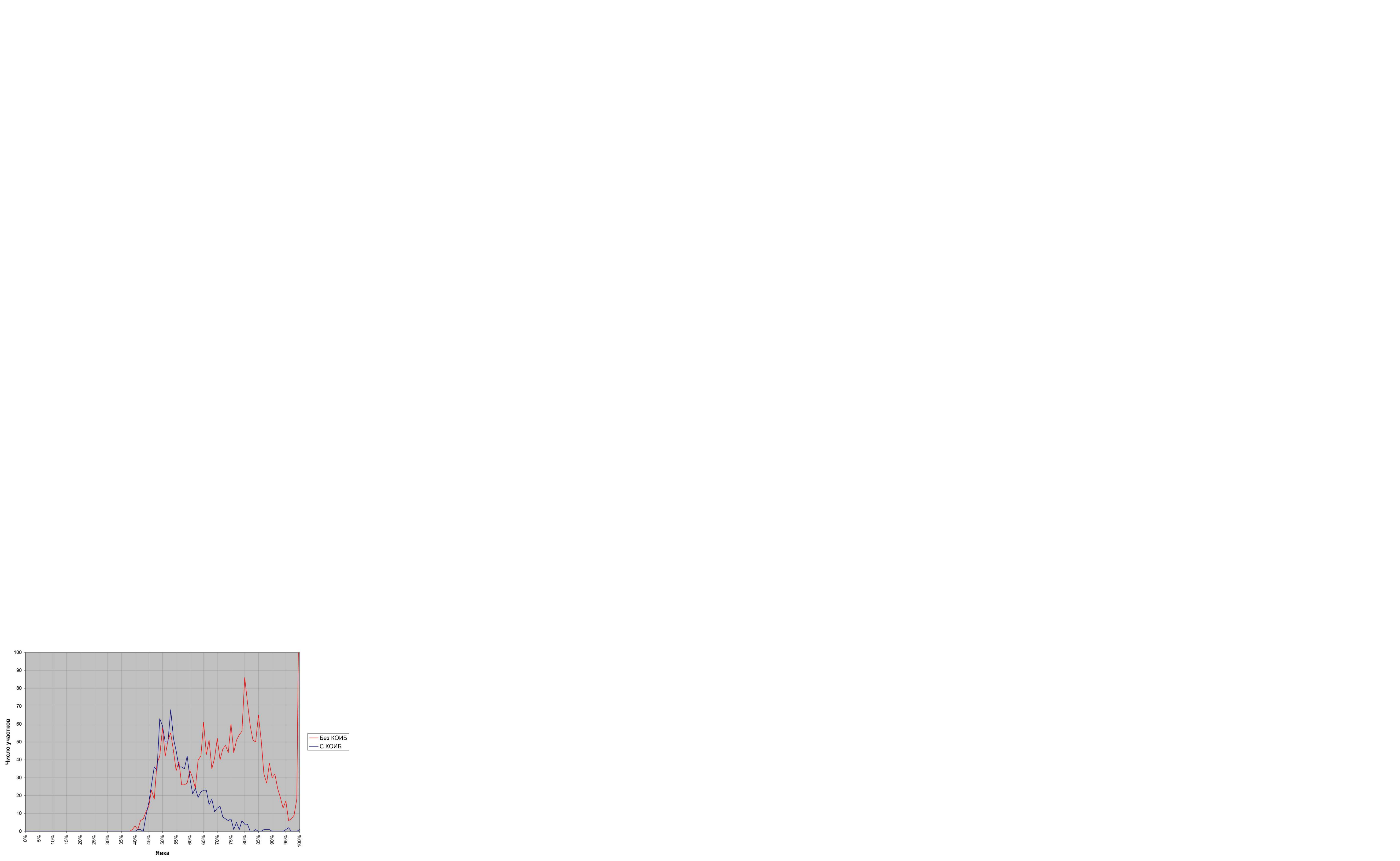}
\end{center}
\caption{Гистограмма явки \emph{по Москве} с КОИБ и без в 2008 г., выборы президента}
\label{2008-KOIB}
\end{figure}

\begin{figure}[!h]
\begin{center}
  \includegraphics[width=0.8\textwidth]{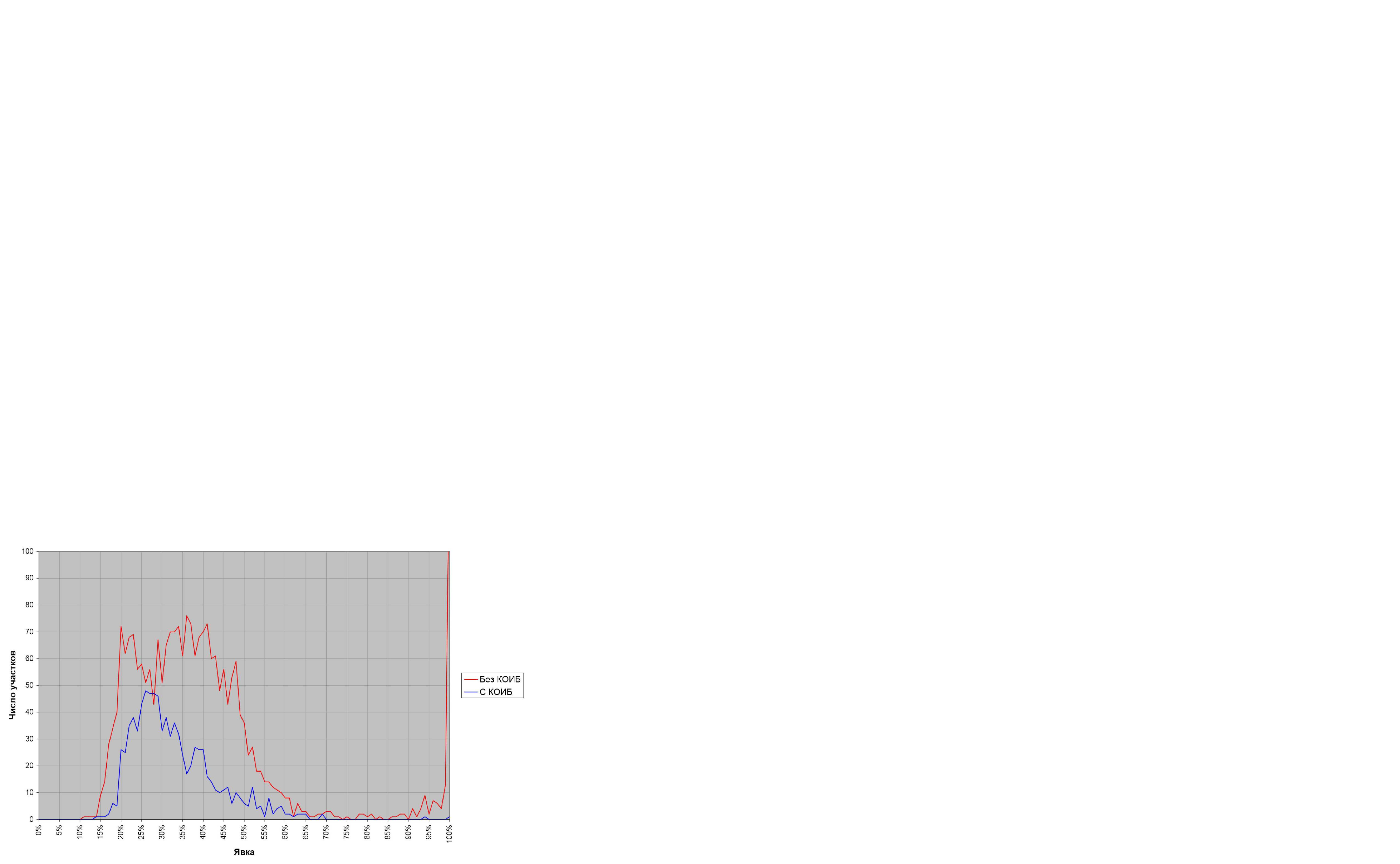}
\end{center}
\caption{Гистограмма явки по Москве с КОИБ и без в 2009 г., выборы Мосгордумы}
\label{2009-KOIB}
\end{figure}

КОИБ~--- это комплекс обработки избирательных бюллетеней, электронное устройство, которое прикрепляется к урне и сканирует бюллетени, избавляя от необходимости ручного подсчёта.  Такие устройства использовались на части избирательных участков в Москве. Поскольку избиратель видит это устройство, уже придя на выборы, само их наличие должно мало влиять на явку (разве что какой-то любитель новой техники пришёл специально на них посмотреть). Тем не менее гистограммы явки на участках с ними и без них существенно отличаются (рис.~\ref{2007-KOIB}--\ref{2009-KOIB}): отсутствие КОИБ сдвигает график в сторону более высокой явки, наиболее это заметно на президентских выборах 2008 года. Это можно объяснить тем, что процедура фальсификации результатов усложняется, если при подсчёте применяется электронное устройство (надо пропускать бюллетени через него, а не просто увеличить цифру или вбросить пачку). Другого объяснения этого эффекта, кажется, не предлагалось; предположение о том, он объясняется концентрацией КОИБ в районах с особым составом жителей, не подтверждается при сравнении данных по участкам с КОИБ и без в одном районе или в соседних районах с близким составом жителей~\cite{buzin}.

\subsection{Сравнение со статистикой других выборов}
В работе~\cite{ordeshook} приводятся гистограммы явки по различным российским выборам, причём отдельно по областям и по национальным республикам. Выясняется, что их вид бывает разным: по областям в 1990-е годы графики были близки к гауссову распределению (<<колокол>>), а по республикам (а также по областям в более позднее время) отчётливо наблюдается аномальный хвост в сторону высоких явок, примерно как на рисунках~\ref{piki-yavki}--\ref{piki-yavki-2}.

\begin{figure}[!h]
\begin{center}
\includegraphics[width=0.8\textwidth]{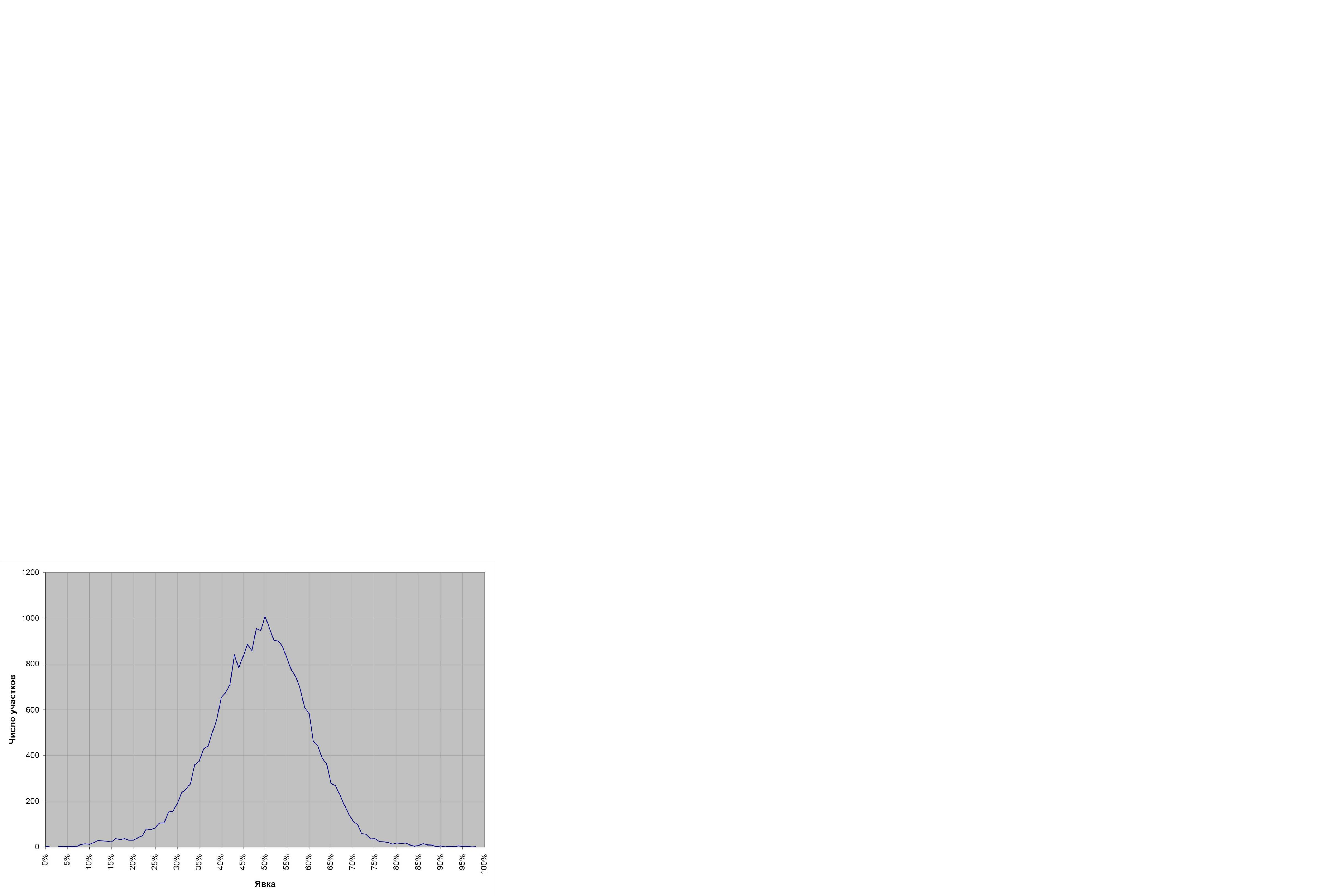}
\end{center}
\caption{Гистограмма явки на выборах 2005~года в Польше}
\label{2005-Poland}
\end{figure}

Для сравнения можно также взять гистограмму явки на выборах 2005~года в Польше (рис.~\ref{2005-Poland}), где тоже нет этого аномального хвоста.

\begin{figure}[!h]
\begin{center}
\includegraphics[width=0.8\textwidth]{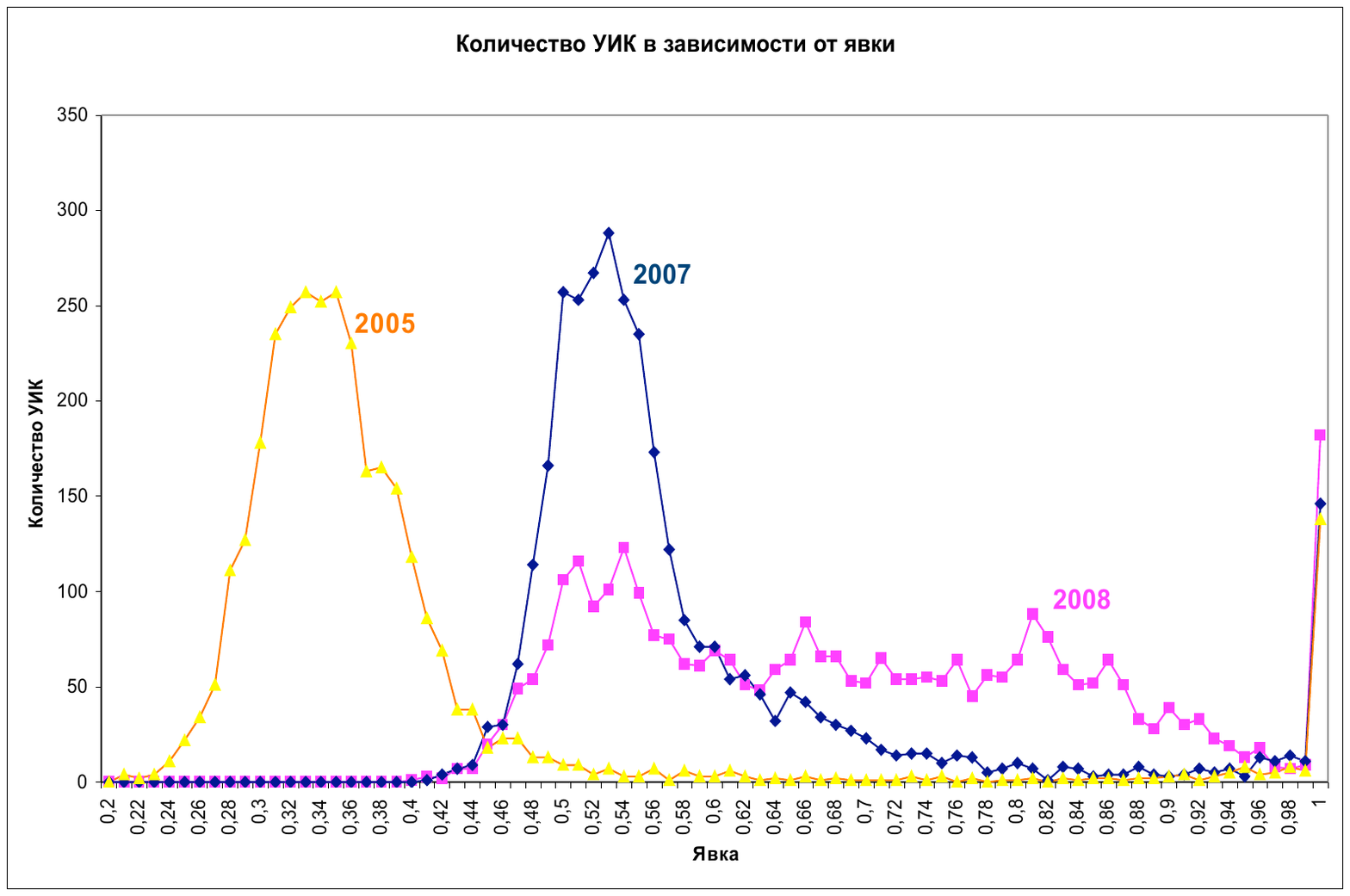}
\end{center}
\caption{Явка на трёх последовательных выборах в Москве}
\label{buzin2005-2008}
\end{figure}

Поучительно сравнение гистограмм явки за разные годы (рис.~\ref{buzin2005-2008} из~\cite{buzin}), которое решительно противоречит гипотезе о разделении участков на группы по социальному составу (выдвигаемой как объяснение в~\cite{cik-arlazarov}, см.~ниже): вряд ли такое разделение могло появиться в 2008~году, если его не было раньше.

\subsection{Отдельные случаи}

Рассматривая эти случаи, нужно иметь в виду, что каждый отдельный рассказ проверить трудно (если вообще возможно) и в принципе его можно считать выдумкой или предвзятым изложением. Тем не менее эти случаи могут рассматриваться как иллюстрации и дополнительное подтверждение гипотез, сделанных при рассмотрении статистических данных. (Здесь отобраны рассказы о конкретных случаях от их непосредственных участников, а не оценки и рассуждения общего характера.)

\begin{itemize}

\item По утверждению Ивана Большакова~\cite{bolshakov},  на участке 1702 партия <<Единая Россия>> получила $192$ голоса (скан копии протокола) при официальных данных в $742$ голоса (сайт избирательной комиссии); результаты остальных партий совпадают.

\item По утверждению Игоря Яковлева~\cite{yakovlev}, на участке 1701 число голосов <<Единой России>> при передаче данных в избирком было увеличено с $482$ до $617$ за счёт уменьшения голосов других партий (в том числе КПРФ: $121\to 81$, <<Справедливая Россия>>: $81\to 61$,  <<Яблоко>>: $25\to 5$).

\item Член избирательной комиссии Андрей Клюкин <<опровергает>> сделанные накануне (как он говорит, спьяну) заявления (пересказать трудно, это надо смотреть)~\cite{klyukin}.

\item Наблюдатель на участке 172 рассказывает о подтасовках при подсчёте числа выданных бюллетеней~\cite{dani444}. В комментариях рассказывается о согласовании результатов с территориальной избирательной комиссией в Хамовниках.

\item Другой рассказ о согласовании в территориальной избирательной комиссии и о голосовании за других~\cite{vreznikov}

\item Это уже не в Москве, а на одновременно проходящих выборах в Азове: видео скандала с обнаружением подготовленных к вбросу бюллетеней~\cite{azov}.

\item Корреспондент журнала <<Русский Newsweek>> Мария Железнова рассказывает о том, как её выгнали из участковой избирательной комиссии после того, как она обнаружила вброс бюллетений за <<Единую Россию>> и написала жалобу~\cite{newsweek}.

\item На участке 192, где голосовал лидер <<Яблока>> Митрохин, не было обнаружено ни одного бюллетеня, поданного за <<Яблоко>>~\cite{yakovlev-mitrokhin}. Как сообщает РИА <<Новости>>~\cite{ria}, <<при повторном пересчете голосов 16 бюллетеней за ``Яблоко''\  были обнаружены в общей папке, где находились избирательные бюллетени, поданные за все партии, кроме ``Единой России''>>.

\end{itemize}

\subsection{Реакция Центризбиркома}

В качестве ответа на критику на сайте Центральной избирательной комиссии (Центризбиркома, ЦИК) размещено несколько статей~\cite{cik}, из которых (хотя бы отчасти) научный характер имеет только одна~\cite{cik-arlazarov} (отметим очевидный конфликт интересов: один из авторов, В.\,Чуров~--- председатель Центризбиркома). Её авторы пишут:

\begin{quote}
Обычно в статьях оппонентов дается анализ статистических данных по двум независимым параметрам, например, распределение количества участковых избирательных комиссий (УИК) в зависимости от явки и распределение голосов за кандидатов (партии) также в зависимости от явки. На основании анализа данных протоколов УИК, находящихся в открытом доступе [4], делается обычно один и тот же вывод: поскольку распределение (например, числа УИК в зависимости от явки) не соответствует нормальному закону распределения (закону больших чисел) относительно средней явки, то результаты выборов являются недостоверными.

Такой вывод можно было бы сделать, если бы мы имели дело с однородным массивом, как избирателей, так и УИК, которые, подчиняясь некоему случайному процессу, участвовали в выборах. Однако это не совсем так. Во-первых, сам процесс выборов отнюдь не случаен, т.к. слишком много факторов на него влияет [так в оригинале], а во-вторых, сам массив, как избирателей, так и УИК отнюдь не однороден.
\end{quote}

\begin{figure}[!h]
\begin{center}
  \includegraphics[width=0.8\textwidth]{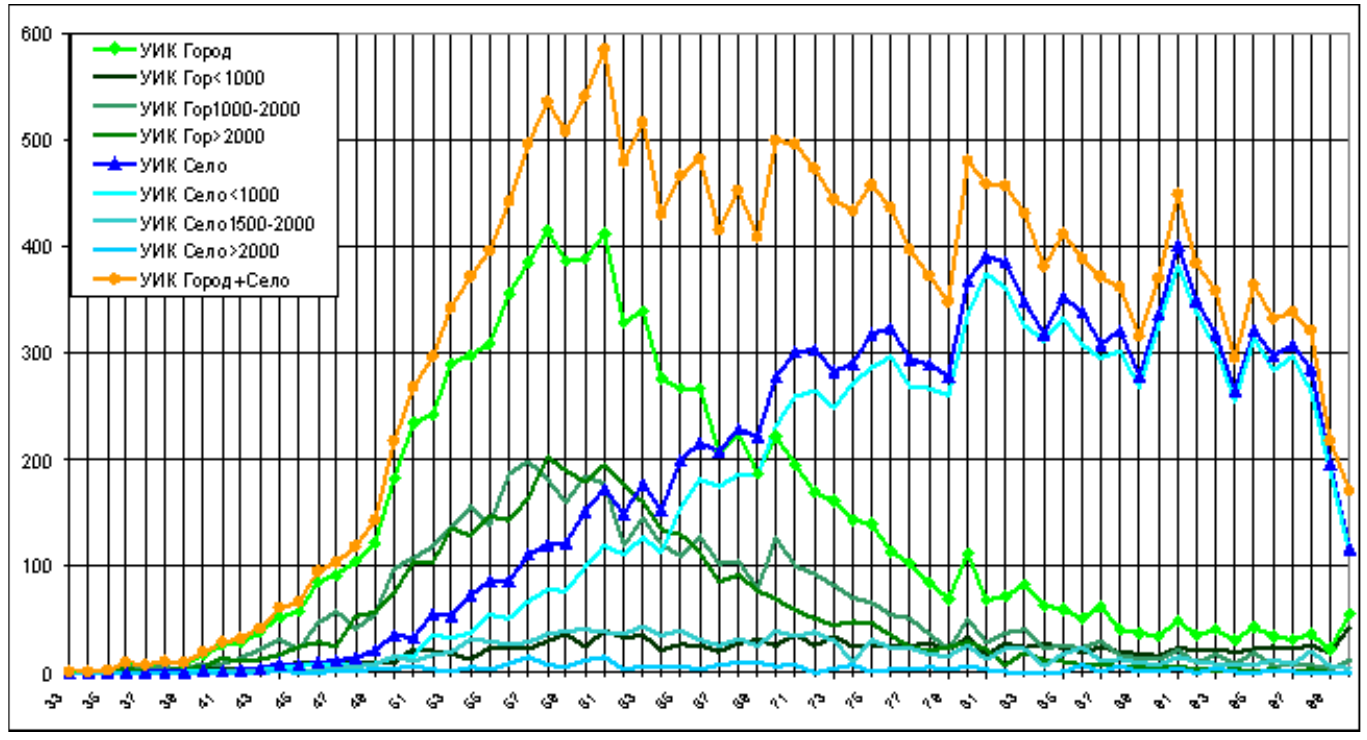}
\end{center}
\caption{Гистограмма явки на президенских выборах 2008 года по 25 регионам Европейской части России из~\cite{cik-arlazarov}, разделённая по типам участков.}
\label{arl1}
\end{figure}

Таким образом, авторы концентрируют внимание на одной претензии (характер распределения участков по явке) и предлагают объяснение: участки по своему месту нахождения делятся на несколько групп, и даже если в каждой группе распределение близко к нормальному, в итоговой гистограмме мы видим сумму. Это объяснение выглядит разумно и подтверждается приводимыми графиками отдельно для городских и сельских избирательных участков (воспроизводится здесь как рис.~\ref{arl1}) по 25 регионам европейской части России (конкретного перечня регионов в статье нет).

Однако авторы ничего не говорят о других странностях:

\begin{itemize}
\item Данные по Москве вряд ли можно объяснить разбиением на город и село, особенно с учётом различий по участкам с КОИБ и без них. Между тем в работах~\cite{buzin,shpilkin-golos,lyubarev}, цитируемых в~\cite{cik-arlazarov}, приводятся и данные по  КОИБ, и сравнительные данные за разные годы по Москве (см. рис.~\ref{buzin2005-2008}). Отсутствие какого-либо упоминания существенных возражений оппонентов и замена их данных на свои (без точного описания последних) производит впечатление научной недобросовестности.

\item Никакое разбиение не может объяснить пики на <<круглых числах>> в общей гистограмме явки (рис.~\ref{piki-yavki-2}); в графиках из~\cite{cik-arlazarov} похожие зубцы немного смещены, но сравнить и перепроверить результаты нельзя из-за отсутствия списка регионов, включённых в гистограмму. Отсутствие упоминания проблемы пиков и замена данных  усиливает впечатление научной недобросовестности.
\end{itemize}

\begin{figure}
\begin{center}
  \includegraphics[width=0.8\textwidth]{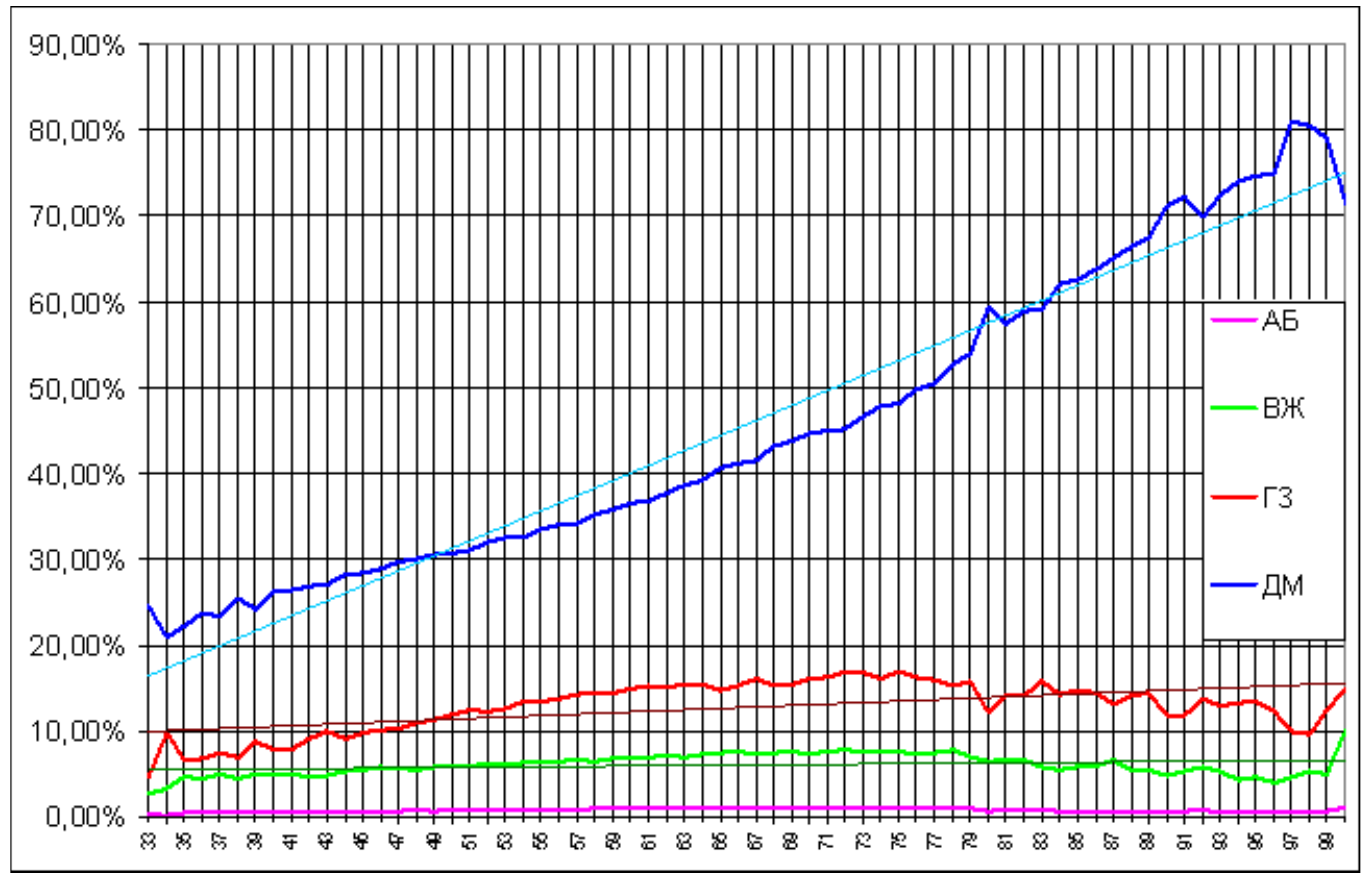}
\end{center}
\caption{График из~\cite{cik-arlazarov}, названный там <<зависимость распределения голосов по кандидатам в процентах от списочного состава УИК в зависимости от явки>>.}
\label{arl2}
\end{figure}

Кроме того, авторы~\cite{cik-arlazarov} без особых комментариев приводят ещё график, который, судя по всему, противоречит их собственным гипотезам (воспроизведённый здесь как рис.~\ref{arl2}). Они называют его <<зависимость распределения голосов по кандидатам в процентах от списочного состава УИК в зависимости от явки (город)>>.  Можно предположить, что он получен усреднением ординат в точках графика типа рис.~\ref{uchastok-pdf}, имеющим абсциссу в данном интервале (хотя от авторов научной статьи стоит ожидать более точного описания, не требующего догадок~--- заметим, что в статьях оппонентов, на которые они отвечают, приводятся корректные графики). Но если так, то это как раз противоречит идее авторов об однородности группы~--- поскольку в пределах этой группы воспроизводится странная закономерность рис.~\ref{uchastok-pdf} (рост по явке вызывает рост голосов за Медведева, но не за остальных, у которых даже видно некоторое падение).

\begin{figure}[!h]
\begin{center}
  \includegraphics[width=0.7\textwidth]{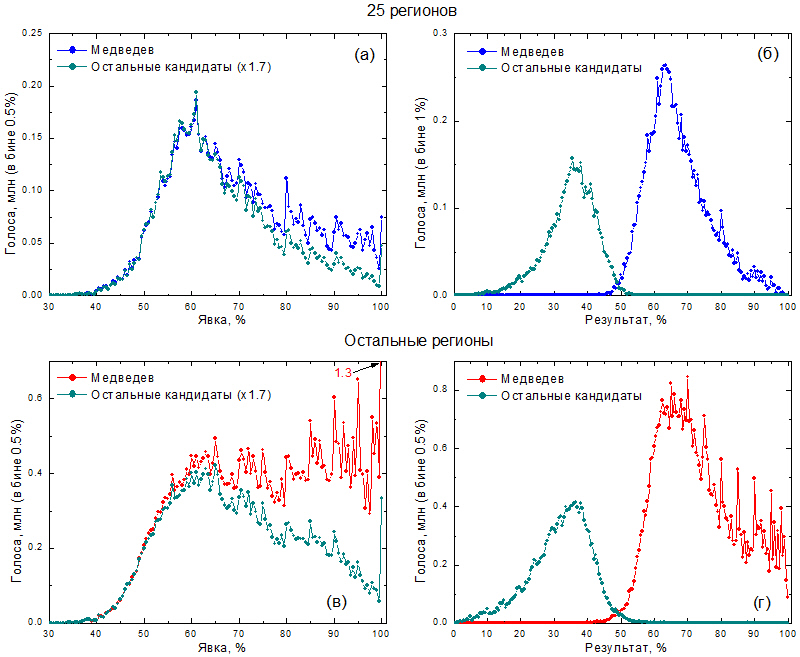}
\end{center}
\caption{Сравнение избранных 25 регионов из~\cite{cik-arlazarov} и остальных регионов России (\cite{pshenichnikov-churov}).}
\label{pshenichnikov-churov.jpg}
\end{figure}

[Дополнение 2018~года. Статья Чурова была подробно разобрана в~\cite{pshenichnikov-churov}. М.\,Пше\-нич\-ни\-ко\-ву удалось получить (у одного из авторов, Соловьёва) список из 25~регионов, которые рассматривались в статье, и он нарисовал гистограммы для этих 25~регионов и для всех остальных (рис.~\ref{pshenichnikov-churov.jpg}).]

\section{Заключение}

Вернёмся к исходному вопросу: доказывают ли статистические данные наличие подтасовок на выборах в Мосгордуму?  Со всеми сделанными оговорками о нематематическом характере вопроса лично я бы ответил на него так:

\begin{itemize}
\item по совокупности статистических и иных свидетельств гипотеза, объясняющая результаты выборов подтасовками, выглядит правдоподобной: \emph{If it looks like a duck, swims like a duck and quacks like a duck, then it probably is a duck};

\item другие гипотезы вызывают сомнение как сильными (и пока никак не обоснованными) предположениями, так и отсутствием объяснений для части данных;

\item в принципе, возможно, что новые данные изменят эту оценку, но пока сторонниками гипотезы честных выборов они не представлены и бремя доказывания лежит на них;

\item судя по статистике, масштабы подтасовок достаточно значительны (десятки процентов);

\item судя по статистике предыдущих выборов в последние годы, ситуация 2009 года не является исключительной, хотя похоже, что ранее подтасовки могли иметь меньший размер.

\item (last but not least) читатель может и должен составить своё собственное мнение по этому вопросу, используя приведенные ссылки на статистические данные и другую информацию.
	
\end{itemize}

\section{Добавление 2011 года}

В связи с <<выборами в государственную думу>> 2011 года возрос интерес к возможностям статистического выявления фальсификаций, а объявленные результаты этих выборов дали интересный материал для анализа. В этом и в следующих разделах мы воспроизведём некоторые наблюдения и замечания, касающиеся анализа этих выборов.

\subsection{Предостережение}

Начнём с предупреждения. В последней главе книги~\cite{myagkov-et-al} её авторы проверяют свои методы на тестовых данных с выборов в США (где, как они считают, фальсификации статистически значимого масштаба крайне маловероятны).

\begin{figure}[!h]
\begin{center}
  \includegraphics[width=0.8\textwidth]{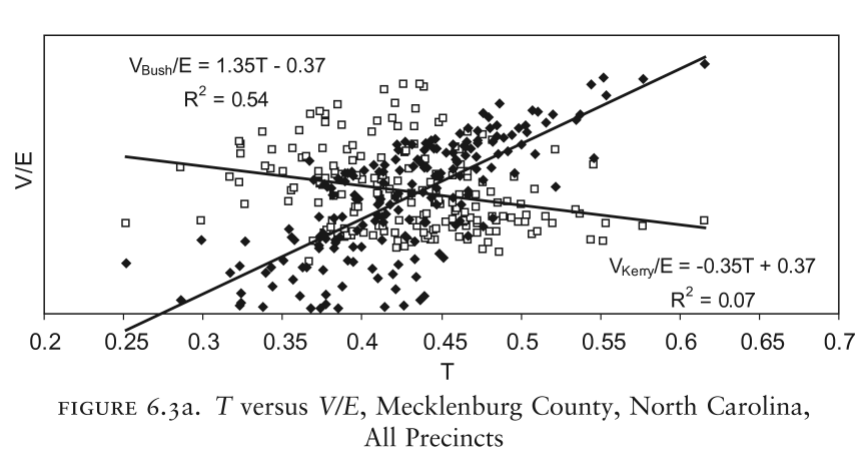}
\end{center}
\caption{(6.3a,~\cite{myagkov-et-al})}
\label{m6-3-a}
\end{figure}

Выясняется, что и здесь в отдельных случаях графики показывают парадоксальную зависимость доли голосов, поданных за одного из кандидатов (по отношению к общему числу избирателей в списках), от явки. Например, такова ситуация на рис.~\ref{m6-3-a} (голоса за Буша и Керри в одном из округов).

\begin{figure}[!h]
\begin{center}
  \includegraphics[width=0.8\textwidth]{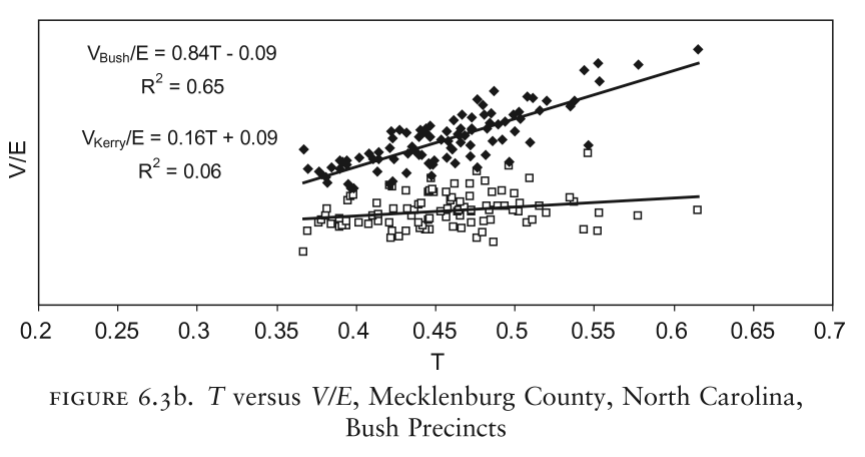}
  \includegraphics[width=0.8\textwidth]{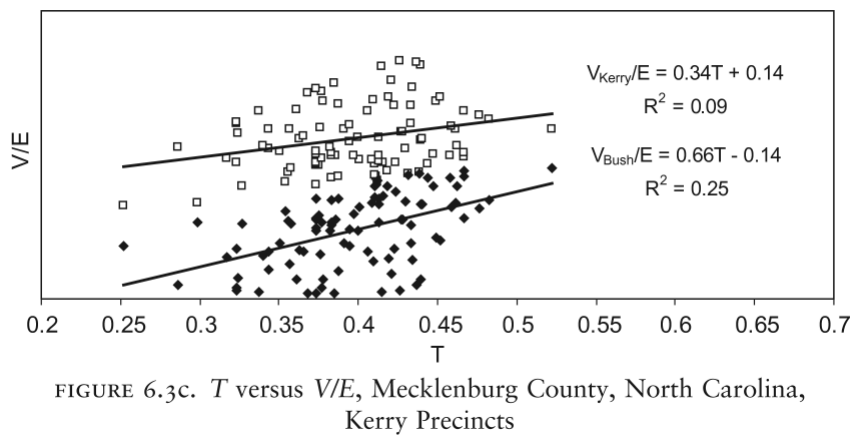}
\end{center}
\caption{(6.3b,c,~\cite{myagkov-et-al})}
\label{m6-3-bc}
\end{figure}

Как пишут авторы, ситуация частично может объясняться географической неоднородностью, и для сравнения приводят аналогичные графики отдельно по участкам, где победил один из кандидатов (рис.~\ref{m6-3-bc}), которые выглядят не так подозрительно.

Объяснение состоит в том, что это разделение предположительно соответствует разделению по каким-то социологическим параметрам, которые одновременно влияют и на явку, и на процент голосов за того и другого кандидатов. Разумеется, как пишут авторы, \eng{``\ldots division of precincts into those carried by one candidate versus another is but an admittedly crude first step in a more comprehensive analysis, especially if we suspect fraud. If, for instance, one candidate fraudulently manipulates votes sufficient to swing precincts from one column to another, then an analysis that merely divides precincts on the basis of who won what is necessarily compromised. It is more appropriate, then, to employ proxy variables --- socioeconomic characteristics --- that we know a priori correlate historically with both turnout and partisan preferences and which the candidates cannot manipulate''}.

\begin{figure}[!h]
\begin{center}
  \includegraphics[width=0.8\textwidth]{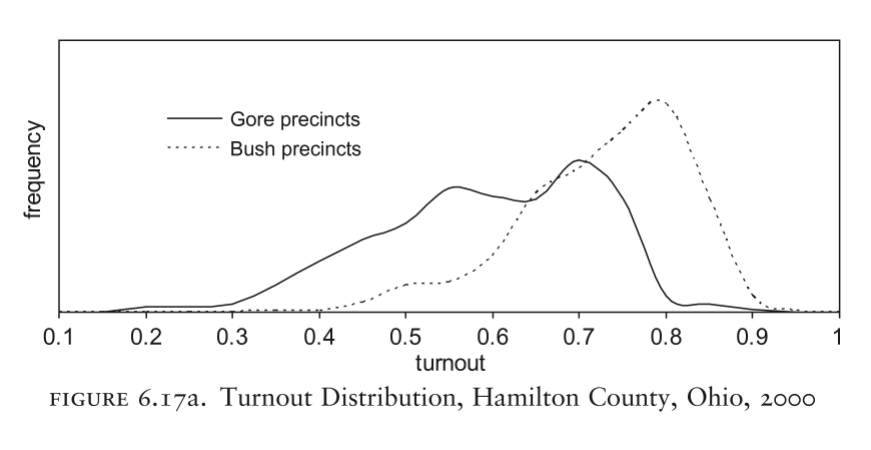}
\end{center}
\caption{(6.17a,~\cite{myagkov-et-al})}
\label{m6-17a}
\end{figure}

Другой разбираемый ими случай --- выборы на одном из участков штата Огайо. Тут (в отличие от предыдущего примера) и распределение по явке далеко от нормального (разве что пиков на круглых числах нет), как видно из рис.~\ref{m6-17a}, где показаны отдельно кривые явки для участков, голосовавших за того и другого кандидата.

\begin{figure}[!h]
\begin{center}
  \includegraphics[width=0.6\textwidth]{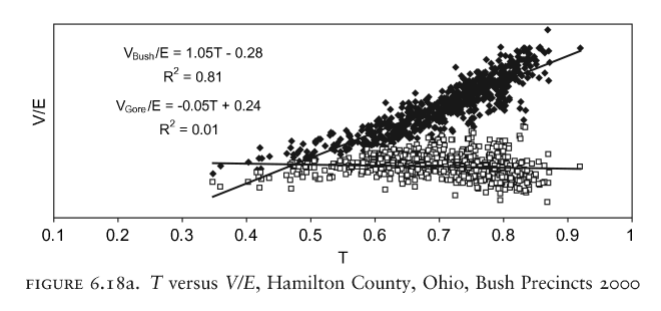}
\end{center}
\caption{(6.18a,~\cite{myagkov-et-al})}
\label{m6-18a}
\end{figure}

Распределение явки и голосов за кандидатов также выглядит подозрительно, даже после того, как выделены участки, голосующие за одного из кандидатов (Буша, см. рис.~\ref{m6-18a}).

Пытаясь понять, в чём тут может быть дело, авторы сравнивают результаты для выборов разных лет (1992, 1996, 2000, 2004) и отмечают, что и кривые явки, и распределения явки и голосов выглядят очень похоже для всех этих выборов. Чем же это можно объяснить? Авторы пишут: \eng{``There are two alternative inferences to be drawn from these figures: either Hamilton County's Democratic Party has been hypnotized, drugged, bought off, or otherwise rendered oblivious to significant and historically consistent fraud, or there is a benign explanation for the patterns our analysis reveals. In fact, there exists such an explanation; namely, the heterogeneity of precincts supporting
Republican candidates''}. Однако и после соответствующего разделения странности графиков остаются, хотя и в меньшей степени, и, как пишут авторы, \eng{``the more reasonable hypothesis is that there exists a further natural refinement of the data\ldots\ The critical word here is ``natural''. Fair and objective analysts are not free to simple regroup data at will in order to eliminate suspicions of fraud\ldots\ Minimally, at least, in any application of our indicators, the analyst needs to verify that there is not some natural and benign demographics explanation for the coefficients he or she recovers before proclaiming fraud''}. Короче, ситуация тут так и остаётся неясной.

\subsection{2011: статистические данные}

\begin{figure}
\begin{center}
  \includegraphics[width=0.7\textwidth]{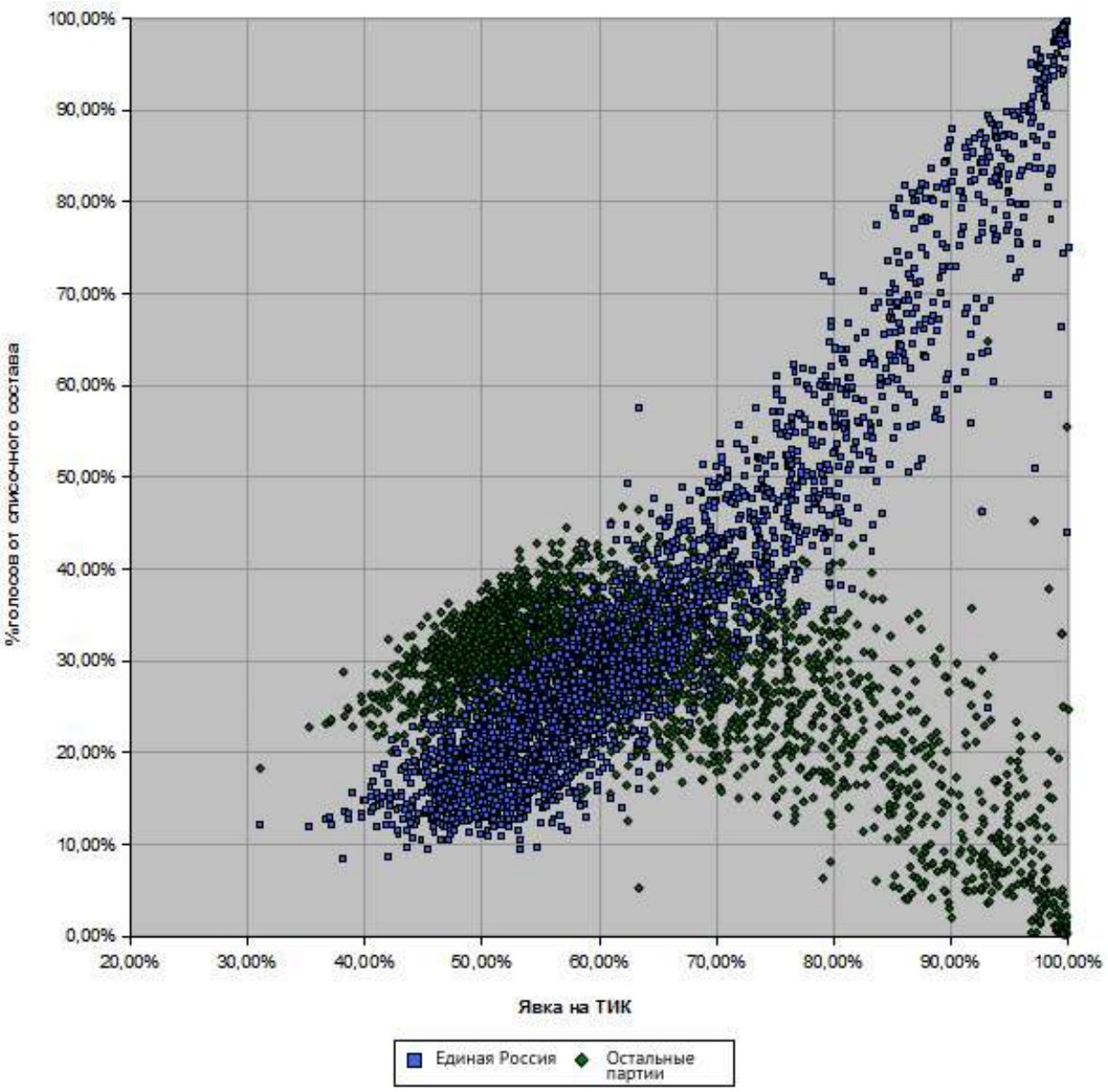}
\end{center}
\caption{Явка и голосование за <<Единую Россию>> в процентах от списочного состава по России}
\label{trv-2011}
\end{figure}

Сергей Шпилькин (тем же способом, что на прошлых выборах) нарисовал данные по территориальным избиркомам в координатах (явка, число голосов за <<ЕР>> в процентах от списочного состава). Получилась картинка, похожая на прошлые выборы (рис.~\ref{trv-2011}, см.~\cite{trv-2011}). Разница с предыдущими графиками в том, что доля голосов за <<ЕР>> растёт быстрее линейной, так что одними <<вбросами>> это объяснить нельзя; можно предположить, что часть голосов за <<ЕР>> была получена изъятием голосов, отданных за другие партии.

\begin{figure}[!h]
\begin{center}
  \includegraphics[width=0.4\textwidth]{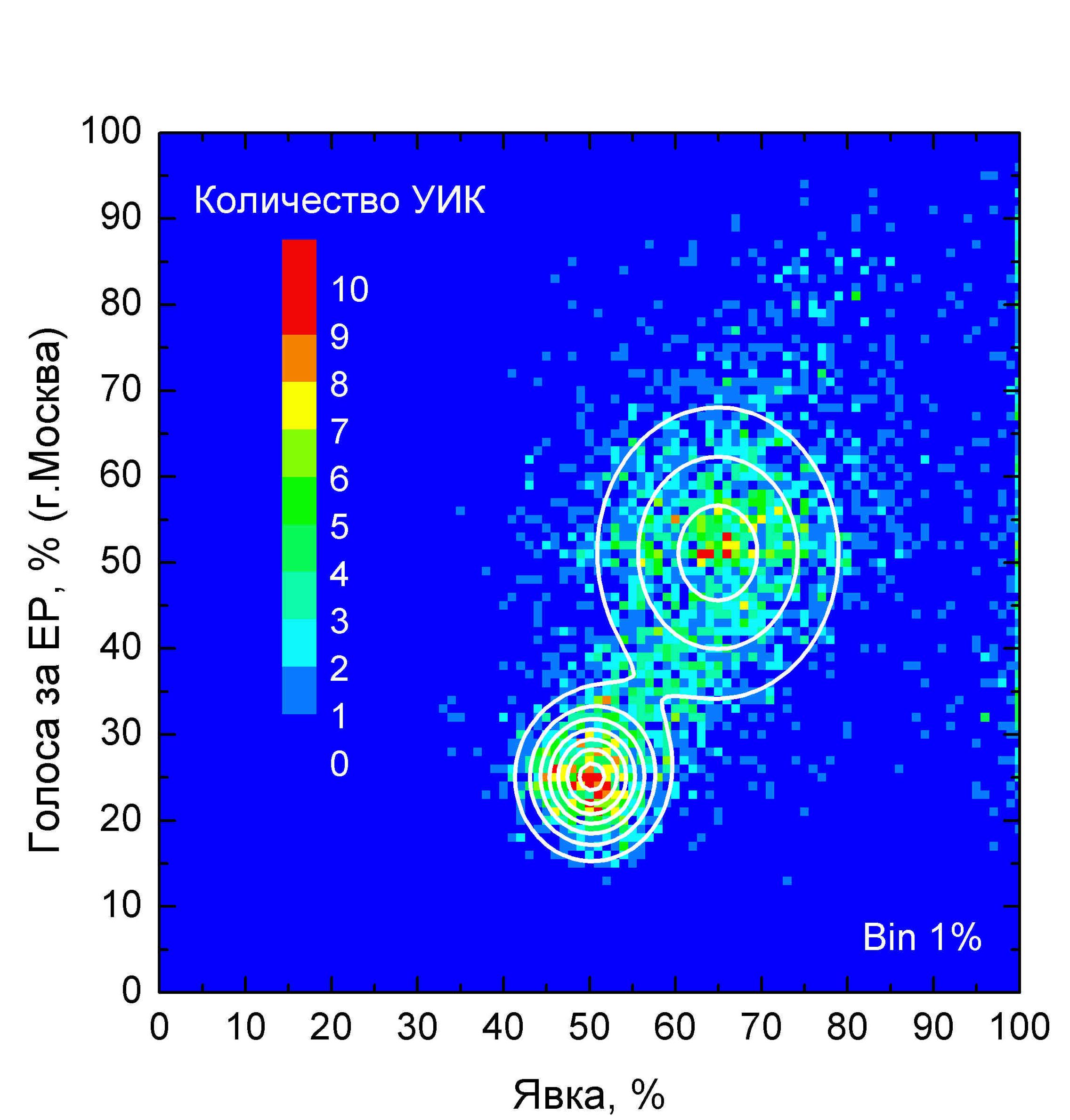}
\end{center}
\caption{Явка и голосование за <<Единую Россию>> по Москве}
\label{oude-moscow.jpg}
\end{figure}

Максим Пшеничников~\cite{oude-rus2011} предложить изобразить официальные результаты выборов в Москве на плоскости (явка, голоса за <<ЕР>> в процентах от общего числа голосов) и заметил, что они явно группируются в два кластера. См. рис.~\ref{oude-moscow.jpg} (белыми линиями показаны приблизительные места этих кластеров). Возникает вопрос: второй (верхний правый) кластер~--- это результат фальсификаций или следствие каких-то неизвестных факторов типа неравномерности распределения сторонников партий по Москве? Сравнение с данными наблюдателей и КОИБ (см. следующие разделы) подтверждает скорее первую гипотезу. [Она подтверждается также и сравнением с голосованием 2012~года, см. ниже.]

\begin{figure}[!h]
\begin{center}
  \includegraphics[width=0.4\textwidth]{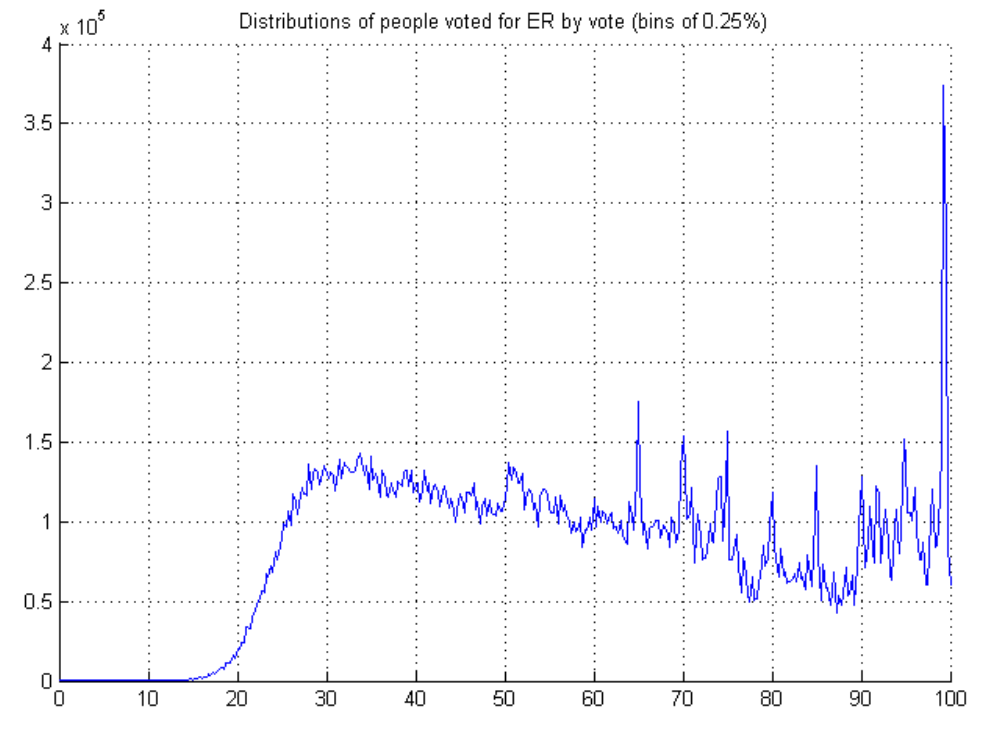}
\end{center}
\caption{Распределение людей по участкам в зависимости от доли голосов за <<ЕР>> на их участке.}
\label{electionpeaks5}
\end{figure}

В журнале Дмитрия Кобака~\cite{kobak} приводится много самых разных графиков, построенных автором журнала и его собеседниками. Вот, например, знакомые пики на круглых числах (рис.~\ref{electionpeaks5}, запись от 12 декабря 2011). [Предупреждение: в некоторых других картинках такого рода виден пик на 50\%; такой пик возникает за счёт дискретности значений простых дробей, если не принять специальных мер, даже в отсутствие фальсификаций. См. подробнее указанную запись и ссылки в ней, где обсуждается, почему так получается, а также~\cite{sergey-k}.]

Сводная таблица разных графиков (включая картинки по отдельным регионам) есть в том же журнале в записи от 6 декабря 2011, в которую добавлены ссылки на более поздние графики.

\subsection{2011: КОИБ}

\begin{figure}[!h]
\begin{center}
  \includegraphics[width=0.5\textwidth]{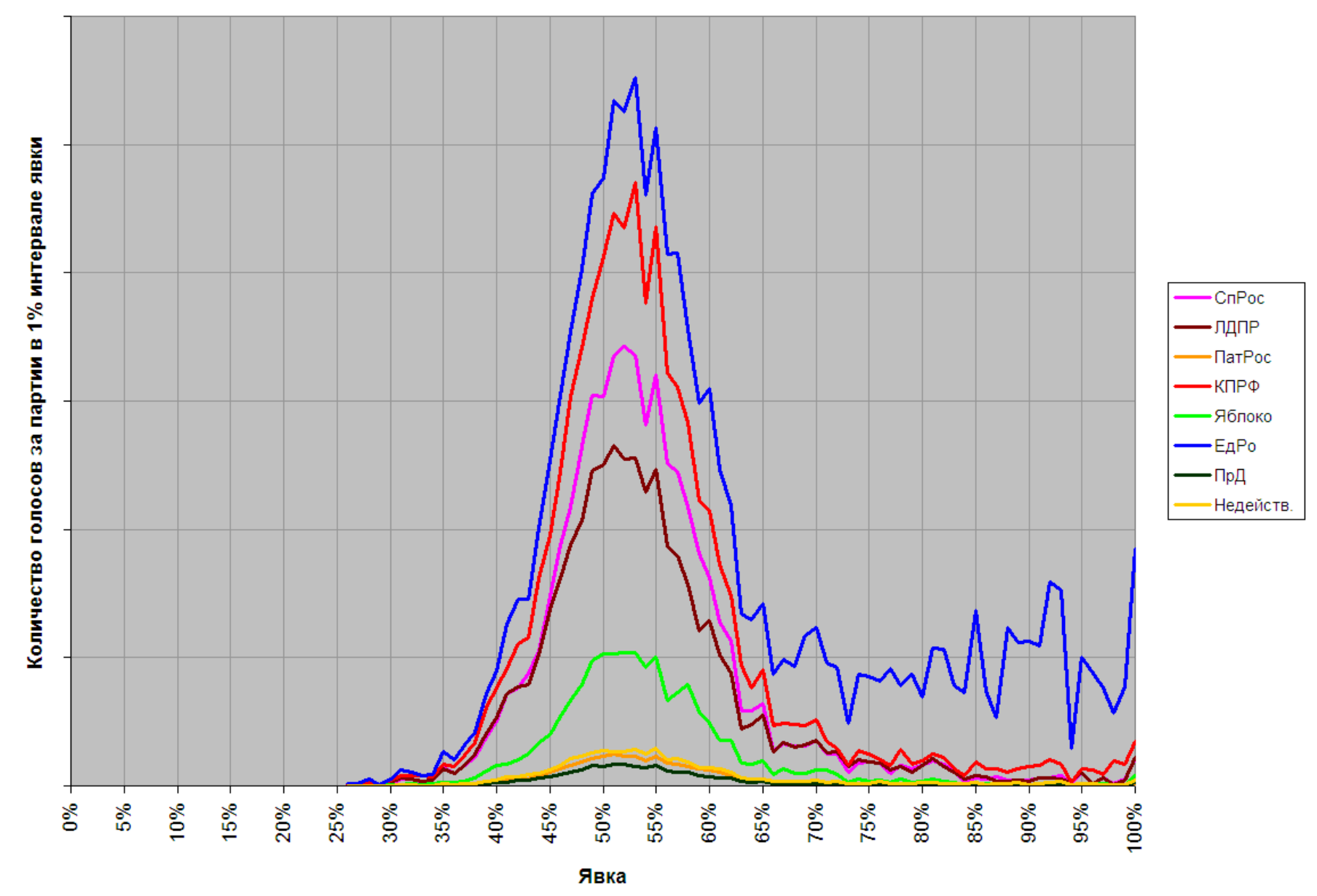}
\end{center}
\caption{Распределение голосов за партии для участков с КОИБ. Суммарные данные в процентах: СР/ЛДПР/ПР/КПРФ/Я/ЕР/ПД/нед. =17/14/1.3/23/5.2/37/0.8/1.6.
Явка 55\%.}
\label{withkoib}
\end{figure}

\begin{figure}[!h]
\begin{center}
  \includegraphics[width=0.5\textwidth]{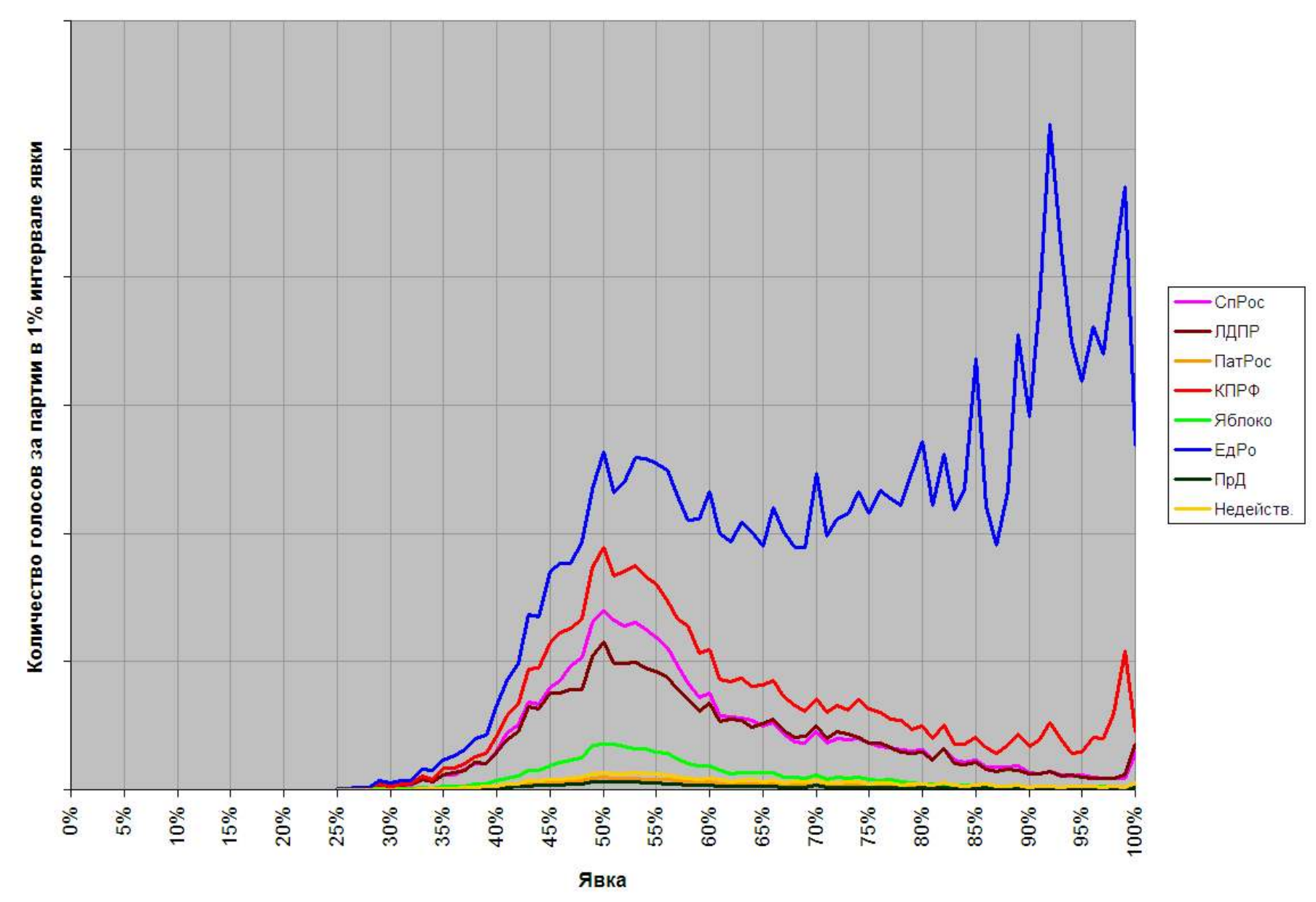}
\end{center}
\caption{Распределение голосов за партии для остальных участков тех же территориальных комиссий. СР/ЛДПР/ПР/КПРФ/Я/ЕР/ПД/нед. =12/10/0.9/18/2.7/54/0.6/1.3. Явка 63\%. Ср. утверждение председателя ЦИК В.\,Чурова: <<Кстати, на известных, опубликованных во многих СМИ графиках, смею вас заверить, все графики всех партий имеют признаки нормального распределения.>>~\cite{churov-echo}.}
\label{nokoib}
\end{figure}

Сергей Шпилькин нарисовал количество проголосовавших за разные партии (в однопроцентных интервалах явки) для участков с КОИБ (тех, которые ему удалось выделить), а для сравнения взял гистограммы для других (без КОИБ) участков тех же территориальных комиссий~\cite{shpilkin-koib}. Логично предположить, что что состав избирателей должен быть близким. Но гистограммы получились разными: гистограмма для участков с КОИБ имеет лишь небольшой <<хвост>> (рис.~\ref{withkoib}), в то время как на гистограмме для участков тех же ТИК без КОИБ хвост очень большой (рис.~\ref{nokoib}).

Надо иметь в виду, что КОИБы распределены неравномерно, поэтому результаты на этих участках нельзя использовать в качестве реконструкции результатов голосования; интересно их соотношение с участками рядом.

Много любопытного обнаруживается и при сравнении аналогичных данных для отдельных территорий. Такие примеры во множестве приведены в~\cite{kireev}. Запись 12 декабря: выясняется, что в Ставрополе на участках с КОИБ явка 45\% и <<ЕР>> получает 34\%, а на участках без КОИБ соответственно 61\% и 55\%. Теоретически это можно пытаться объяснять особенностями распределения КОИБ по участкам (высказывались даже забавные предположения о том, что <<редкие аппараты наверняка установили в районах, где меньше вероятность, что их попортят~--- то есть там, где как раз и живёт основной яблочный электорат>>~\cite{sl-lopatnikov}), но это можно проверить, сравнив результаты с выборами 2007 года: на \emph{всех} участках с КОИБ результат уменьшился по сравнению с прежним на 23\%--36\%, в среднем на 29\%, а на участках без КОИБ~--- в среднем на 7\%.

В Москве КОИБы стояли в некоторых районах (не во всех участках, но во многих), в этих районах <<ЕР>> по официальным данным получила 30\% при среднем по Москве 47\%. Чтобы понять, не связано ли это со специальным выбором районов, можно сравнить с прошлыми выборами~--- видно, что дело не в этом (см. таблицу в~\cite{kireev}, запись от 11 декабря 2011). Ещё интересно, что в районах без КОИБ <<Яблоко>> потеряло 23\% прежнего результата, а в районах с КОИБ увеличило свою долю.

В Астраханской области официальное среднее для <<ЕР>> 60\%, а на участках с КОИБ~--- всего лишь 37\%~\cite{faereas}.

В журнале~\cite{kireev} приведены многие другие подтверждения фальсификаций, основанные на анализе результатов с КОИБ (рост явки в течение дня в Магнитогорске, данные по Юргинскому району Тюменской области, а также данные по Нижнему Новгороду, к которым мы ещё вернёмся). Там приводятся также замечательные результаты по Стерлитамаку (не имеющие отношения к КОИБ): на большинстве участков явка и доля голосов за <<ЕР>> чудесным образом оказываются близкими к 75\% (отклонение в пределах 1--2\%).

\begin{figure}[!h]
\begin{center}
  \includegraphics[width=0.6\textwidth]{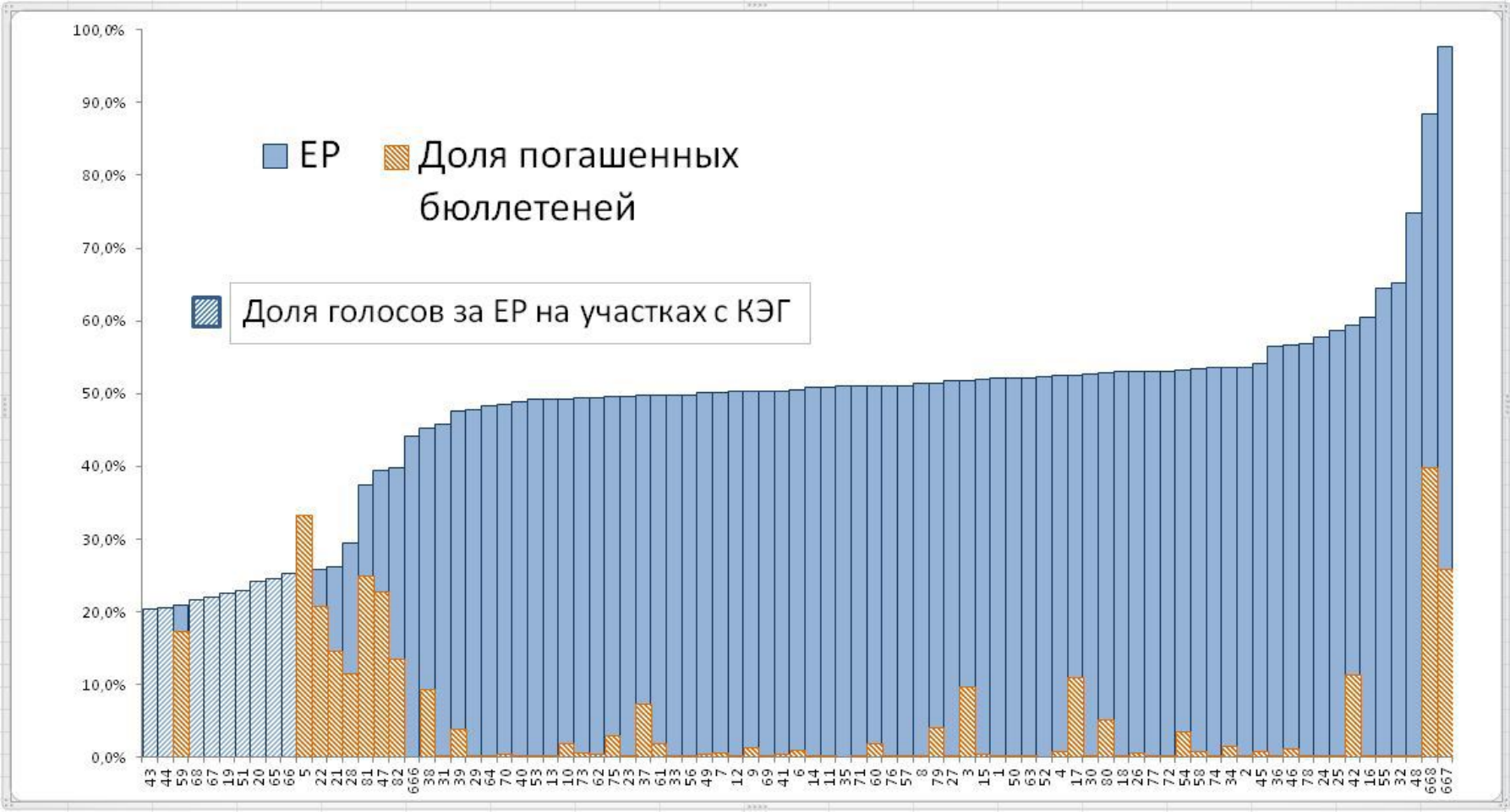}
\end{center}
\caption{Доля голосов за <<ЕР>> на различных участках в Сыктывкаре. <<Бледные голубые столбцы~--- автоматизированные УИКи>>~\cite{buzin-komi}.}
\label{syktyv}
\end{figure}

Помимо КОИБ, на некоторых участках встречались <<комплексы электронного голосования>> (КЭГ), где вообще не было бюллетеней, а голосование было электронным. Интересный график~\cite{buzin-komi}  построил А.~Бузин по Сыктывкару (рис.~\ref{syktyv}): на нём видно, что на автоматизированных участках доля голосов за <<ЕР>> от 20\% до 30\% от общего числа избирателей в списке, что существенно (несколько десятков процентов) меньше, чем в большинстве остальных.

\subsection{2011: наблюдатели}

На выборах 2011 года по призыву ассоциации <<Голос>>~\cite{golos} и проекта <<Гражданин наблюдатель>>~\cite{nabludatel} многие записались в наблюдатели на избирательные участки и попытались предотвратить нарушения и получить копию протокола~--- среди них как минимум пять человек, которых я лично знаю (и которые никогда раньше такого не делали). Вот отчёты моих знакомых из Москвы (в случае отсутствия жалоб указан процент <<ЕР>> в итоговом протоколе с участка).

\begin{itemize}
\item
\url{mccme.livejournal.com/11963.html} (В.М.\,Слободкин, вбросы и др.)
\item
\url{mccme.livejournal.com/11182.html} (В.\,Радионов, нет жалоб, 20\%)
\item
\url{motimatik.livejournal.com/119571.html} (Г.\,Колюцкий, вбросы и др.)
\item
\url{a-shen.livejournal.com/24589.html} (А.\,Чернов, автор, нет жалоб, 21\%)
\item
\url{burivykh.livejournal.com/64708.html} (В.\,Клепцын, подсчёт без
показа наблюдателям)
\end{itemize}

Большее количество данных от наблюдателей, подтверждающих правильность подсчёта (нет жалоб), может позволить оценить реальные результаты голосования. В проекте <<Гражданин наблюдатель>> была сделана попытка получить более или менее представительную выборку по Москве, послав наблюдателей в заранее отобранные участки (по 1--2 в каждом округе). Будем надеяться, что подробный отчёт о результатах появится. Пока что (на 16 декабря) на их сайте есть результаты обработки 131 протокола по Москве и 223 по остальной России. (Из этих данных не ясно, включены ли в них протоколы по участкам с жалобами.) Есть также более ранние данные по 36 участкам без нарушений в Москве~\cite{slon-zakharov}.

\begin{itemize}
\item
по Москве: СР--16\%, ЛДПР--12\%, ПР--1.7\%, КПРФ--23\%, Я--13\%, ЕР--31\%, ПД--1\%, недействительные--2\%. Явка--54\%.
\item
по 36 участкам без нарушений в Москве (на следующий день после выборов): СР--18\%, ЛДПР--13\%, ПР--1.8\%, КПРФ--25\%, Я--18\%, ЕР--23\%, ПД--1.1\%. Явка--49\%.
\item
по остальной России: СР--21\%, ЛДПР--14\%, ПР--1.5\%, КПРФ--23\%, Я--8\%, ЕР--30\%, ПД--1\%, недействительные--2\%. Явка--54\%. (Надо иметь в виду, что по остальной России выборка очень далека от представительной за счёт разного количества наблюдателей в разных регионах)
\end{itemize}

\begin{figure}
\begin{center}
  \includegraphics[width=0.5\textwidth]{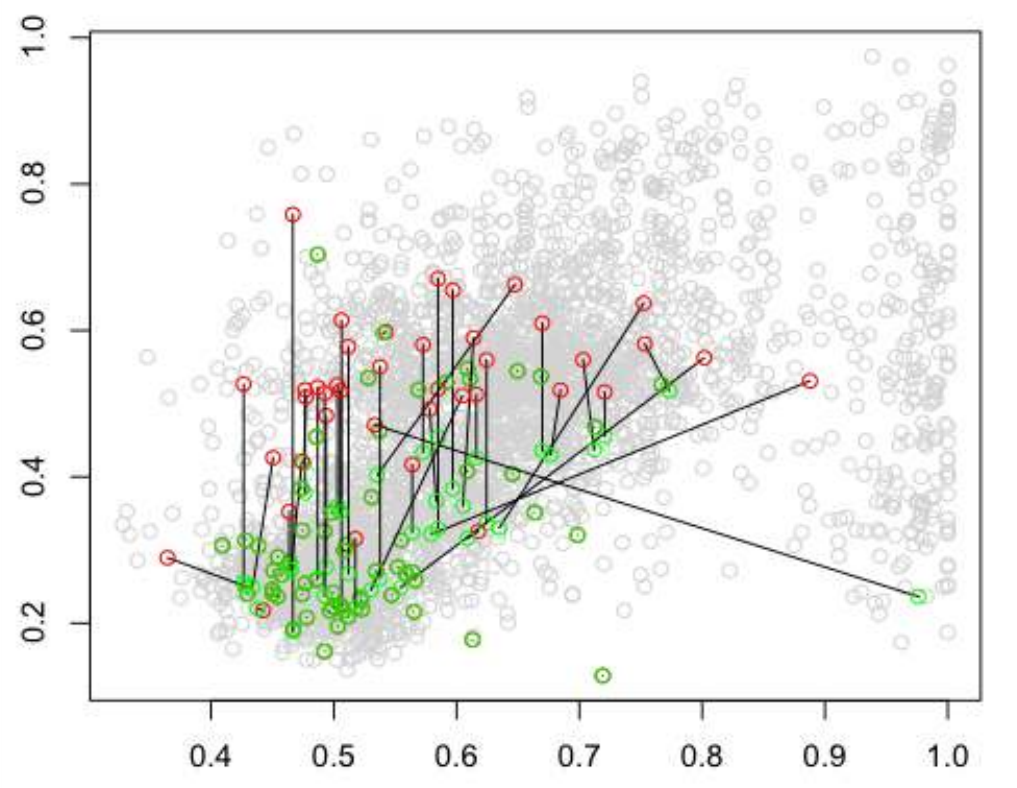}
\end{center}
\caption{Изменения явки и доли голосов за <<ЕР>> в Москве на участках, где протоколы не совпадают с официальными данными. Серые кружки~--- официальные данные по другим участкам.}
\label{change} \end{figure}

В некоторых случаях выданные наблюдателям протоколы расходятся с официальными данными. Сергей Слюсарев~\cite{jemmybutton} взял данные с сайтов~\cite{ruelect,karta} по Москве и изобразил изменения (протокол, зелёные кружки $\to$ официальные данные, красные кружки) в тех же координатах, что на рис.~\ref{oude-moscow.jpg} (рис.~\ref{change}); видно, что изменения происходят в направлении от первого кластера ко второму. Аналогичные данные по Нижнему Новгороду обсуждаются ниже.

\subsection{2011: соседние участки}

Как оценить правдоподобность гипотезы о том, что избиратели разных партий распределены по Москве неоднородно, и это объясняет статистические аномалии в результатах выборов по Москве (см. обсуждение выше)?

\begin{figure}[!h]
\begin{center}
  \includegraphics[width=0.9\textwidth]{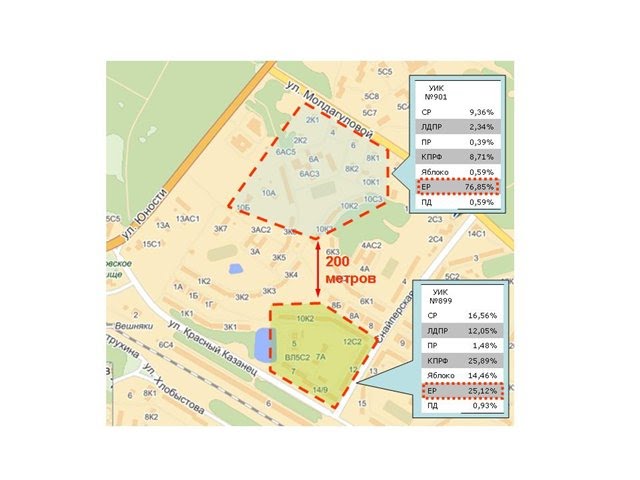}\qquad
\end{center}
\caption{Сравнение результатов соседних московских участков.}
\label{compare-2}
\end{figure}

\begin{figure}[!h]
\begin{center}
  \includegraphics[width=0.7\textwidth]{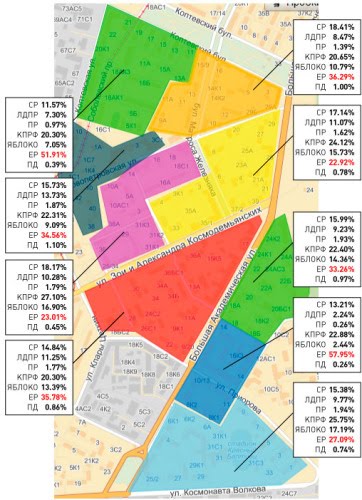}\qquad
\end{center}
\caption{Сравнение результатов соседних московских участков (продолжение).}
\label{compare-3}
\end{figure}

Можно попытаться сравнить данные по соседним участкам. Некоторые примеры такого рода можно найти на сайте~\cite{za-chestnye}, см. рис.~\ref{compare-2} и рис.~\ref{compare-3}. (Конечно, теоретически это может объясняться и какими-то экстраординарными различиями в составе жителей.)

\begin{figure}[!h]
\begin{center}
  \includegraphics[width=0.3\textwidth]{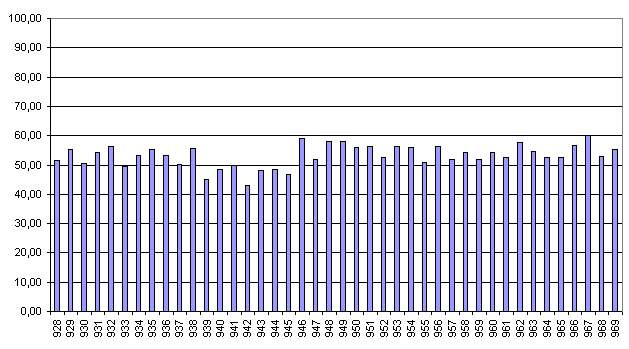}\qquad
  \includegraphics[width=0.3\textwidth]{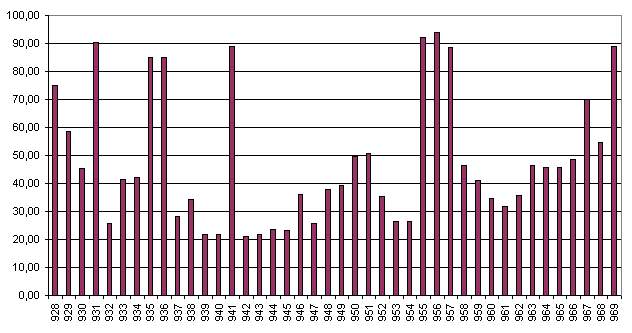}
\end{center}
\caption{Сравнение выборов в Гольяново: 2007 (слева) и 2011 (справа).}
\label{gol}
\end{figure}

Чтобы оценить влияние возможной неоднородности, можно сравнить результаты нескольких выборов в одном и том же районе (поскольку неоднородность связана с составом жителей, который меняется сравнительно медленно). Например, в~\cite{anderson-mike} приводятся результаты в одном и том же районе Гольяново (рис.~\ref{gol}); видно, что неоднородность в 2011 сильно больше, чем в 2007.

Более систематически такой анализ можно проводить, изобразив результаты на картах. Такой <<электоральной географией>> систематически занимается Александр Киреев~\cite{kireev}. В его журнале (запись от 10 декабря 2011) приведены карты голосования за <<Яблоко>> на выборах 1999 года и 2005~года, а также на нынешних~--- и ясно видно, что на прежних выборах картина одна, а на нынешних совсем другая. При этом границы, где результат <<Яблока>> резко меняется, часто совпадают с границами районов, что может объясняться различными уровнями фальсификации в районах с разной администрацией.

Убедительный пример последнего обнаружился в Санкт-Петербурге~\cite{golos-spb}. Там по техническим причинам вышло так, что участок мог относиться к одной территориальной комиссии по федеральным выборам и к другой~--- по местным. Список партий на федеральных и местных выборах был одни и тот же, так что особых причин для сильных отклонений при голосовании на двух выборах нет, и средние по всему Петербургу близки (35.4\% и 37,0\% для <<ЕР>>). Но образовалось 58 участков, где результат <<ЕР>> в госдуму более чем на 300~голосов превышает результат в местное законодательное собрание, и 47~участков с обратным соотношением. Примеры разницы в ту и другую сторону (в формате явка/ЕР/КПРФ/СР/ЛДПР/Я/ПР/ПД) $$1193/262/231/384/111/162/17/19$$ на федеральных против $$1688/1576/30/31/30/5/3/2$$ на местных (участок 1445); $$1314/852/104/201/118/9/8/1$$ на федеральных против $$1302/272/202/379/111/281/19/16$$ на местных (участок 302). Можно предположить, что в разных районах фальсификации происходили по-разному (для каждого из участков вышестоящие комиссии по тем и другим выборам различаются); кроме того, надо иметь в виду, что кандидаты на местных выборах больше заинтересованы в их результатах, чем в голосовании по федеральному списку.

\subsection{2011: отдельные случаи}

Повышенный интерес к выборам привёл к тому, что различных свидетельств фальсификаций (рассказов, видео, фото и пр.) так много, что привести ссылки на сколько-нибудь заметную долю их невозможно. Отсылая читателя к поиску в интернете (в частности, в YouTube), отметим лишь несколько случаев:

\begin{itemize}
        \item
Владимирские единороссы~--- и те возмущены масштабами фальсификаций \cite{vladimir-er}. (Очень впечатляющий ролик, его герои выглядят искренне и вызывают уважение, на мой взгляд.)
        \item
Публикация в Новой газете о <<каруселях>> в Санкт-Петербурге~\cite{novaya} выглядит уже совсем запредельно: она воспроизводит жалобы студентов (в адрес Александра Салаева, баллотировавшегося по 18-му округу в Колпинском районе) на то, что им не заплатили деньги за участие в <<карусели>>, голосовании за заказчика на нескольких участках~(!). Салаев в ответном интервью отрицает эти обвинения, но не очень убедительно~\cite{novaya-2}. В~\cite{mladovesti} автор утверждает, что обнаружил в социальных сетях (vkontakte) предположительных авторов этих признаний, в комментариях утверждают, что знают некоторых из героев публикации, автор предлагает опубликовать опровержение, если его пришлют, 17 декабря опровержения ещё не было.
        \item
Оксана Дмитриева, одна из лидеров <<Справедливой России>>, пишет на своём сайте~\cite{dmitrieva}: <<\ldots по данным сводного протокола ТИК №\,4 избирательной территории 21 (Красногвардейский район) СР получила 10186 голосов. А в систему ГАС `` Выборы''\ введено 5030~голосов. То есть у нас украдено 5156~голосов~--- более 50\% голосов. По округу, где баллотировалась бывшая судья Уставного суда СПб Людмила Кулешова, расхождение составляет 1922~голоса~--- 63\% голосов украдено\ldots>>.
        \item
Олег Козырев, направивший жалобу в прокуратуру по поводу того, что <<избирательная комиссия Московской области изготовила информационные материалы, совпадающие до степени смешения с агитационными материалами Московского областного отделения политической партии ``Единая Россия''>>, получил ответ~\cite{kozyrev}, в котором говорится, что эти заявления <<носят клеветнический характер, $\langle\ldots\rangle$ порочат репутацию Избирательной комиссии Московской области\ldots\ Ваше заявление направлено в Прокуратуру Московской области, Следственный Комитет Российской Федерации по Московской области для проведения соответствующей проверки и привлечения Вас к уголовной ответственности по статье 129 УК РФ>> (кстати, недавно отменённой).
        \item
О рассмотрении обращения <<Яблока>> по поводу фальсификации выборов (несовпадения данных с копии протокола, выданной наблюдателю, и официальных)~\cite{sud}: <<В 4 часа утра наблюдатель от ``Яблока''\ получил заверенный печатью и подписями членов комиссии протокол: за ``Единую Россию''\ проголосовало 259~человек\ldots\ Днём 5 декабря на интернет-сайте Мосгоризбиркома появились совсем другие цифры: $\langle\ldots\rangle$ у ``Единой России''\ оказалось на 456 голосов больше\ldots\

Чтобы объяснить, как такое возможно, в суд пришли представитель Мосгоризбиркома Дмитрий Реут, заместитель председателя Территориальной избирательной комиссии района Перово Ирина Шевцова и 7 членов участковой избирательной комиссии. Дмитрий Реут\ldots\ подверг сомнению само право партии ``Яблоко''\ обращаться в суд в защиту нарушенных избирательных прав граждан\ldots\ Выданный наблюдателю от ``Яблока''\ протокол он назвали ``технической ошибкой''\ и передал слово коллеге из ТИКа. ``То, что выдали наблюдателю,~--- это нельзя назвать протоколом, это черновик'',~--- пояснила И.~Шевцова. ``Просто наблюдатель вынудил комиссию подписать этот черновик.''\ Суд отказал ``Яблоку''\ в отмене результатов выборов по этому участку.>>
\end{itemize}

\subsection{2011:  Нижний Новгород}

\begin{figure}
\begin{center}
  \includegraphics[width=0.7\textwidth]{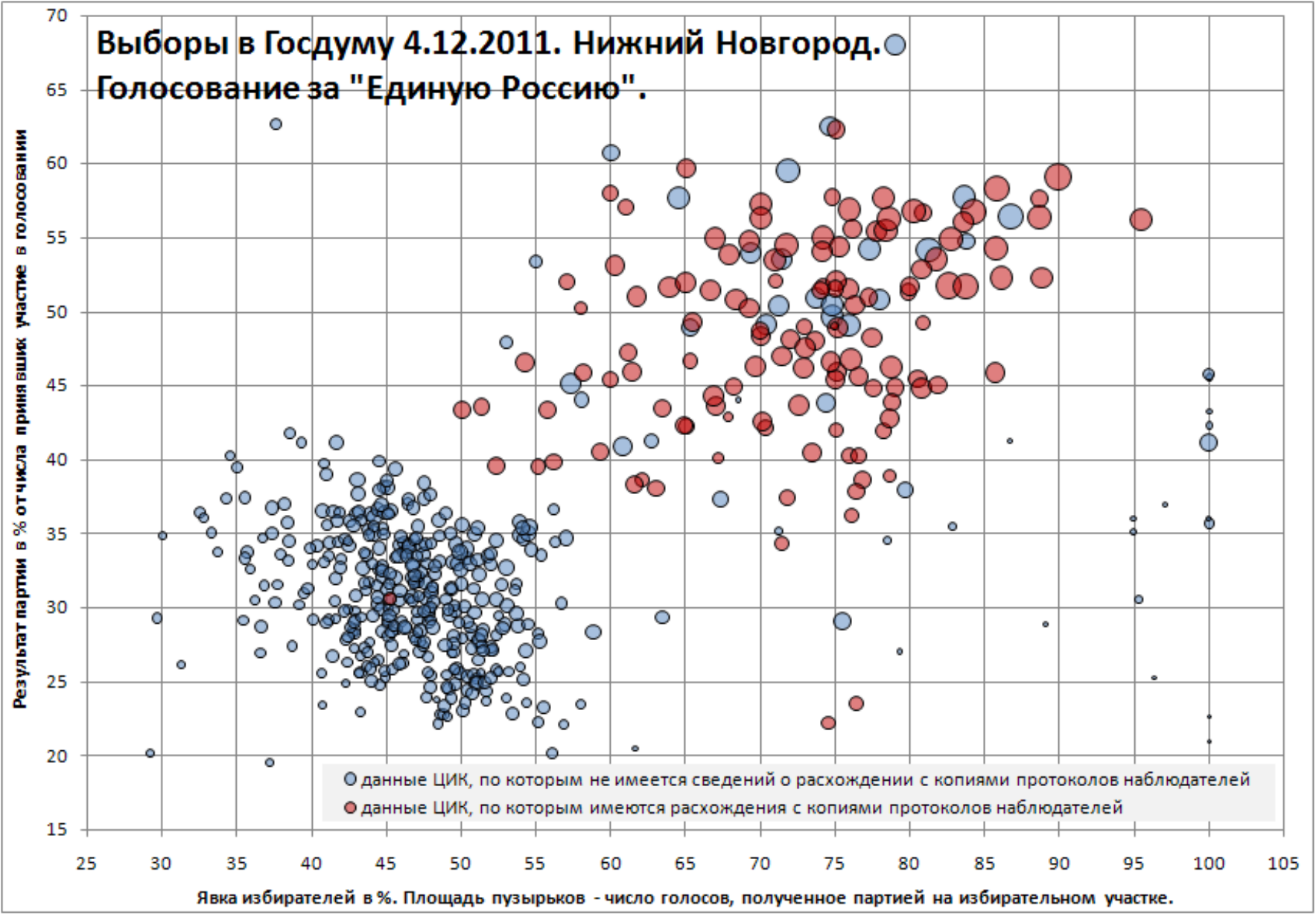}
\end{center}
\caption{Распределение участков в Нижнем Новгороде}
\label{nn1}
\end{figure}

На примере Нижнего Новгорода хорошо видно, как согласуется
информация из различных источников. Прежде всего посмотрим
на общую картину~\cite{vladislav-01} распределения участков в координатах
(явка, голосование за <<ЕР>>), см. рис.~\ref{nn1}, на котором пока не надо обращать внимание на цвета точек.

Видно, что участки делятся на две группы, и возникает вопрос, с чем это связано. Одно из возможных объяснений~--- неоднородность участков. Как можно проверить эту гипотезу? Александр Киреев сравнил результаты выборов по разным районам с 2007~годом~\cite{kireev-701143} (можно предположить, что неоднородность уже и тогда должна была как-то проявиться). Он пишет:

\begin{quote}
<<Я сделал вот такую табличку [таблица~\ref{tab1}], где показаны результаты голосования за <<ЕР>> и явку на этих и на прошлых думских выборах $\langle\ldots\rangle$ В 5 районах <<ЕдРо>> потеряло около 25\%, но в других районах [их $4$] гораздо меньше, и [в этих же $4$~районах] резко возросла явка. В 5 районах, где ЕдРо потеряло много голосов, явка снизилась на 5~--  6\%, как и в остальной России\ldots\ [напротив,] в Канавинском, Московском, Приокском и Советском районах она возрасла, причем в Канавинском районе сразу на 20\%!>>
\end{quote}

\begin{table}
{\scriptsize
\begin{tabular}{|c|c|c|c|c|c|c|}
\hline
{\large\strut} Районы & ЕР 2011 & ЕР 2007 & Явка 2011& Явка 2007 &$\Delta$ЕР &$\Delta$явка\\
\hline\hline
{\large\strut}Канавинский& 50.93\%&54.61\%& 69.47\%&49.65\%&-3.68\% &+19.82\% \\
\hline
{\large\strut}Московский&39.78\% & 54.08\%&63.79\% & 48.83\%& -14.30\%&+14.96\%\\
\hline
{\large\strut}Сормовский& 30.10\%&56.54\% &45.89\% &51.73\% &-26.44  &-5.84\%\\
\hline
{\large\strut}Автозаводской Север&33.74\% &59.48\% &47.83\% &53.39\% &-25.74\% &-5.56\% \\
\hline
{\large\strut}Автозаводской Юг& 34.56\%&58.89\% &45.78\% &51.39\% &-24.33\% &-5.61\% \\
\hline
{\large\strut}Ленинский& 30.20\%&55.02\% &43.77\% &49.84\% &-24.82\% &-6.07\% \\
\hline
{\large\strut}Нижегородский& 27.49\%&49.71\% &50.15\% &51.63\% &-22.22\% &-1.48\% \\
\hline
{\large\strut}Приокский&46.38\% &53.01\% &63.23\% &52.63\% &-6.63\% &+10.60\% \\
\hline
{\large\strut}Советский& 38.06\%&49.61\% &56.91\% &52.83\% &-11.55\% &+4.08\% \\
\hline

\end{tabular}}
\caption{Данные по выборам 2007 и 2011 годов в Нижнем Новгороде}\label{tab1}
\end{table}

Из этой таблицы видно, что районы, в которых в 2007~году результаты были
близки, теперь ведут себя по-разному (и это различие примерно соответствует
различию между двумя группами точек на рис.~\ref{nn1}). Это можно пытаться
объяснить либо внезапно возникшими географическими аномалиями, либо
неправильным подсчётом голосов.

		Для оценки правдоподобности этих гипотез можно посмотреть также на данные по участкам, оборудованным КОИБ. Оказывается, что в районах с повышенной явкой и результатами <<ЕР>> участки с КОИБ явно  выделяются: в них явка и результаты <<ЕР>> существенно ниже. Александр Киреев приводит~\cite{kireev-koib} соответствующие данные по одному из четырёх районов Нижнего Новгорода, где явка и результаты за <<ЕР>> и явка были выше: в целом по району явка $63\%$, <<ЕР>> получила $46\%$; если ограничиться 9 участками с КОИБ, то явка будет $52\%$ и результат <<ЕР>>~--- лишь~$32\%$.
		
\begin{figure}[!h]
\begin{center}
  \includegraphics[width=0.7\textwidth]{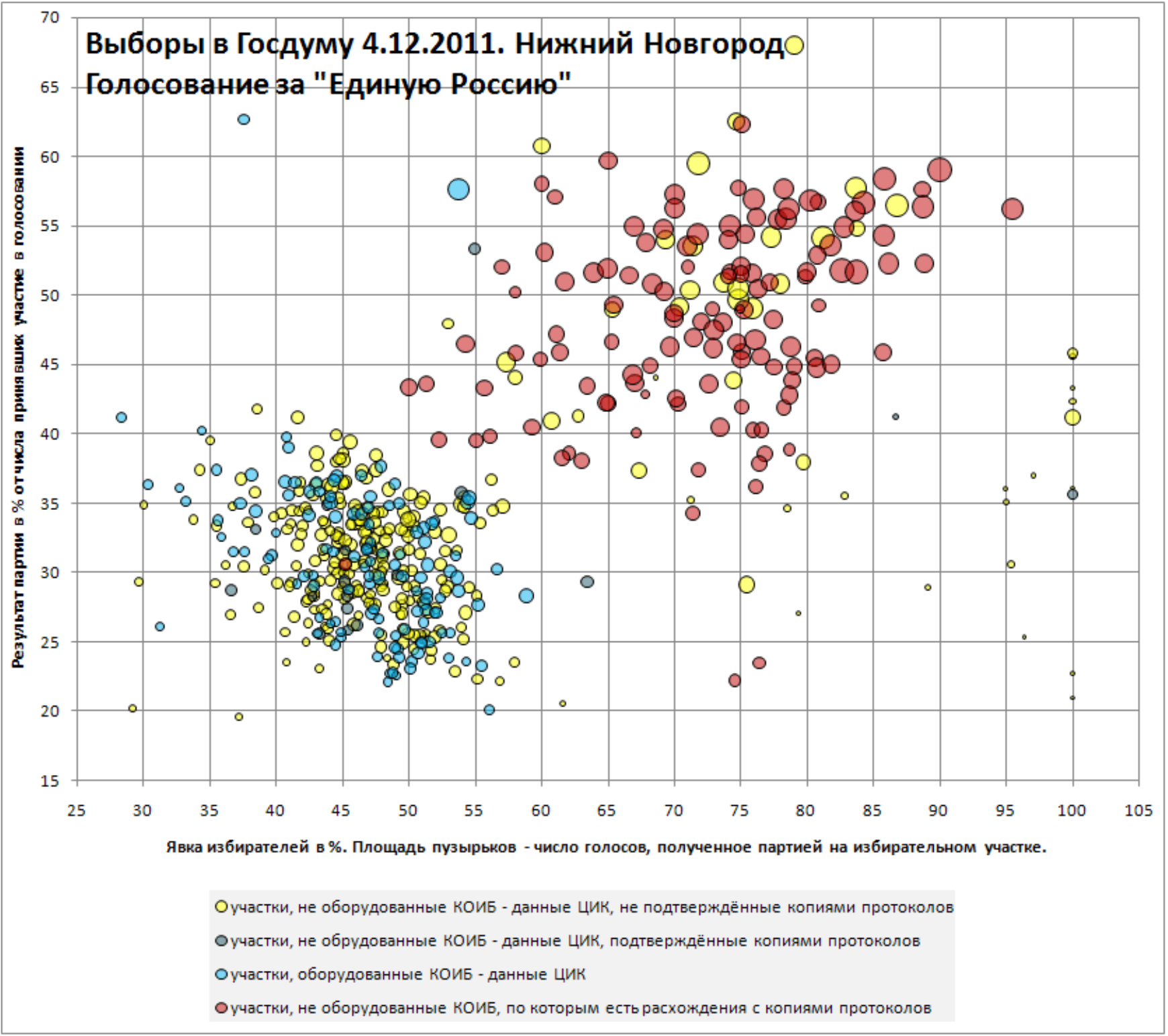}
\end{center}
\caption{Участки с КОИБ среди других в Нижнем Новгороде.}
\label{nn-koib}
\end{figure}

Этот эффект иллюстрируется также рис.~\ref{nn-koib}~\cite{vladislav-01}: участки с КОИБ в Нижнем Новгороде показаны на рисунке голубым цветом; они в основном оказываются в левой нижней группе (с меньшей явкой и меньшим результатом <<ЕР>>).

\begin{figure}[!h]
\begin{center}
  \includegraphics[width=0.7\textwidth]{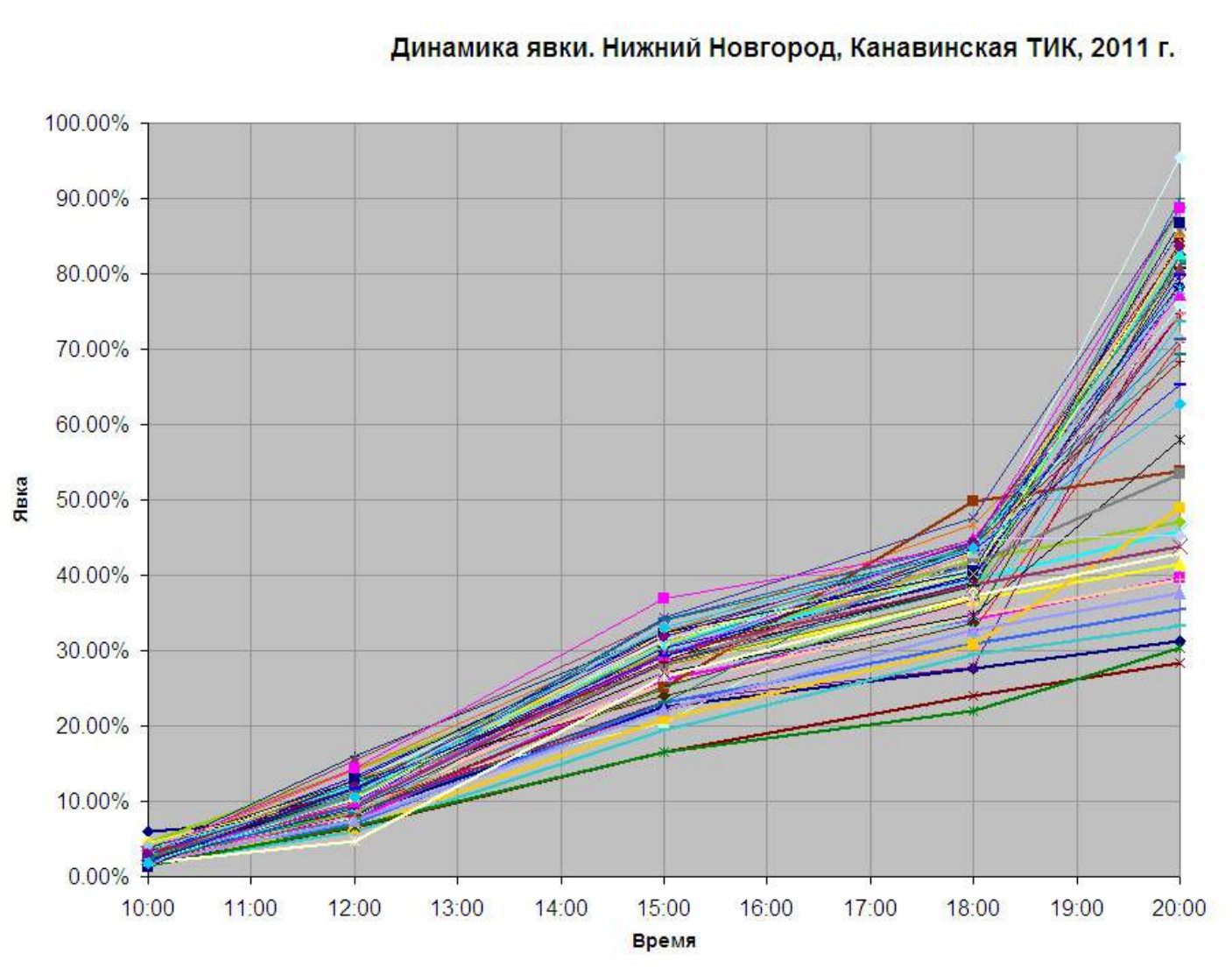}
\end{center}
\caption{Изменение явки со временем в течение дня голосования в Канавинском районе Нижнего Новгорода.}
\label{dynamics}
\end{figure}

Какая ещё информация может прояснить ситуацию? Среди опубликованных ЦИК данных есть сведения об изменении числа проголосовавших избирателей со временем (участковые избирательные комиссии сообщают о числе проголосовавших несколько раз в течение дня). Сергей Шпилькин~\cite{podmoskovnik-dynamics} изобразил эти данные по Канавинскому району (одному из четырёх <<аномальных>>, см. выше) на графике, отложив по горизонтальной оси время дня, а по вертикальной~--- число проголосовавших к этому времени (рис.~\ref{dynamics}).

Обнаруживается, что участки чётко делятся на две группы: в одних явка достаточно плавно возрастает в течение дня, а во вторых в последний момент возникает скачок: объявленное число проголосовавших существенно выше, чем можно ожидать на основе предыдущих данных. Это можно объяснить фальсификацией, если предположить, что данные по явке в течение дня были настоящими, а в официальных результатах явка была искусственно увеличена. 

\begin{figure}[!h]
\begin{center}
  \includegraphics[width=0.5\textwidth]{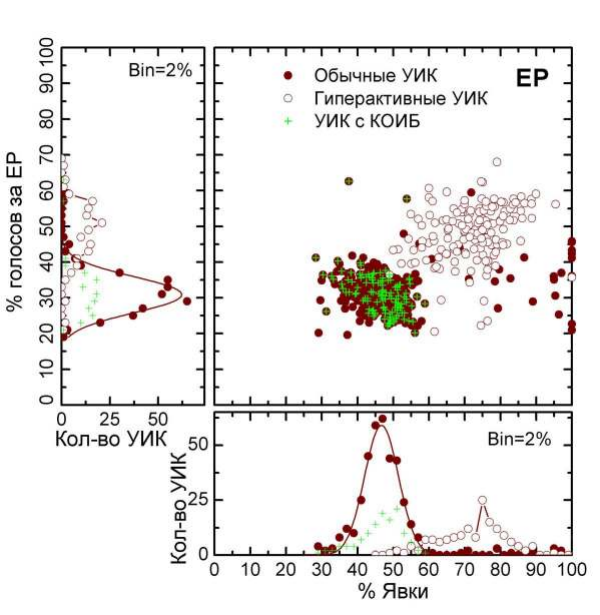}
\end{center}
\caption{Результаты голосования на <<гиперактивных участках>>.}
\label{dynamics-res}
\end{figure}

При этом предположении действительно ожидать увеличения доли голосов за <<ЕР>> в этих <<гиперактивных>> участках, и она действительно наблюдается, см. рис.~\ref{dynamics-res} из~\cite{dynamics-res}~--- практически вся верхняя правая группа оказывается составленной из <<гиперактивных>> участков (таковыми считались участки, в которых явка за последний интервал подсчёта превысила некоторый порог, а именно, 13\%).

\begin{figure}[!h]
\begin{center}
  \includegraphics[width=0.7\textwidth]{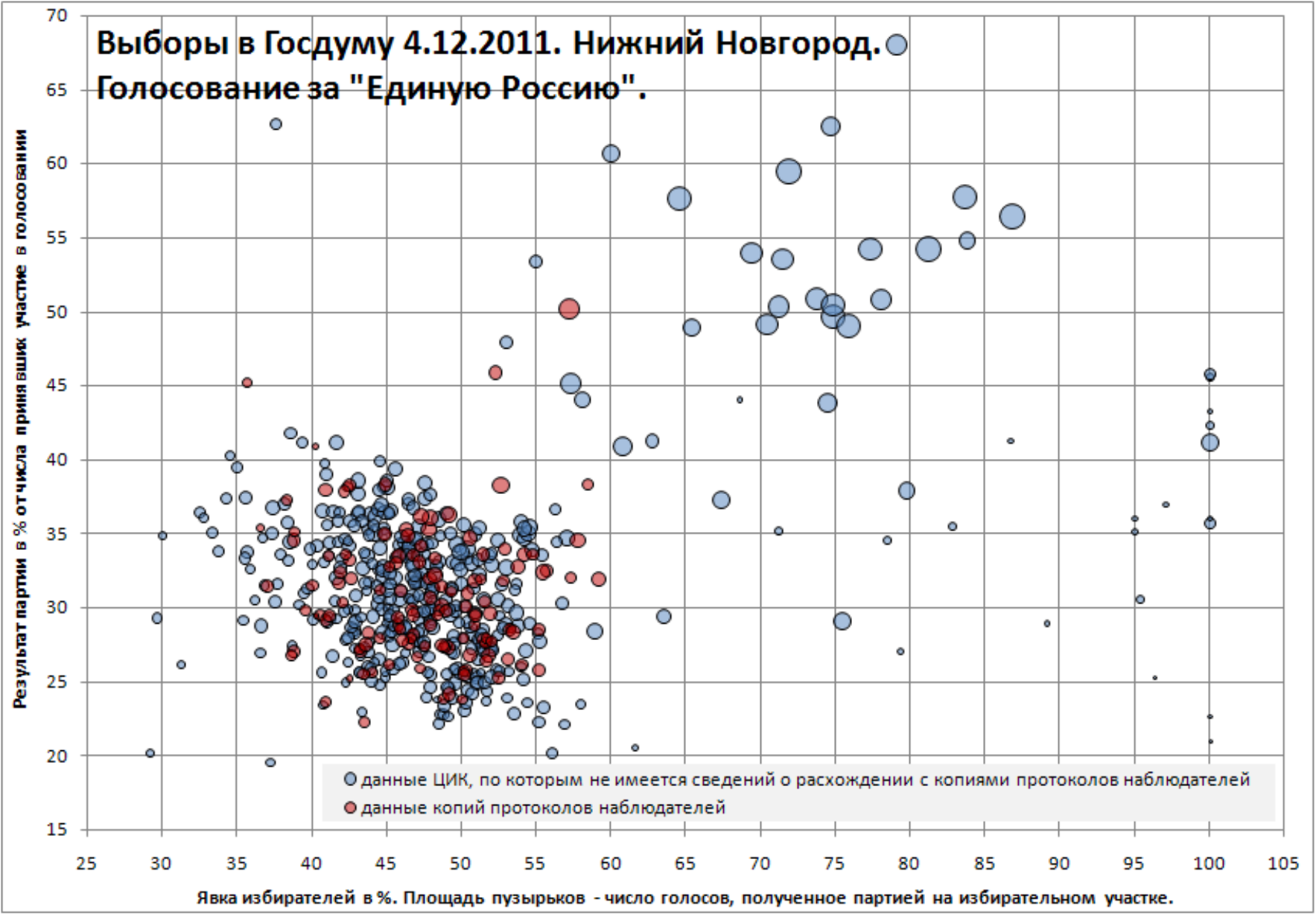}
\end{center}
\caption{Данные по явке и голосованию за <<ЕР>> в Нижнем Новгороде, скорректированные по копиям протоколов.}
\label{nn-corr}
\end{figure}

До сих пор мы ограничивались лишь данными из официальных источников. Возникает вопрос, можно ли эти <<косвенные улики>> подкрепить какими-то прямыми? Оказывается, что да: есть данные протоколов, полученных наблюдателями Нижнего Новгорода на участках~\cite{ruelect,vovikbo}, и эти данные (по фотографиям копий протоколов) во многих случаях не совпадают с официально объявленными. Эти расхождения выделены цветом на приведённом выше рис.~\ref{nn1}. Красным цветом показаны официальные данные по участкам, где они расходятся с данными наблюдателей. Видно, что красные точки составляют большую часть верхней группы. Если в случаях расхождений заменить официальные данные на  данные с копий протоколов, то от верхней группы останется лишь небольшая часть (рис.~\ref{nn-corr};  отметим, что не со всех участков есть копии протоколов).

Расхождения в данных стали предметом обсуждения на заседании законодательного собрания Нижегородской области (см. видеозапись~\cite{nn-zayavl}). Точнее говоря, представители <<Единой России>> по рекомендации своего предводителя уходят с заседания, когда представитель КПРФ делает заявление о несоответствии данных протоколов и официальных данных (по его словам, например, в Канавинском районе $20869$ голосов были занесены как $46725$).

\subsection{2011: обсуждение}

Выборы 2011 года стали предметом оживлённого обсуждения, в том числе на митингах и в интернете. К сожалению, значительная часть обсуждений основана на недоразумениях: многие сторонники гипотезы фальсификаций не понимают сложностей статистического анализа и приводят доводы типа <<раз распределение не по Гауссу~--- значит жулики>>. С другой стороны, многие их оппоненты, справедливо указывая на необоснованность таких доводов, ограничиваются этим и не знакомятся с реальными аргументами в пользу гипотезы фальсификации.

Различных обсуждений в интернете очень много, в том числе по указанным выше ссылкам; для полноты приведу несколько ссылок с аргументами против гипотезы массовых фальсификаций. Я старался выбрать в первую очередь те, где авторы проявляют математическую грамотность, а также те, в которых авторы предпринимали самостоятельные попытки обработать данные: \cite{mib,neretin,tutubalin,jemmybutton,sergey-k}. В последней заметке видна эволюция взглядов автора: начав с заявления <<Сеть заполонили спекуляции на тему аномального распределения голосов\ldots\ При этом эксперты-математики, пытающиеся объяснить народу весь ужас того, как нагло его обманывают власти, на самом деле в лучшем случае никакие не эксперты, т.к. допускают грубые математические ошибки, а в худшем~--- сами нагло врут и подтасовывают цифры>>, после нескольких дней экспериментов с данными тот же автор пишет уже гораздо спокойнее: <<Итого, если принимать гипотезу о фальсификации, то примерно в трети регионов проводилась централизованная накрутка, ещё в одной трети~--- нескоординированная накрутка по собственной местной инициативе, а [в] оставшейся трети~--- всё чисто. Но это, повторю, при условии принятия гипотезы о массовой фальсификации.>> (Далее приводится гипотетический список регионов трёх указанных типов, довольно близкий к оценкам других авторов.) Другие часто цитируемые: \cite{lqp,svshift}.

\noindent
Комментируя обсуждения результатов выборов, Леонид
Волков\footnote{Он столкнулся с избирательной системой на собственном опыте: его сняли с местных выборов, забраковав подписи, со следующей (замечательной в своём роде) формулировкой: <<в отличие от экспертов, сообщивших суду, по каким основаниям подписи избирателей были признаны недостоверными, пояснения избирателей о собственноручном выполнении ими подписи и проставлении даты её внесения какими-либо другими доказательствами не подтверждены>>.} пишет~\cite{lv}:

\begin{quote}

   <<Посмотрим ещё раз на наши две группы соображений.

   1. Даже несмотря на то, что независимых наблюдателей было (увы) не очень много, выявлено большое количество грубейших нарушений. Сам факт [того], что фальсификации были, этим доказан. Однако оценить их масштаб только на основании видеороликов и переписанных протоколов~--- нельзя.

  2. Математические методы позволяют грубо оценить масштаб фальсификаций. Безусловно, только оценить~--- никак точно отсечь участки с вбросами от территорий, где все поголовно любят Путина, на этих графиках невозможно. Благодаря математическим методам мы получаем общее представление о том, что были украдены миллионы голосов (не сотни тысяч, но и не десятки миллионов)~--- реальный результат ``ЕР''\ (с отброшенными фальсификациями, но по-прежнему с учётом административного ресурса) должен быть где-то между 32 и 40 процентами. Любые более точные оценки пусть остаются на совести их авторов, но даже очень грубой оценки достаточно, чтобы понимать~--- узурпация власти имела место, парламентское большинство ``ЕР''\ основано исключительно на фальсификациях.

   А в ответ говорят следующее:

  1. Наблюдений~--- мало. Даже если мы отменим результаты выборов на всех тех участках, где были сняты видеоролики о вбросах, результаты выборов изменятся не более чем на 0{,}5\%.

  2. Матметоды ничего не доказывают. Глядя на кривую Гаусса с предательски задранным хвостом, вы не вправе утверждать, что нарушения были.

    Вы уже поняли, что в этом споре что-то не так? Мы говорим:

        1. Видео доказывают, что нарушения были.

        2. Математика позволяет оценить их масштаб.

   А нам говорят:

        1. Ваши видео ничего не оценивают.

        2. Ваша математика ничего не доказывает.>>

\end{quote}

В заключение попытаюсь сформулировать собственный ответ на вопрос: можно ли оценить размеры фальсификаций и восстановить более или менее точно реальные результаты голосования? Ответ мне представляется двояким (первый пункт отчасти повторяет Волкова, хотя написан до знакомства с его текстом):

\begin{itemize}
        \item грубо оценить с достаточным правдоподобием можно~--- различные методы оценки показывают, что, с одной стороны, доля <<ЕР>> завышена не на доли процента; с другой стороны, видно, что заметная часть голосов, засчитанных за <<ЕР>>, существовала в реальности. (Грубые прикидки разными способами дают цифры доли реальных бюллетеней за <<ЕР>> в районе 30--40\%, но их надёжность весьма условна. Если же ограничиваться судебными решениями, то судя по сложившейся практике, их не наберётся и на доли процента~--- но к рассматриваемому вопросу это отношения не имеет.)

        \item в такой оценке вообще нет особого смысла, на мой взгляд. Скажем, если у шулера нашли два туза в рукаве, то странно было бы пытаться пересчитать его выигрыш, оценив, сколько бы он мог выиграть, играя честно. Другой пример того же рода: спортсмена, который во время кросса срезал путь по кустам, обычно дисквалифицируют, а не пытаются пересчитать его результат с учётом замеченных сокращений пути.
\end{itemize}

\section{Post Scriptum: о связи теории с практикой}

Приведу (с минимальной орфографической правкой) фрагмент из записи в <<Живом журнале>> уже упомянутого выше~\cite{shmelev} Александра Шмелёва~\cite{shmelev-2}; отмечу лишь, что опасения про мартовские голосования оказались сильно преувеличенными (см. ниже).

\begin{quote}

<<Так, не далее чем вчера я общался с одной своей знакомой, работающей в районной управе. Женщиной средних лет, бесконечно далекой от всей политики и воспринимающей выборы как одно из ``массовых мероприятий'', проводящихся управами и префектурами, в которых надо участвовать ``по работе''\ (в общем, типичный сотрудник районной управы). В частности, за время этого разговора мне удалось узнать следующее:

\begin{enumerate}
\item
Фальсификации готовились и организовывались строго по вертикали. Сотрудников управ, отвечающих за УИКи (по одному человеку на УИК, обычно в должности зампреда комиссии, председатель при этом в курсе, но сам, типа, ни в чем не виноват), неоднократно собирали на совещания, куда приезжали представители префектуры и объявляли процент, который должна получить ЕР на каждом участке, а также подробно и в деталях обсуждали вбросы, карусели и пр. Как их проводить, как не попастся, как нейтрализовывать наблюдателей и т.п.~(в частности, было объявлено, что наблюдатели от СР~--- ``свои'', их можно не бояться, а вот яблочники и коммунисты могут напакостить).

\item
Причём это происходит не в первый раз. Ровно по такой же модели проводились выборы в Мосгордуму-2009 и президентские выборы (+муниципалитеты) в 2008-м. Единственное отличие~--- за это время из управ поувольнялось много специалистов, знающих избирательное законодательство и умеющих мухлевать технично, поэтому сейчас <<работать было сложнее>>.

\item
По итогам выборов те сотрудники, которые с фальсификациями успешно справились, были награждены: им выписаны специальные премии по итогам года. Тем, кто давал больше всех (в её районе~--- 59\% за ЕР)~--- аж 5 окладов. Остальным меньше. Тех же, кто сфальсифицировать ничего не смог, все эти 2~недели очень жёстко прессовали и лишили всех премий (в т.ч.~и регулярных). Среди последних оказалась и моя респондентка. Не потому, что она решила проявить гражданскую сознательность, конечно (думаю, если бы я заговорил с ней на эту тему, она очень бы удивилась), но потому что на её участке до 4 утра дежурили корреспонденты ``Новой газеты'', и она до последнего хотела вбросить, но не смогла. Тем не менее глава управы не принял это в расчёт и уже несколько раз вызывал ее на ковёр, грозя увольнением: не справляешься ты, дескать, с работой. В т.ч.~доводя до слёз.

\item
На весенние выборы уже объявлена планка: надо будет вбрасывать аж по 1400 бюллетеней на участок. 700~--- за муниципалов и 700~--- за Путина. КОИБы~--- не панацея, для участков, оборудованных КОИБами (как у моей респондентки) обещают выделять по ``полчаса спокойствия'', за которые надо будет успеть эти 1400 бюллетеней в КОИБ опустить. Сразу после праздников начнутся тренировки (это непросто, надо будет опускать практически по бюллетеню в секунду).>>

\end{enumerate}
\end{quote}

\section{Голосование в марте 2012 года}

Результаты этого голосования интересны не только сами по себе, но и в сравнении с результатами голосования 2011~года.

\begin{itemize}
\item
А.~Киреев~\cite{kireev-dagestan} обнаружил интересную закономерность в Дагестане. Например, в Сулейман-Стальской территориальной комиссии всего $42$ участка, из них $6$ с ненулевым числом голосов за Жириновского ($4,3,2,2,5,4$). Такое распределение (либо $0$, либо $2$--$5$) уже само по себе странно: естественным образом провал в единице объяснить сложно. Но к тому же Киреев обнаружил, что это ровно те $6$~участков, на которых стояли КОИБ.  (Случайное совпадение $6$ выбранных чисел из $42$~--- это почти что главный приз в Спортлото, вероятность этого около $2\cdot 10^{-7}$.)  Аналогичные совпадения (ненулевые результаты ровно там, где КОИБ) ещё в 11 территориальных комиссиях (Агульская, Бабаюртовская, Дагогниская городская, Дахадаевская, Дербентская городская, Кулинская, Лакская, Левашинская, Магарамкентская,  Тляратинская, Южно-Сухокумская; надо иметь в виду, что на некоторых из них меньше КОИБов и соответственно ненулевых результатов. Всего участков с КОИБ (они же~--- с ненулевыми результатами Жириновского) $33$, и на двух из них результат $1$, а на остальных~--- больше~$1$).

\item
Он же (рис.~\ref{kireev-vladik}) обнаружил замечательную стабильность результатов в левобережной части Владикавказа~\cite{kireev-vladik} (справа приведены данные про голосование 2011~года, выделены участки, где в 2011~году данные почему-то переставились). Колонка со звёздочкой добавлена Киреевым~--- это доля Путина, если не учитывать недействительные бюллетени.
\begin{figure}
\begin{center}
  \includegraphics[width=0.85\textwidth]{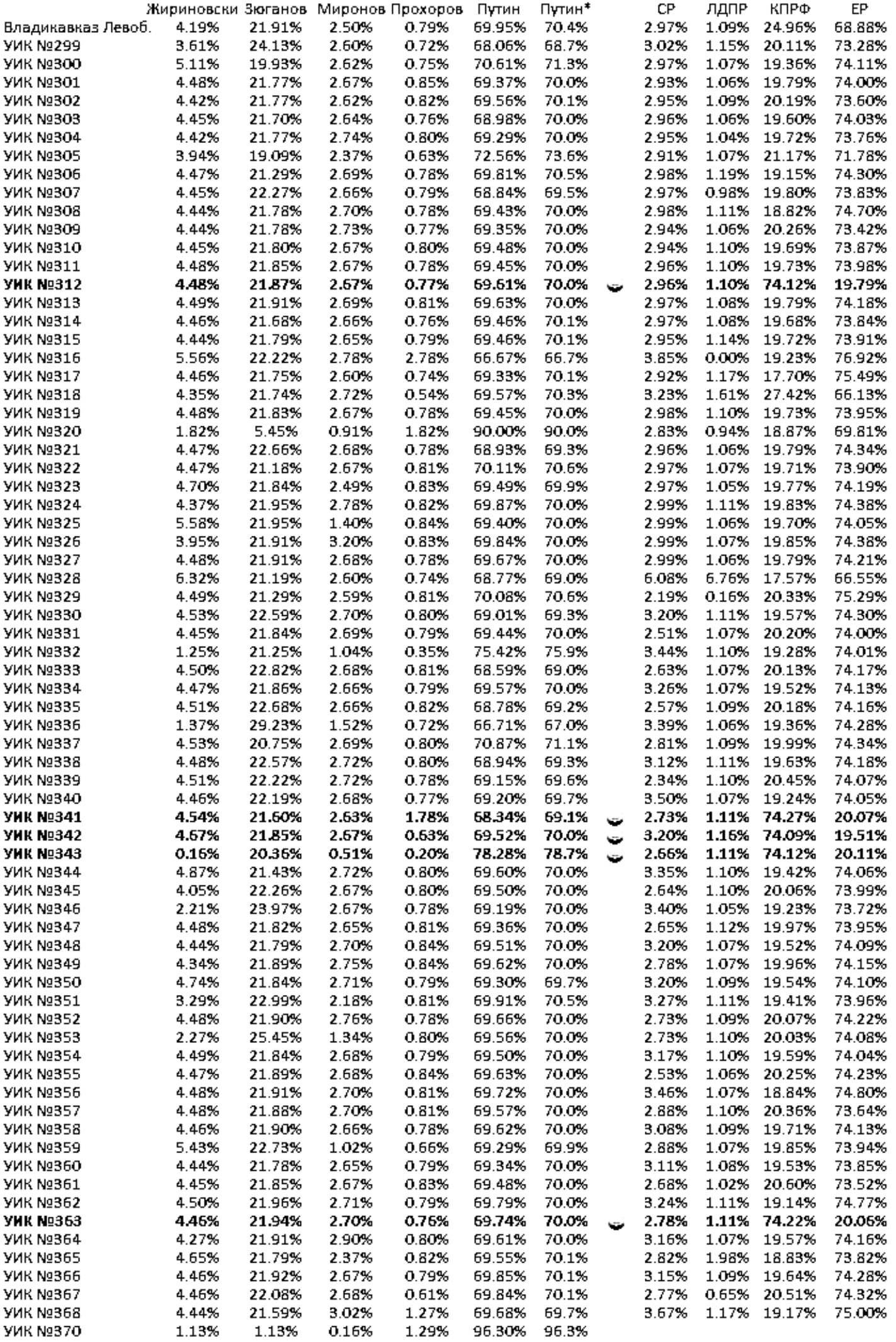}
\end{center}
\caption{Результаты во Владикавказе (левобережная часть),  \cite{kireev-vladik}.}
\label{kireev-vladik}
\end{figure}

\item
Пики явки и процента голосов за <<лидера>> видны и на этих выборах; для разнообразия можно посмотреть на двумерную картинку, где каждая точка соответствует участку: в правом верхнем углу видна сетка кратных $10$ (а также $5$), рис.~\ref{grid}.

\begin{figure}[ht]
\begin{center}
  \includegraphics[width=0.6\textwidth]{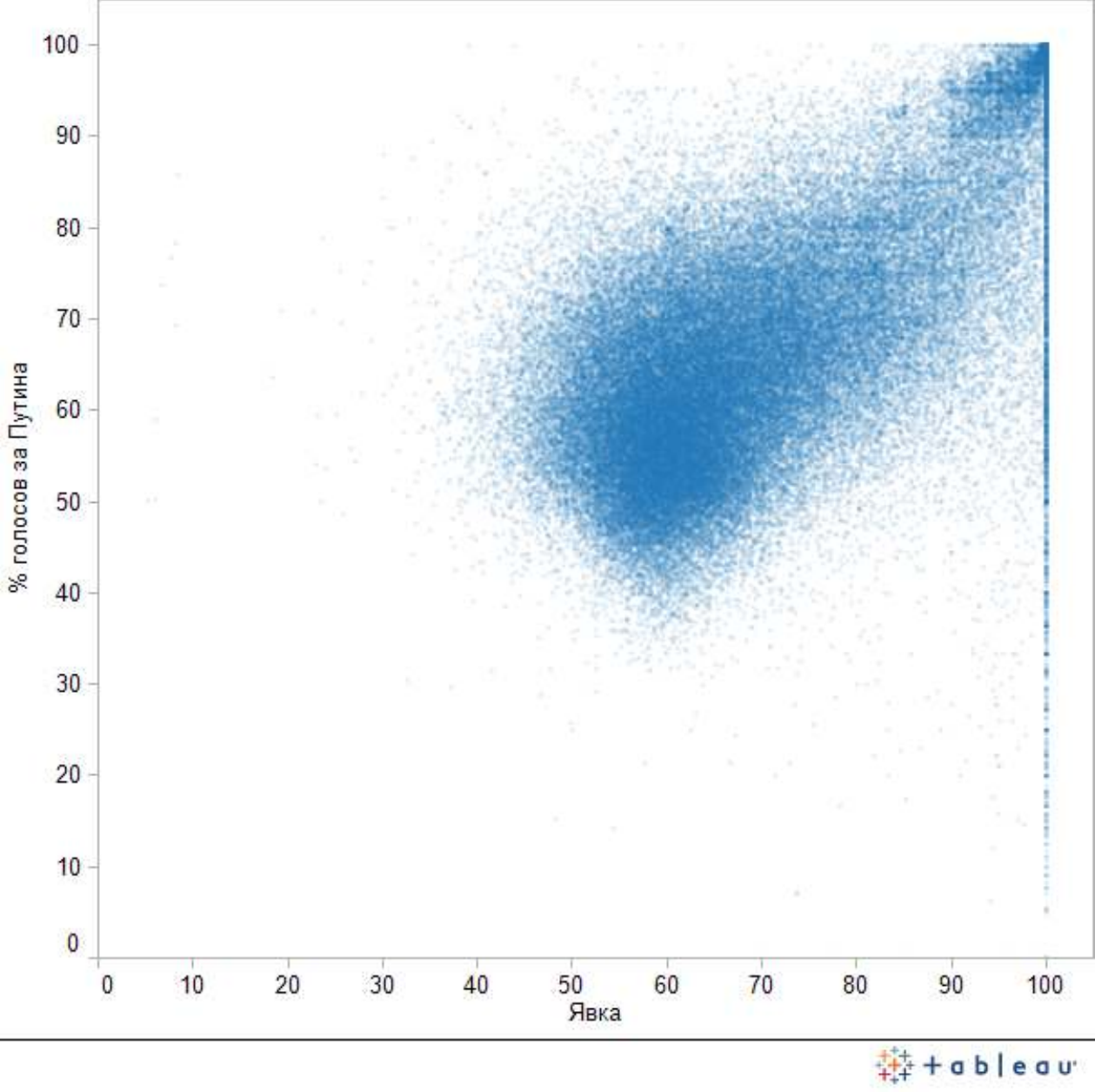}
\end{center}
\caption{Данные по явке и голосованию за <<лидера>>, \cite{ilyabo}.}
\label{grid}
\end{figure}

\item
М.~Пшеничников~\cite{oude-rus12a} сравнил аналогичные диаграммы для голосования 2011 и 2012 годов (рис.~\ref{grid-compare}); видно, что аномальный  <<хвост кометы>> уменьшился.

\begin{figure}[ht]
\begin{center}
  \includegraphics[width=0.6\textwidth]{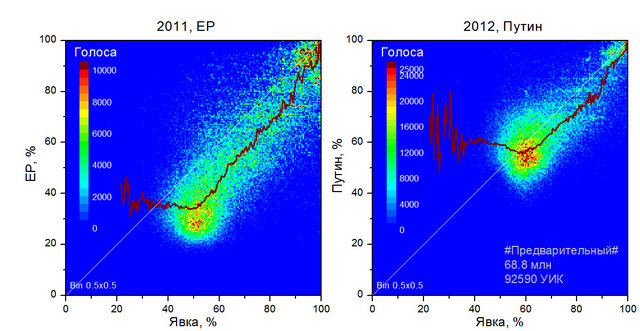}
\end{center}
\caption{Сравнение голосований 2011 и 2012 годов, \cite{oude-rus12a}.}
\label{grid-compare}
\end{figure}

\clearpage

\item
Сравнение <<аномальных хвостов>> на гистограмме тоже показывает, что <<ненормальная>> (далёкая от нормального распределения) часть сильно уменьшилась (рис.~\ref{compare-hist}, \cite{oude-rus12b}).
\begin{figure}[ht]
\begin{center}
  \includegraphics[width=0.9\textwidth]{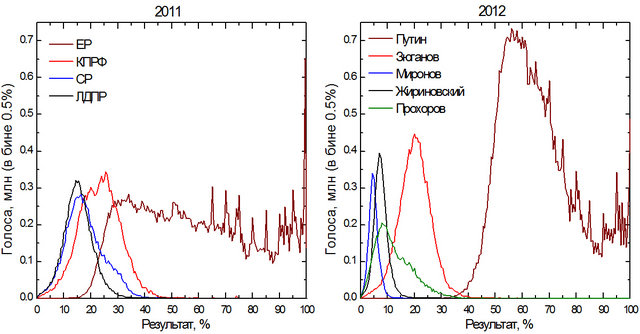}
\end{center}
\caption{Сравнение гистограмм 2011 и 2012 годов, \cite{oude-rus12b}.}
\label{compare-hist}
\end{figure}

\item
Для сравнения можно привести схематическое изображение гистограммы 2011 года, сделанное В.\,Е.\,Чуровым~(рис.~\ref{churov-version},\cite{churov}) ; <<аномальный хвост>> не показан.
\begin{figure}[ht]
\begin{center}
  \includegraphics[width=0.5\textwidth]{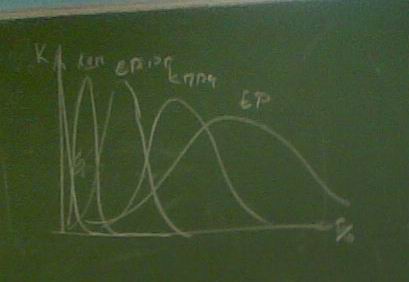}
\end{center}
\caption{Гистограмма 2011 года, фотография рисунка, сделанного В.Е.~Чуровым на доске во время его доклада~\cite{churov}.}
\label{churov-version}
\end{figure}

\item
Возможные причины больших отклонений распределения явки по Москве от нормального (географическая неоднородность, вбросы) легко оценить по  гистограммам явки в Москве разные годы (А.~Бузин~\cite{buzin-curves}), рис.~\ref{buzin-curves}.

\begin{figure}[!h]
\begin{center}
  \includegraphics[width=0.6\textwidth]{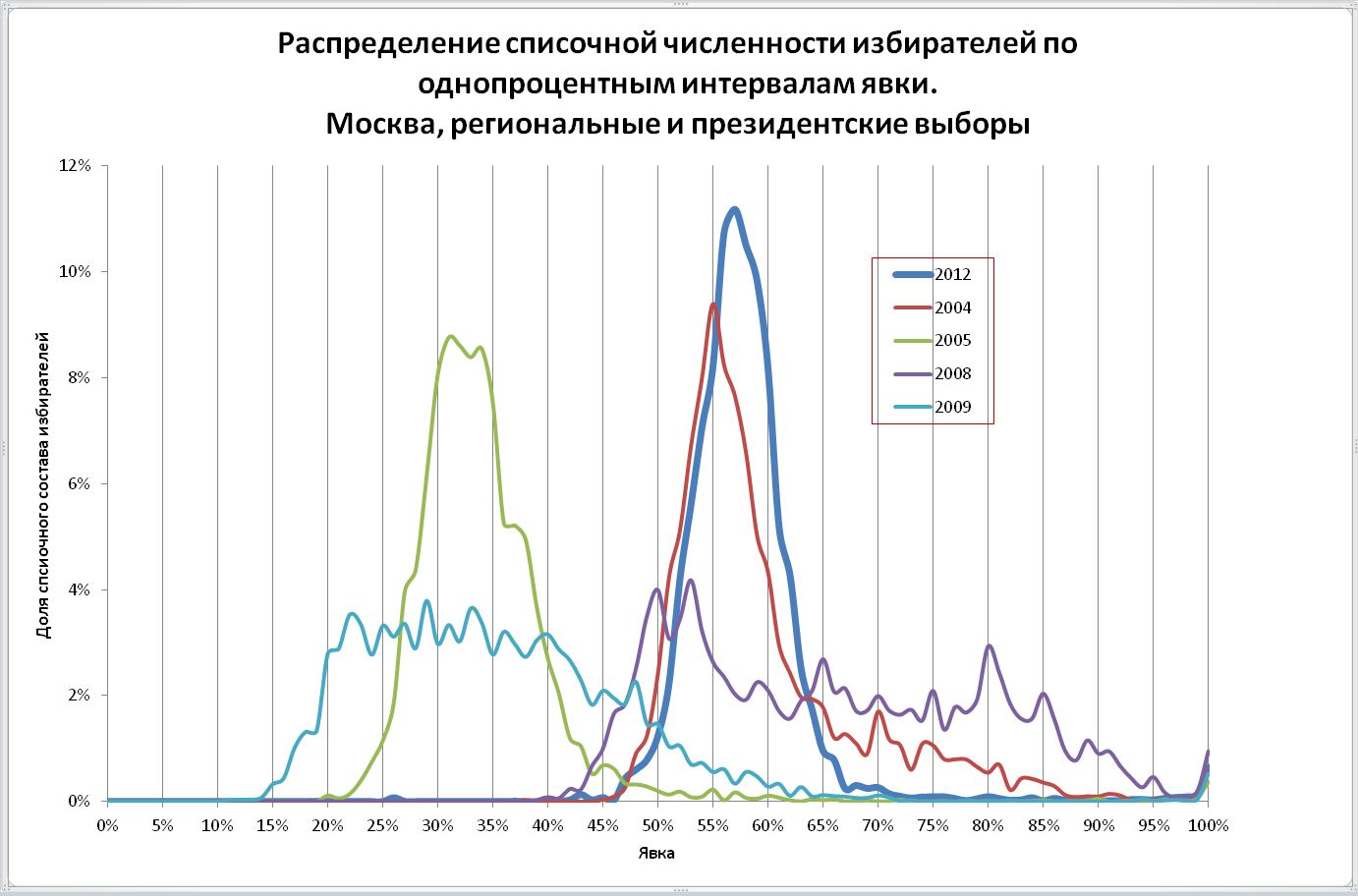}
\end{center}
\caption{Гистограммы явки разных лет по Москве, \cite{buzin-curves}.}
\label{buzin-curves}
\end{figure}

\item
Интересно сравнить более подробно результаты голосований 2011 и 2012 года по Москве; на втором было больше наблюдателей (рис.~\ref{compare-mow}, \cite{oude-rus12moscow}).

\begin{figure}[!h]
\begin{center}
  \raisebox{15mm}{\hbox{\includegraphics[width=0.5\textwidth]{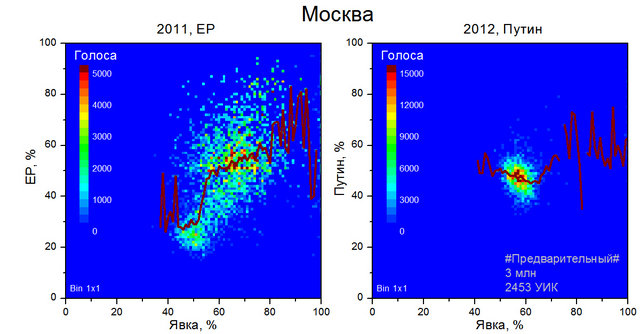}}}\qquad \includegraphics[width=0.4\textwidth]{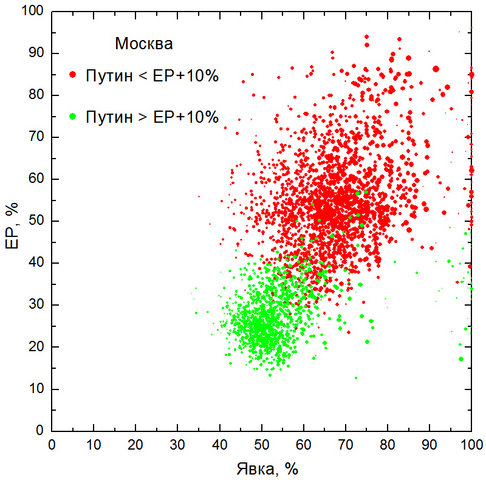}\\
  \includegraphics[width=0.7\textwidth]{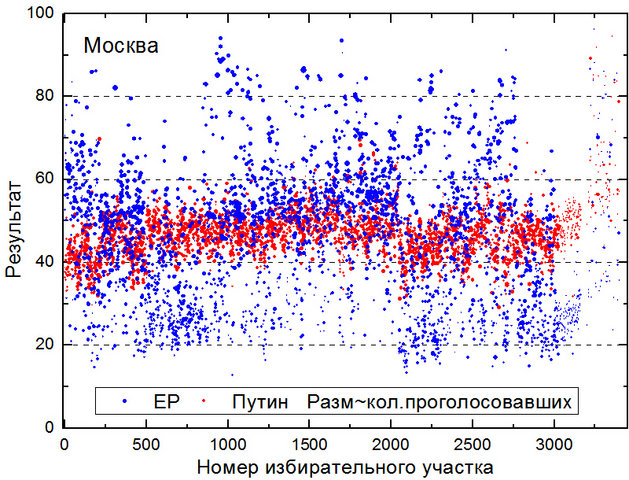}\\
\end{center}
\caption{Сравнение  разброса по Москве в 2011 и 2012, \cite{oude-rus12moscow,oude-rus12moscow2,oude-rus12moscow3}.}
\label{compare-mow}
\end{figure}
\item
Интересно, что в Петербурге такого изменения картинки не произошло, и жалоб наблюдателей тоже было много : (рис.~\ref{compare-spb}, \cite{oude-rus12spb}).

\begin{figure}
\begin{center}
  \includegraphics[width=0.8\textwidth]{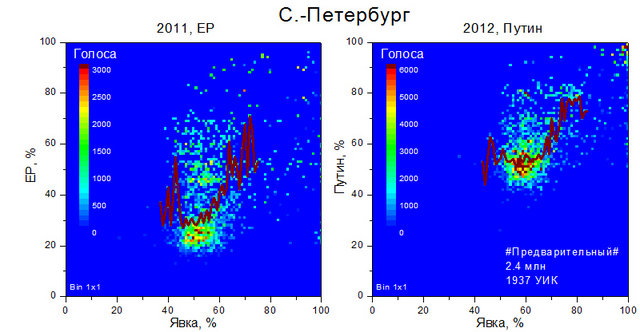}
\end{center}
\caption{Сравнение  разброса по Петербургу в 2011 и 2012, \cite{oude-rus12spb}.}
\label{compare-spb}
\end{figure}

\item В одной из комиссий (ТИК 27) в Петербурге имеется удивительное скопление в точке $(60,80)$ в координатах (явка, результат), которое разрешается на отдельные участки лишь в большем увеличении (рис.~\ref{compare-spb27}).
\begin{figure}
\begin{center}
  \includegraphics[width=0.55\textwidth]{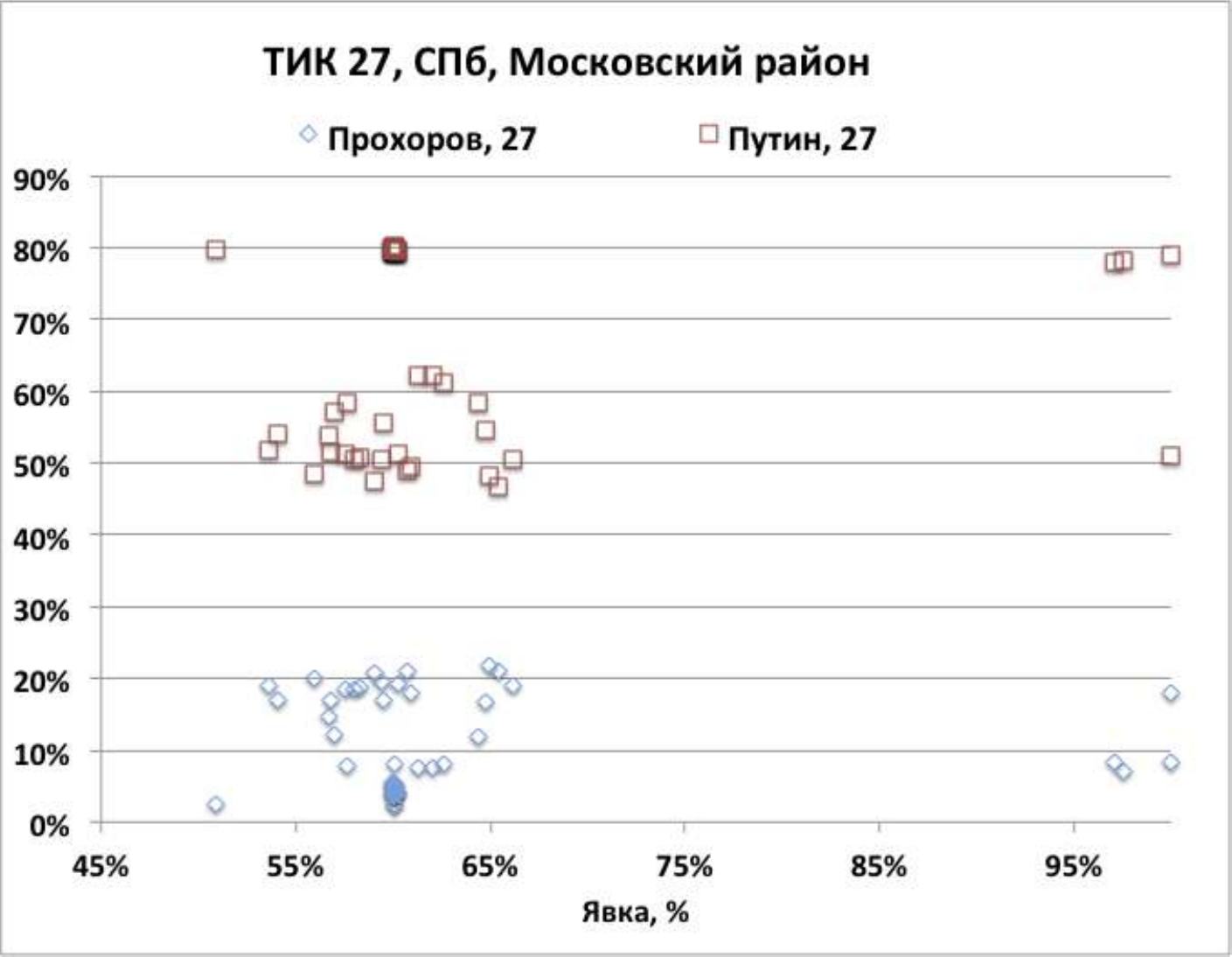}\qquad
  \includegraphics[width=0.35\textwidth]{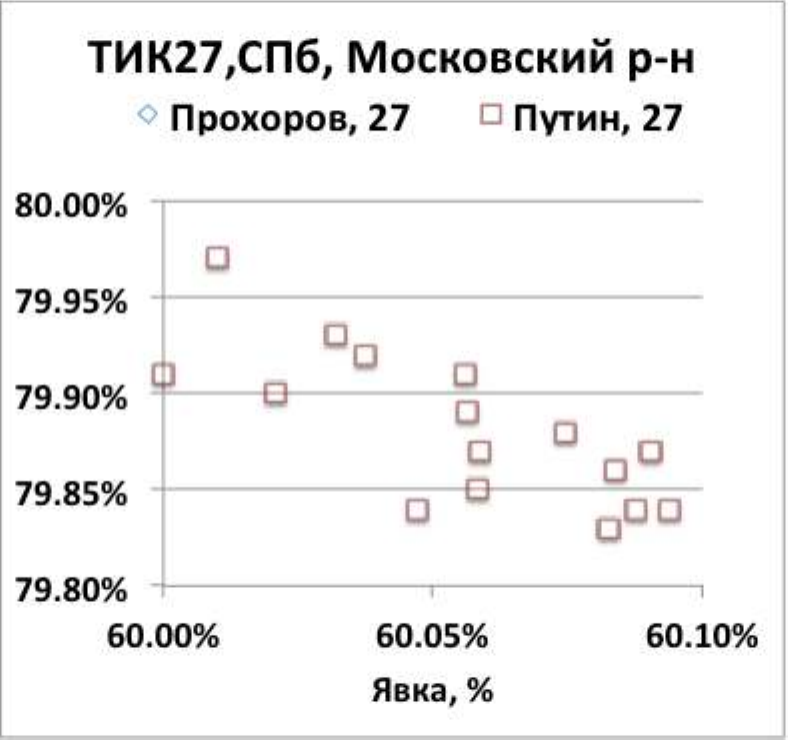}
\end{center}
\caption{Петербург, ТИК 27, \cite{kireev-spb}.}
\label{compare-spb27}
\end{figure}

Интересно сравнить это с публикацией в <<Аргументах и фактах>>  (Петербург):
\begin{quote}
Наблюдатель от КПРФ: в ТИК №\,27 протоколы заносят с <<черного хода>> [Опубликовано : 2012-03-05 00:10:56]

Об этом корреспонденту <<АиФ-Петербург>> сообщила наблюдатель в ТИКе от КПРФ Анна Смирнова.

По её словам, лишь небольшая часть председателей УИКов, которые приносят в ТИК протоколы и бюллетени, заходят в здание с центрального входа и передают результаты выборов в присутствии наблюдателей. Большая часть заходит с <<чёрного хода>>, охраняемого полицией. Туда наблюдателей вопреки требованиям закона не пускают. При этом в протоколах, занесённых с <<чёрного хода>>, значатся крайне высокие цифры поддержки кандидата в президенты Владимира Путина~--- в некоторых случаях около 90\%. [Последнее утверждение не подтверждается данными.]
\end{quote}

\item
Некоторые наиболее удивительные изменения (2012 год в сравнении с 2011) собрал Александр Киреев~\cite{kireev-records}:
\begin{quote}
Явка больше всего выросла в Кисловодске: на 35,54\%, с 36,66\% до 72,22\%. А в Моздокском районе Северной Осетии она на столько же упала: с 88,75\% до 52,19\%. При этом в Кисловодске процент Путина по сравнению с ЕР возрос в 2 раза с 34,67\% до 68,96\%, а в Моздокском районе <<необъяснимо>> упал с 74,36\% до 64,38\%.

Больше всего Путин потерял по сравнению с Единой России в Новоульяновском районе Ульяновской области: $-25\%$! Было 79,42\%, стало 54,58\%. При этом явка снизилась на 21 пункт с 82,71\% до 61,77\%. На втором месте Арбат $\langle\ldots\rangle$ $-23{,}11\%$ c 63,16\% до 40,05\%. Тут без учёта фальсификаций на думских выборах нам никак не справиться с объяснением этого феномена. [В той же записи комментирует один из наблюдателей: <<Я тоже была на участке (2618), где в декабре было 77\%, в этот раз получили 41\%. Допущены были ко всем этапам по полной программе.
А в декабре на участке 1541 мы пытались заставить соблюсти процедуру, но разнарядка была делать фальсификацию, и нас тупо выгнали.>>]

\ldots Немножко сложнее подсчитать динамику Жириновского и Миронова. [В целом] Жириновский набрал почти в $2$ раза меньше, чем ЛДПР или же 55\% от электората ЛДПР. В ряде республик у него просто сумашедший рост. Скажем, в Эльбрусском и Баксанском районах Кабардино-Балакарии у ЛДПР было вообще ноль голосов, а сейчас в Эльбрусском районе 3,24\%, а в Баксанском вполне правдоподобные для республики 3,13\%. То есть, скажем, в Баксанском районе было ноль, а теперь вдруг откуда-то взялись 822 голоса! То есть из 822 человек, проголосовавших за Жириновского в Баксанском районе, \emph{ни один} не голосовал за ЛДПР на думских выборах! По другим районам Кабардино-Балкарии было 0,01\%, а сейчас до нескольких процентов. Из русских областей выделяется всё тот же Новоульяновский район, в котором у Жириновского 10,59\%, а у ЛДПР было 3,86\%.

\end{quote}

\end{itemize}


\section{2018: дополнения}

В этом разделе (написано в мае 2018 года) приведены ссылки на некоторые публикации и работы, не вошедшие в ранее написанный текст (в частности, появившиеся позже). С тех пор материала накопилось довольно много, так что этот обзор будет по необходимости фрагментарным.

\subsection{Возражения Ю.\,Неретина и эксперимент 2012 года}

Известный математик Юрий Неретин в декабре 2011 года опубликовал запись~\cite{neretin-15-12-2011}, в которой высказал свою оценку выборной статистики:
\begin{quote}
\ldots
в интернете и некоторых mass-media я столкнулся с массовыми обличениями председателя избиркома Чурова в существовании негауссовых вероятностных распределений. [Так у Неретина.]

Приведу один пример, самый массовый, и служащий своего рода символом веры\ldots\ Вот две (почти одинаковых) диаграммы, автор первой~--- Сергей Шпилькин, автор второй~--- Максим Пшеничников. [Мы воспроизводим только первую (рис.~\ref{neretin-15-12-2011})~--- они похожи.]  С одной из этих картинок люди ходили на митинг\dots\ На основании этой диаграммы и исходя из того, что распределение должно быть гауссовым, производится вычисление <<действительных процентов>> проголосовавших за разные партии.
\begin{figure}[h]
\begin{center}
\includegraphics[width=0.6\textwidth]{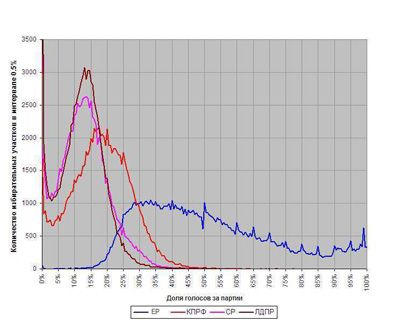}
\end{center}
\caption{График С.\,Шпилькина, приводимый Ю.\,Неретиным~\cite{neretin-15-12-2011}.}\label{neretin-15-12-2011}
\end{figure}
На обеих картинках изображена общероссийская статистика по избирательным участкам. По горизонтали~--- проценты голосов, отданных за партии, по вертикали~--- число избирательных участков, на которых достигнут данный процент. Эти диаграммы позволяют уверенно сказать о наличии фальсификации за счет <<зубчиков>> на значениях $50, 60, 65,\ldots$.\footnote{Как показывает более тщательный анализ, <<зубчик>> около $50\%$ является результатом наличия небольших участков с чётными знаменателями, где ровно половина в числителе. Чтобы избежать этого артефакта, лучше строить графики общего числа избирателей, попавших в ту или иную группу явки, тогда малые участки не вносят большого вклада. ~--- А.Ш.} На этом основании можно сделать нижнюю оценку значения фальсификаций (автор не вычислял, думаю, порядка $2$--$3$ процентов или, скорее, чуть больше). Эти проценты вовсе не безобидны с точки зрения результатов выборов. Понятно, что общий размер фальсификаций мы так не вычислим. Здесь спорить не о чем, и речь не об этом\ldots\

Чуров обвиняется в негауссовости нарисованных на графиках распределений. В этом, простите, он невинен. Потому что \textbf{эти распределения не могут быть гауссовыми}. Чуров мог лишь их шевелить в ту или иную сторону.

$\langle\ldots\rangle$
Представим себе идеальный российский регион с идеально однородным населением. $\langle\ldots\rangle$ Давайте сочтем полученное распределение (то, которое рисовалось на картинках) гауссовым. Теперь представим себе Россию, состоящую из двух существенно разных идеальных регионов. Нам придется сложить два графика.

Вполне может получится волнистая линия [рис.~\ref{neretin-sum}]

\begin{figure}[h]
\begin{center}
\fbox{\quad \includegraphics[width=0.4\textwidth]{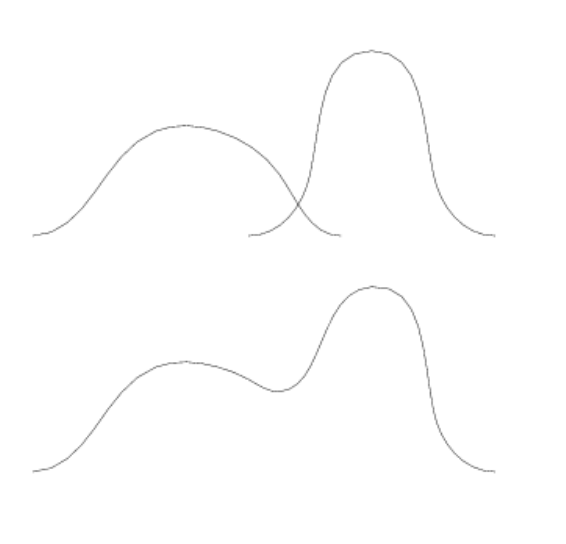}}
\end{center}
\caption{Сумма двух распределений}\label{neretin-sum}
\end{figure}

И даже если линия будет куполообразной, гауссианой она не будет, за редчайшими исключениями.

При сложении многих таких графиков могут получаться самые разнообразные картинки, в зависимости от того, какие плотности мы складываем. Есть люди, которые могут подумать, что если сложить много гауссовых плотностей (в случае России~--- 100), то получится гауссова кривая. Но это лишь предмет мистической веры.

$\langle\ldots\rangle$
В действительности все чуть сложнее, потому что многие регионы неоднородны (даже Москва), и никаких оснований ожидать везде гауссовых распределений нет.

$\langle\ldots\rangle$
Меня могут спросить, а как же фальсификация влияет на график? Отвечаю~--- график шевелится с примерным сохранением формы\ldots\ Читатель может воспринимать это как задачу. На мой взгляд, на реальном графике ЕдРа впадина должна еще прогнуться, а левый купол должен расшириться вправо и вырасти в \hbox{высоту}.

Если бы меня, не показывая мне эти графики, попросили бы нарисовать примерные формы этих 5 графиков, каковыми они должны были бы быть без воли Чурова\dots\ я бы их примерно такими и нарисовал (испытывая некоторые сомнения о форме загибов в крайнем левом участке). <<Проценты>> же зависят от точного расположения графика, а не от его примерной формы.

Не буду пытаться реферировать других сочинений на ту же тему, их очень много, уровень их вызывает недоумение, эти исследователи завязаны между собой, очень редко кто-нибудь дает отпор абсурдным или необоснованным заявлениям. В частности, картинки, приведенные выше, являются общепризнанным слоганом этой отрасли.

Добавлю, что очень рисковано делать глобальные выводы из анализа статистики ЦентрИзбиркома без подключения социологических исследований. Потому что неоднородность выборок (то, что производит кажущиеся экзотичными картинки)~--- это проблема социологии. Так же, как не очевиден внутренний <<случайный процесс>>, заставляющий человека не голосовать или голосовать тем или иным образом (причем в разных стратах общества он может быть разным).
\end{quote}

Неретин безусловно прав в том, что не всегда гистограммы различных данных по голосованиям близки к гауссовому распределению (см. выше о выборах в США).  Во многих случаях кривая имеет колоколообразную форму, но гауссово распределение (график $e^{-x^2}$, с точностью до множителей по обеим координатам)~--- это не просто похожая на колокол кривая, а некоторое совершенно конкретное распределение. Дополнительную путаницу вносит то, что гауссово распределение называют <<нормальным>> и это слово понимается в нематематическом смысле как <<то, как должно быть>>, в противоположность <<ненормальному>>.\footnote{В других языках это тоже так: по-английски это распределение называют ``normal''. Это не просто совпадение~--- действительно, такое распределение часто встречается в природе, и на то есть причины, отчасти покрываемые <<центральной предельной теоремой>>. Она утверждает, технически говоря, что сумма большого числа независимых случайных величин близка к нормальному распределению при некоторых предположениях. Конечно, эта теорема не является математическим доказательством того, что то или иное экспериментальное распределение должно быть нормальным, но в какой-то мере объясняет, почему нормальное (гауссово) распределение часто встречается.}

Неретин объясняет (совершенно верно), как может возникнуть кривая другой формы при сложении нескольких гауссовых распределений с разными параметрами. В случае голосования в Москве эти распределения соответствовали бы разным группам жителей, разделённым какими-то параметрами типа уровня дохода, места жительства и пр. Вопрос о том, действительно ли это так, не математический, а экспериментальный, и голосование 2012 года, которое произошло через несколько месяцев после обсуждаемого Неретиным, дало ответ на этот вопрос: форма кривой оказалась совсем другой, колоколообразной (рис.~\ref{buzin-curves}).\footnote{Теоретически можно предположить, что как раз в 2012 году колоколообразная форма получилась при искажении <<истинного>> распределения 2011 года. Но это противоречит и здравому смыслу (почему предположительные искажения сложно устроенной кривой дали простую колоколообразную кривую?), и свидетельствам наблюдателей, которые в 2012 году в Москве почти не зафиксировали нарушений, в отличие от 2011 года.} Аналогичное сравнение для всей России (правда, тут сложнее найти контрольную группу) см. выше на рис.~\ref{compare-hist}.

\begin{figure}[!h]
\begin{center}
\includegraphics[width=0.99\textwidth]{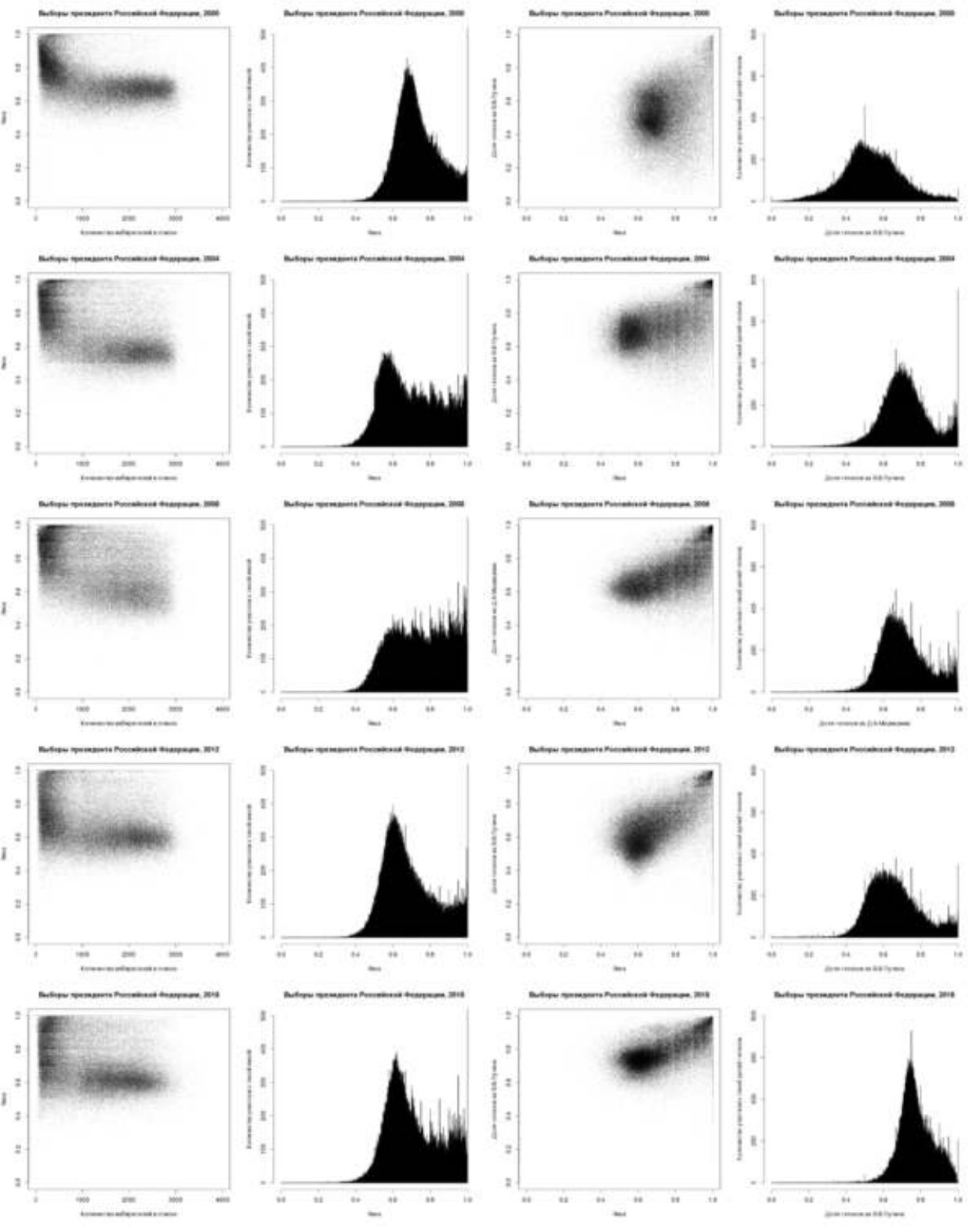}
\end{center}
\caption{Графики для голосований 2000~-- 2018, \cite{kupriyanov-2018}.}\label{kupriyanov-2018-3pic.pdf}
\end{figure}

Алексей Куприянов~\cite{kupriyanov-2018} опубликовал более подробные <<диагностические графики>> для голосований разных лет (рис.~\ref{kupriyanov-2018-3pic.pdf}), где изменение формы кривой (а также размера пиков, см.~ниже) хорошо заметно. Он пишет:

\begin{quote}
На рис.~7 [рис.~\ref{kupriyanov-2018-3pic.pdf} в нашей нумерации] представлены диагностические графики для президентских выборов в России за 2000~--  2018~гг. Сверху~--- вниз: годы 2000, 2004, 2008, 2012, 2018. Слева~--- направо: диаграмма рассеяния в осях <<количество зарегистрированных избирателей>> / <<явка на участке>>; гистограмма явки на участках; диаграмма рассеяния в осях <<явка на участке>> / <<доля голосов за лидера на участке>>; гистограмма долей голосов за лидера на участке.

Обратите внимание на <<волны>> явки на <<круглых>> процентах, которые становятся заметны с 2004 года (в 2004 году особенно хорошо видна <<волна>> на отметке 50\%~--- в этом году ещё сохранялось требование о наличии кворума) и аналогичные <<волны>> доли голосов, отданных за лидера, наиболее заметные в 2008 году, но наблюдаемые с 2004 года. Заметно улучшение ситуации с явкой в 2012 году и ухудшение в 2018 году.
\end{quote}

\subsection{Севастополь: оценка $p$-value}

Интересные официальные результаты были получены на референдуме\footnote{Термин <<референдум>> используется для краткости и не подразумевает какой-либо юридической или моральной оценки события.} в Севастополе. А именно, как обнаружил один из читателей блога Александра Киреева в <<Живом журнале>>, данные из официальных результатов оказываются <<круглыми числами>>:
\begin{quote}
В комментах к посту о нарисованных результатах в ДНР, некий аноним заметил то, что пока никто из нас до сих пор не заметил: результаты референдума о присоединении Севастополя тоже совпали с точностью до человека! Мы все внимательно смотрели на Крым, но там проценты не совпали с точностью до человека, хотя, на мой взгляд, они все равно там как минимум частично рисовались. Итак:

Официальные данные по городу Севастополю

Вы за воссоединение Крыма с Россией на правах субъекта Российской Федерации? $262041$  $\langle\ldots\rangle$

Вы за восстановление действия Конституции Республики Крым 1992 года и за статус Крыма как части Украины? $9250$ $\langle\ldots\rangle$

Действительных голосов $271291$  $\langle\ldots\rangle$

Недействительных голосов $2810$ $\langle\ldots\rangle$

Всего голосов $274101$ $\langle\ldots\rangle$

Общее число зарегистрированных избирателей $306258$ $\langle\ldots\rangle$

Видим нарисованную цифру проголосовавших~--- $274101/306258=0{,}895000$

цифра <<за>>~---  $262041/274101=0,95600$

Попадание с точностью до человека сразу по двум позициям!
$\langle\ldots\rangle$

update:  Процент <<за Украину>> у них видимо тоже посчитанный ~--- только криво: $9250$~--- это ровно $3,53\%$ ($3,52998\%$)от\ldots\ $262041$. То есть от количества голосов <<за Россию>>. Не на ту строчку умножили. Ну а недействительные просто написали по остаточному принципу\ldots\ Вот и вся разгадка.~\cite{kireev-sebastopol}
\end{quote}

Это замечание описывает некоторую статистическую аномалию в официально объявленных результатах~\cite{2014-sebastopol-official}. Она состоит в том, что если вычислить три отношения, то они оказываются <<круглыми>>:
\begin{itemize}
\item $274101/306258=0{,}895{\color{red}000}294$ (явка)
\item 262041/274101=0,956{\color{red}00}162 (голоса <<за>>)
\item 9250/262041 = 0.0352{\color{red}99}819 (отношение <<за>>/<<против>>)
\end{itemize}
Более формально, в каждом случае числитель дроби получается округлением результата умножения знаменателя на десятичную дробь с тремя значащими цифрами:
\begin{itemize}
\item $306258\cdot 0.895=274100.91 \approx 274101$;
\item $274101\cdot 0.956=262040.556 \approx 262041$;
\item $262041\cdot 0.0353 =9250.047\approx 9250$.
\end{itemize}

Возникает вопрос, насколько правдоподобно, что такая аномалия возникла <<случайно>>. Статистическая практика говорит, что надо оценить вероятность такой аномалии при нулевой гипотезе (<<на выборах производился подсчёт бюллетеней>>). Это значение называется $p$-value, и чем оно меньше, тем сильнее аргумент против нулевой гипотезы.

Наивный подсчёт $p$-value  можно было бы произвести так: посмотреть, как часто встречаются числа с аналогичным свойством среди возможных числителей. Следующие числа будут:
\begin{itemize}
\item $306258\cdot 0.896\approx 274407$ ($+307$);
\item $274101\cdot 0.957\approx 262314$ ($+273$);
\item $262041\cdot 0.0354 \approx 9276$ ($+26$).
\end{itemize}
Если теперь перемножить $1/307$, $1/273$ и $1/26$, то получится примерно одна двухмиллионная. Однако это вычисление требует коррекции сразу по нескольким причинам.

\begin{itemize}
\item Наша <<нулевая гипотеза>> представляет собой на самом деле класс гипотез: мы не знаем соответствующих распределений (вероятности того или иного значения числителя). Мы можем ожидать, что они меняются достаточно медленно между соседними круглыми значениями, но нет причины считать их постоянными. Поэтому надо ввести запас~--- множитель, оценивающий сверху такие колебания. Консервативная оценка: $1{,}5$ для каждого из трёх отношений. Надо ещё отметить, что эти отношения не связаны никаким соотношением (нельзя восстановить одно из них по остальным, исходя из правил предположительного подсчёта).

\item
Как отмечал ещё Борель, множество различных статистических аномалий, не определённых до опыта, может быть очень велико. В данном случае гипотеза была сформулирована до знакомства с данными (это наблюдение было сделано при обсуждении аналогичных аномалий в других референдумах). Но можно было бы рассматривать и другие отношения: если первые два представляют собой два традиционно считаемых главными параметра (явка и голоса <<за>>), то третье отношение обычно не вычисляется и ничем не лучше других отношений. Чтобы учесть это с запасом, введём поправочный коэффициент $10$ (Bonferroni correction).

\item Одновременно был объявлен референдум в остальной части Крыма, и аналогичная аномалия в его результатах (не наблюдаемая) была бы, пожалуй, не менее удивительной. Тут трудно выбрать поправочный коэффициент, поскольку непонятно, считать ли другую часть Крыма равноправным районом или делить на части. Учитывая соотношение между количеством жителей, оценим поправочный коэффициент как $2$--$6$.

\item Для надёжности, чтобы покрыть не учтённые нами факторы, дополнительно возьмём множитель $10$.
\end{itemize}
Теперь вычисление даёт верхнюю оценку для $p$-value:
$$
\frac{1}{307}\times \frac{1}{273}\times\frac{1}{26} \times 1{,}5\times 1{,}5\times 1{,}5\times 6\times 10\times 10 \approx \frac{1}{1000}=0{,}1\%
$$

Распространённая ошибка состоит в том, чтобы назвать $p$-value вероятностностью и сказать, что <<вероятность нулевой гипотезы (получения результатов подсчётом) не превосходит одной десятой процента>>. Это неправильно: во-первых, вообще говорить о вероятности нулевой гипотезы в условиях отсутствия естественного распределения на гипотезах не имеет смысла, во-вторых, если пытаться придать этому выражению смысл и применять формулу Байеса, то в неё входит априорная вероятность. Так что выводы из такого значения $p$-value по необходимости субъективны (зависят от наличия и правдоподобности альтернативных гипотез, важности вопроса, наличия каких-либо свидетельств в пользу нулевой гипотезы и др.)~--- можно отметить лишь, что подсчитанная нами оценка существенно ниже традиционных порогов ($5\%$, $1\%$), используемых при оценке значимости в статистических экспериментах~---  так что результаты с такими $p$-values обычно считают надёжно установленными.

\subsection{Анализ пиков и сетки}

\begin{figure}[!h]
\begin{center}
\includegraphics[width=0.5\textwidth]{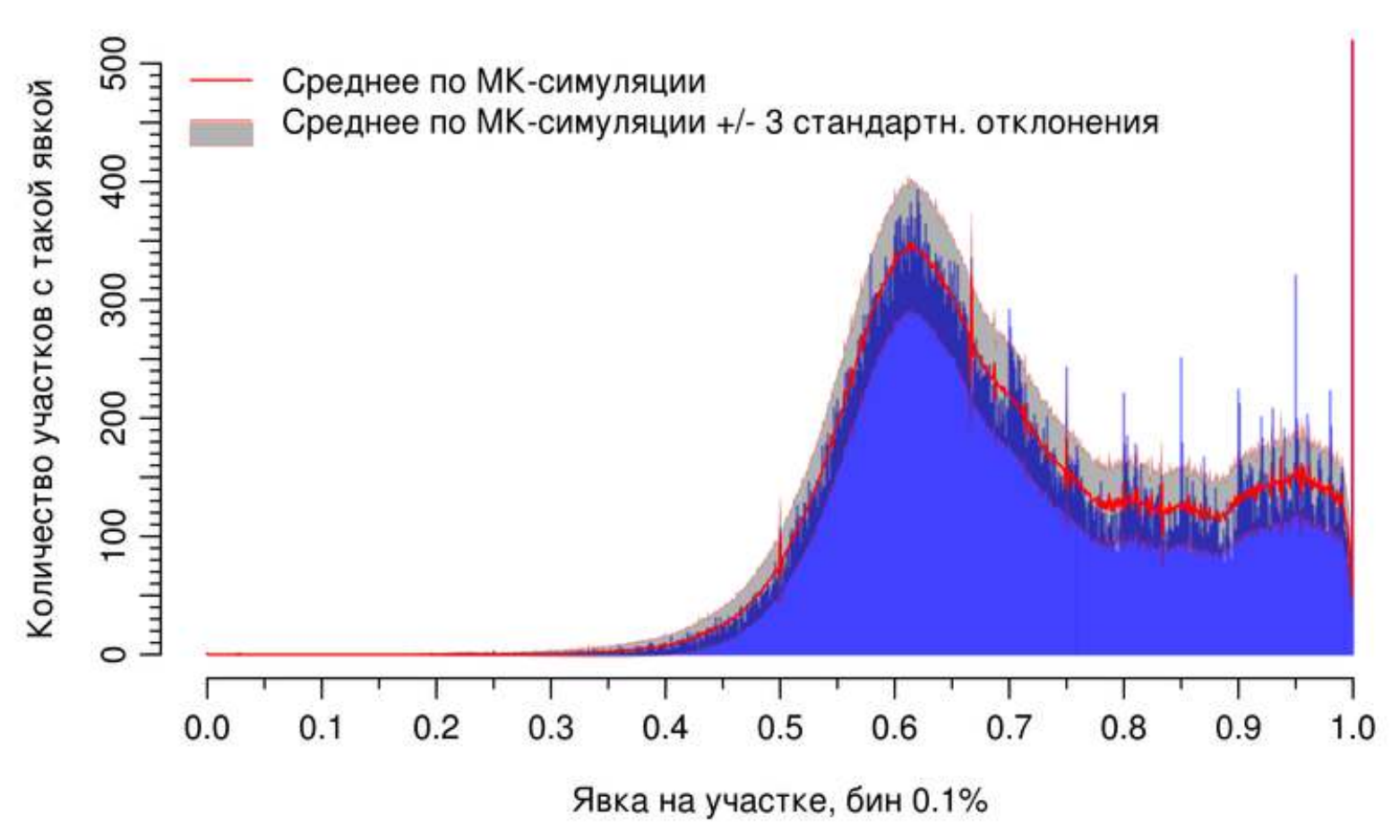}
\end{center}
\caption{Пики 2018 года и их анализ с помощью метода~\cite{KSP-integer-percentages}, рисунок из~\cite{kupriyanov-2018}.}\label{2018-kupriyanov-2}
\end{figure}

\begin{figure}[!h]
\begin{center}
\includegraphics[width=0.8\textwidth]{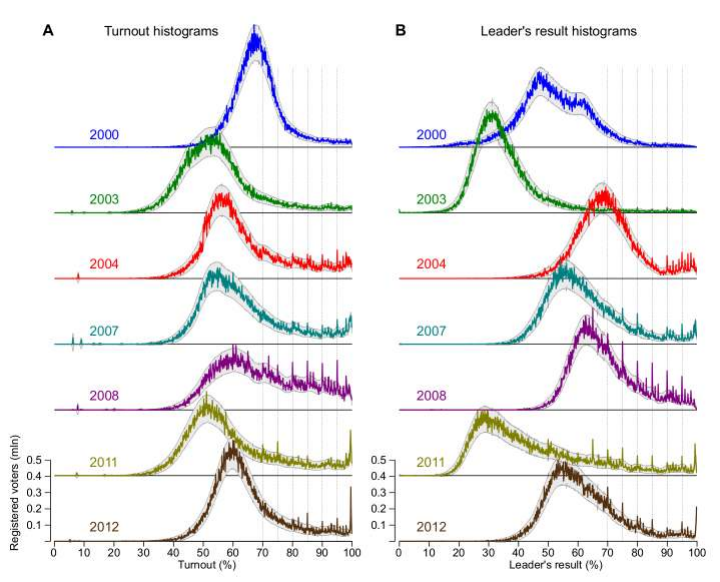}
\end{center}
\caption{(A) Гистограммы явки для всех выборов 2000--2012 (сверху вниз). Все гистограммы показывают общее число зарегистрированных избирателей на всех участках с данной явкой плюс-минус $0.05\%$ (например, значение на $70\%$ соответствует явкам от $69.95\%$ до $70.05\%$). Отмечены границы, соответствующие $99\%$ интервалам при моделировании по методу Монте-Карло с $10000$~повторениями. Значения с $100\%$ явкой не показаны. (B) Аналогичные гистограммы для результатов <<лидера>>, значения в $100\%$ также опущения для удобства сравнения.~\cite{KSP-integer-percentages}.}\label{KSP-integer-percentages.pdf}
\end{figure}

\begin{figure}[!h]
\begin{center}
\includegraphics[width=0.4\textwidth]{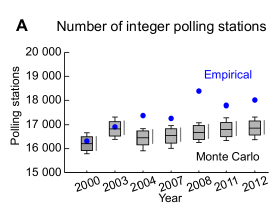}
\end{center}
\caption{(A) Числа участков с близким к целому ($\pm 0.05\%$) числом процентов явки или результата <<лидера>>~\cite{KSP-integer-percentages} показаны синими точками. Прямоугольники показывают $0.5\%$ и $99.5\%$ квантили по результатам моделирования, вместе с математическим ожиданием (горизонтальная линия), вверх и вниз вынесены максимальные и минимальные значения, наблюдавшиеся при моделировании. }\label{KSP-integer-percentages-2.pdf}
\end{figure}

\begin{figure}
\begin{center}
\includegraphics[width=0.5\textwidth]{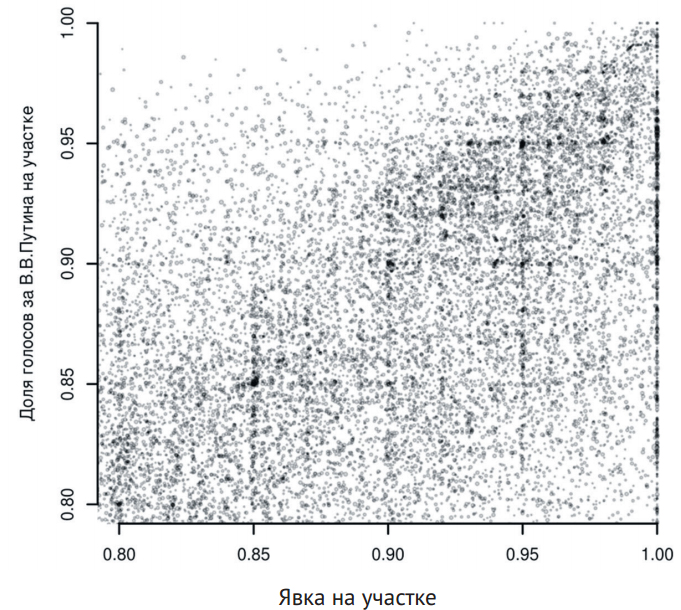}
\end{center}
\caption{Сетка в 2018 году,  рисунок из~\cite{kupriyanov-2018}.}\label{2018-kupriyanov-1}
\end{figure}

В связи с анализом $p$-value возникает вопрос~--- можно ли получить какие-то количественные оценки для (пожалуй, наиболее очевидной) ситуации с пиками на <<круглых числах>>? Несмотря на очевидную уязвимость такого рода действий, такие пики на наблюдались и при голосовании 2018 года. (рис.~\ref{2018-kupriyanov-2}, см.~\cite{kupriyanov-2018}).
 На этом рисунке, помимо самих пиков, показаны результаты моделирования, предложенного в~\cite{KSP-integer-percentages}. При таком моделировании на каждом участке берутся официальные результаты явки (или доли голосов <<за>>) и предполагается, что каждый избиратель независимо принял решение о явке (о голосовании <<за>>) c такими вероятностями. Дальше считается, сколько голосов пришлось на участок с данным <<круглым>> процентом. В полученном массиве данных (в~\cite{KSP-integer-percentages} моделирование проводилось $10000$ раз, так что для каждого участка получалось $10000$ чисел. Затем были отброшены верхние $1\%$ чисел, и взято наибольшее из оставшихся. Превышение этого уровня означает, что произошло событие, которое при моделировании наблюдалось менее чем в $1\%$ случаев. Авторы нарисовали кривые для разных лет, где видно систематическое превышение этого уровня на <<круглых>> процентах (рис.~\ref{KSP-integer-percentages.pdf})
Интегральная оценка (по всем целым процентам) даёт ещё более заметное превышение (начиная с 2004 года), авторы оценивают соответствующее $p$-value как $0.0001$.  Зависимость от года показана на рис.~\ref{KSP-integer-percentages-2.pdf}.
Отметим в заключение раздела, что и <<сетка>> на результатах продолжает быть хорошо заметной (рис.~\ref{2018-kupriyanov-1}, см.~\cite{kupriyanov-2018}).

\subsection{Графики и видео}

Интересные результаты удалось получить с помощью видеонаблюдения. В 2018 году (как и в некоторые предыдущие годы, начиная с 2012) на многих участках были установлены видеокамеры, сигнал с которых транслировался в интернет, и его можно было записать (или иногда получить официальные записи). Получилась довольно большая коллекция интересных записей, см., например,~\cite{video-vbros2016} с участка 174 в Казани (напомним, что правилами выборов не предусмотрено добавление каких бы то ни было бюллетеней в стационарные ящики для голосования). Но более интересно, что можно, не искажая происходящего на участке процесса, произвести достаточно точный подсчёт явки, просто посмотрев видео и подсчитав подходивших к урне (конечно, кто-то мог подойти и не опустить бюллетень, или кто-то мог опустить не только свой бюллетень, но и бюллетень члена семьи, и это не является нарушениями, но погрешность тут невелика).

\begin{figure}[t]
\begin{center}
\includegraphics[width=0.95\textwidth]{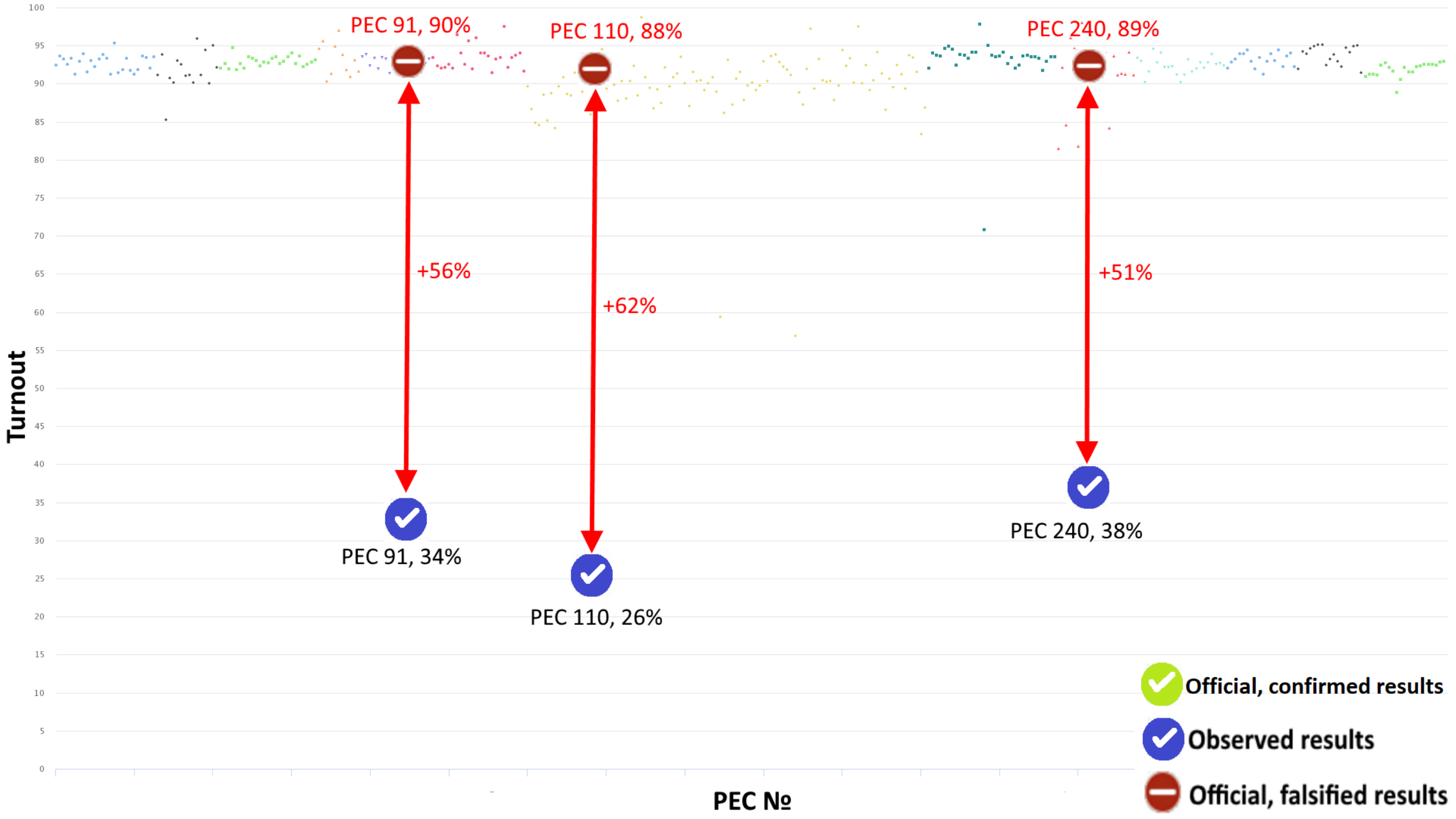}
\end{center}
\caption{Данные наблюдений в Кабардино-Балкарии и официальные результаты~\cite{mateusz-bajek-talk}. По горизонтали отложены номера участков, по вертикали~--- явка, разноцветные мелкие точки сверху показывают официальные результаты. Стрелки связывают официальные результаты на трёх участках с тем местом, где они должны были бы быть по результатам подсчёта, выполненного наблюдателями.}\label{bajek-1.pdf}
\end{figure}

Такой опыт был проделал наблюдателями из Польши~\cite{mateusz-bajek}, которые выбрали для наблюдения несколько наугад взятых участков в трёх регионах (Пенза, Татарстан, Кабардино-Балкария) на голосовании~2018 года. Они зафиксировали нарушения во всех трёх регионах, но доля их была разной. Они пишут (привожу русский перевод текста в~\cite{mateusz-bajek}):

\begin{quote}
В избирательной комиссии номер 857 в Пензенской области официальная явка в $18$ часов вечера была $80\%$. Мы наблюдали лишь $47\%$.

Столь же странная ситуация была в комиссиях с номерами 2029 и 1792 в Татарстане. В отличие от наших подсчётов, которые показали явку в $52$-$54\%$, официальные данные были $80\%$ и $82\%$, примерно на $30\%$ больше.

В то же время в двух других комиссиях в Татарстане и трёх других комиссиях в Пензенской области, [где велось наблюдение, официальная] явка в точности соответствовала нашим наблюдениям.

Что можно сказать про явку в Кабардино-Балкарии? Во всех комиссиях, где мы вели наблюдение (номера 91, 110 и 240) официальная явка сильна отличалась от наших оценок. По нашим оценкам, явка была в диапазоне от $21\%$ до $34\%$, в то время как официальные данные были от $76\%$ до $86\%$.





\end{quote}
Результаты такого рода наблюдений поучительно поместить в контекст официальных результатов, как сделал Mateusz Bajek в своём докладе~\cite{mateusz-bajek-talk}, рис.~\ref{bajek-1.pdf}.

\begin{figure}[!h]
\begin{center}
\includegraphics[width=0.7\textwidth]{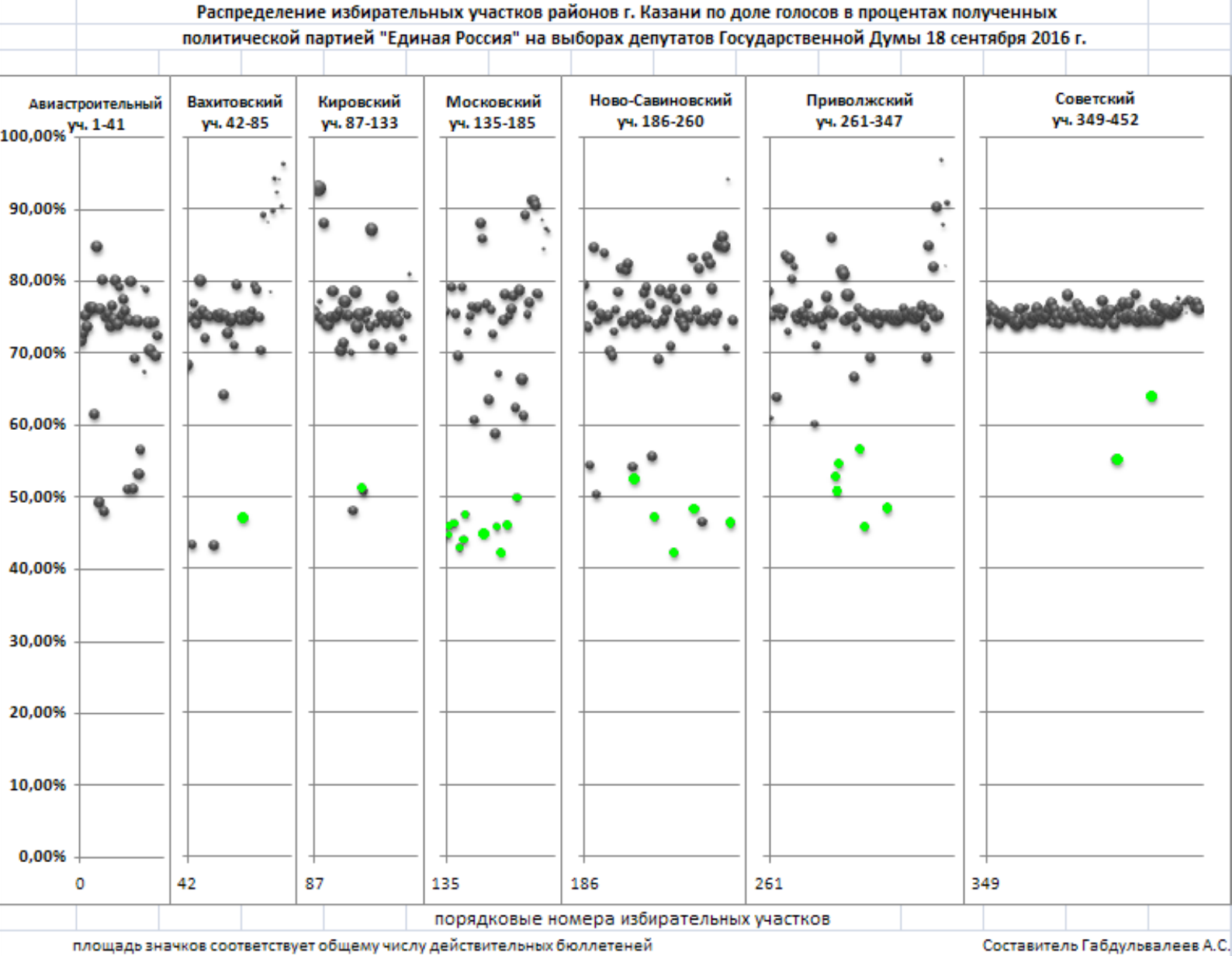}
\end{center}
\caption{Официальные результаты голосования 2016 года (Татарстан) из~\cite{tatobservers-ER2016}. По горизонтали~--- номера участков, по вертикали~--- официальные результаты <<Единой России>>. На участках, отмеченных зелёным цветом, присутствовали наблюдатели.}\label{tatobservers-ER2016.pdf}
\end{figure}

На рис.~\ref{tatobservers-ER2016.pdf} показана аналогичная диаграмма (Азат Габдульвалеев, Ассоциация наблюдателей Татарстана, голосование 2016 года~\cite{tatobservers-ER2016}). На участках, отмеченных зелёными точками, присутствовали наблюдатели.

\subsection[Оценки размера фальсификаций: метод Шпилькина]{Оценки размера фальсификаций:\\метод Шпилькина}

Часто задают вопрос о том, можно ли оценить <<размер фальсификаций>>, <<число фальшивых голосов>> и пр. по данным статистики. Стоит сразу же отметить, что само это понятие не вполне точно определено (скажем, как нужно учитывать принудительное голосование?), и смысл гипотетически восстановленных результатов не вполне ясен, поскольку голосование является лишь частью избирательного процесса, и нарушения в других частях меняют его смысл.\footnote{Проводя аналогию со спортом, можно было бы спросить, надо ли пытаться вычесть из результата прыгуна в длину то расстояние, на которое он заступил начальную черту, или же аннулировать этот результат. Другая аналогия: надо ли корректировать число взяток в карточной игре, если у игрока обнаружен туз в рукаве?} Тем не менее с довольно большой точностью, видимо, этот вопрос всё же технически корректен, и такие попытки предпринимались разными авторами.

\begin{figure}
\begin{center}
\includegraphics[width=0.5\textwidth]{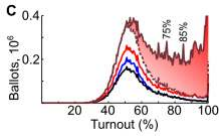}
\end{center}
\caption{Доли различных партий при голосовании 2011 года как функция от явки на участке}\label{KSP-anomalies-page2.pdf}
\end{figure}

Один из подходов основан на наблюдении из~\cite{KSP-anomalies}, показанном на рис.~\ref{KSP-anomalies-page2.pdf} (фрагмент рис.~1 по нумерации статьи).
По горизонтали отложена явка, по вертикали~--- доля голосов, которую получил каждый из кандидатов на участках с такой явкой. Наблюдение состоит в том, что при небольших значениях явки результаты кандидатов пропорциональны друг другу с довольно высокой точностью, а после этого результаты <<лидера>> отклоняются вверх (и имеют пики), в то время как результаты остальных кандидатов остаются примерно пропорциональными. Одно из возможных объяснений этому состоит в том, что к реальным бюллетеням добавляются бюллетени (физические или при переписывании результатов) за <<лидера>>, отчего эти участки, во-первых, сдвигаются в зону высокой явки, а, во-вторых, на них результат <<лидера>> оказывается непропорционально высоким. Если принять такое объяснение, то количество этих добавленных бюллетеней можно оценить по графику, выбрав область невысоких явок и скорректировав справа от неё результат <<лидера>>, исходя из гипотезы пропорциональности.

Но какие есть причины считать, что это объяснение правильно (в большинстве случаев или по крайней мере в значительной доле случаев)? Почему не предположить, следуя Неретину, что есть какие-то особые участки или регионы, где и явка больше, и доля <<лидера>> выше?

На это есть нескольо возражений. Во-первых, сам факт пропорциональности (с хорошей точностью) результатов всех участников в диапазоне небольших явок и не-лидеров по всему диапазону служит некоторым достаточно информативным методом внутренного контроля: если есть разброс параметров по участкам, нет никаких причин ожидать такой пропорциональности в области низких явок (и везде для остальных кандидатов). Поэтому это уже является аргументом в пользу объяснения с помощью добавок.

\begin{figure}
\begin{center}
\includegraphics[width=0.95\textwidth]{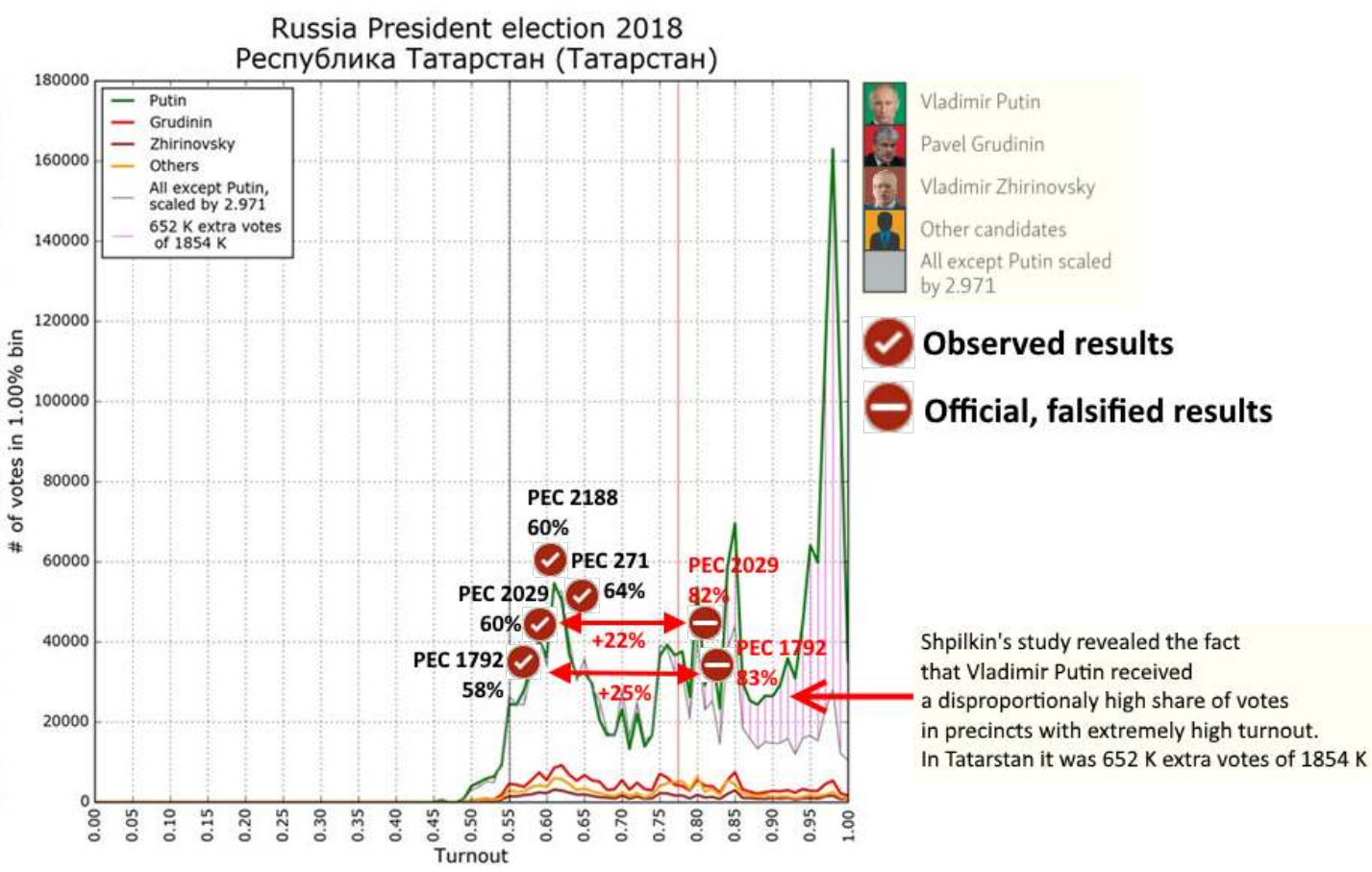}
\end{center}
\caption{Данные наблюдений и перемещение точек на гистограмме явки.}\label{bajek-2.pdf}
\end{figure}

Во-вторых, есть возможность сравнить диаграммы с результатами на отдельных участках, полученными наблюдением. Такого рода картинки приведены в уже цитировавшемся докладе~\cite{mateusz-bajek-talk}, см. рис.~\ref{bajek-2.pdf}
Здесь снова видна пропорциональность в диапазоне небольших явок и видно, что в тех двух случаях из четырёх проверенных, когда явка выходила из диапазона пропорциональности, она не соответствовали измерениям наблюдателей (интерес представляют горизонтальные координаты точек).

\begin{figure}[!h]
\begin{center}
\includegraphics[width=0.95\textwidth]{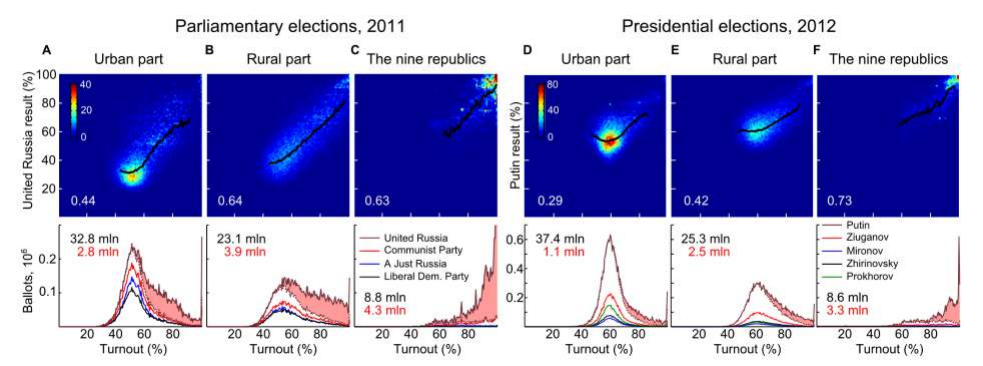}
\end{center}
\caption{Доли различных партий при голосованиях 2011 и 2012 года как функция от явки на участке,  для трёх групп участков: условный <<город>>, условное <<село>> и отдельно посчитанные девять республик.}\label{KSP-anomalies-page12.pdf}
\end{figure}

В-третьих, если предположить, что появление непропорционально больших результатов для <<лидера>> объясняется географической неоднородностью, то это должно проявляться при переходе к отдельным регионам или к отдельным категориям участков (<<город vs. село>>). Такая оценка была сделана в~\cite{KSP-anomalies}, см. рис.~\ref{KSP-anomalies-page12.pdf} (рис.~12 по нумерации статьи).
Видно, что эффект сохраняется и при разделении (по крайней мере в первых двух группах: в третьей участок малых явок, где можно было бы <<поймать>> пропорциональность, заметен плохо).

Из этих же графиков видно и четвёртое возражение к гипотезе неоднородности: хотя парламентские и президентские выборы имеют свои особенности, всё же сильное изменение неоднородности за год (при сохранении общего вида кривых) выглядит маловероятным.

Что же касается внутренного контроля для всей России и для отдельных регионов, то соответствующие графики очень наглядно представлены на анимациях Кирилла Калинина~\cite{kalinin-shpilkin}, где можно (для отдельных регионов и для всей России) наблюдать, как выбор порога <<нормальных явок>> влияет на пропорциональность.

Несмотря на все эти подтверждения, ограничения этого метода очевидны. Во-первых, если в регионе почти нет достоверных данных (ср. рис.~\ref{bajek-1.pdf}), то информацию извлечь просто неоткуда. Во-вторых, модель добавления голосов за <<лидера>>~--- не единственно возможная, голоса могут и переписываться при сохранении явки, и явка может увеличиваться изменением общего числа избирателей в списке, и так далее. Поэтому к конкретному значению числа <<добавленных голосов>>, полученному таким методом, не следует относиться слишком серьёзно: хотя так можно отличить, условно говоря, $10\%$ от $1\%$, просто посмотрев на график на приведённых рисунках, но точные оценки тут  сомнительны.

\subsection{Оценки Мебане~--  Калинина}

\begin{figure}[!h]
\begin{center}
\includegraphics[width=0.75\textwidth]{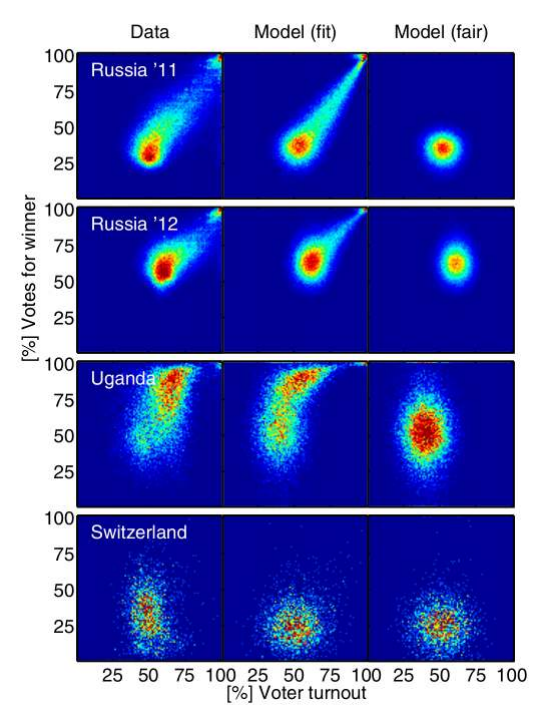}
\end{center}
\caption{Сравнение наблюдаемых и модельных результатов голосований: Россия в 2011 году, Россия в 2012 году, Уганда и Швейцария. Левая колонка: объявленные результаты. Средняя: построенное приближение в рамках модели. Правая: ожидаемые по модели результаты, если исключить два механизма фальсификаций. (Рисунок с номером 4 из работы~\cite{klimek-et-al}.)}\label{klimek-et-al-page4.pdf}
\end{figure}

Чтобы повысить надёжность оценок, полезно сравнивать оценки, получаемые разными методами, по возможности независимыми. Такая возможность возникает благодаря работе~\cite{klimek-et-al} \eng{(Klimek, Egorov, Hanel, Thurner)}, где предложен другой метод оценки <<восстановленного результата>> голосования, хотя и исходящий из тех же данных, что и метод из~\cite{KSP-anomalies}, но действующий иначе. Авторы рассматривают двумерную гистограмму и пытаются приблизить её некоторой моделью, включающей в себя некоторые доли добавленных голосов (смещение результатов) и <<рисованных>> (не связанных с оригинальными результатами) участков. Результаты такого подбора модели показаны на рис.~\ref{klimek-et-al-page4.pdf}.
Левая колонка показывает объявленные результаты в координатах (явка, доля <<лидера>>). Она приближается трёхкомпонентной моделью, и результат приближения (наилучшее приближение в рамках модели) показано в средней колонке. Правая колонка показывает, что было бы, если в модели убрать фальсификации.  Как пишут авторы, <<для Швейцарии приближающая модель почти совпадает с её ``честной''\ частью. Результаты для России и Уганды можно объяснить моделью с большим числом участков с фальсификациями.>> (\eng{``For Switzerland, the fair and fitted models are almost the same. The results for Russia and Uganda can be explained by the model assuming a large number of fraudulent units.''})

Результат третьей колонки можно считать <<восстановленным>> результатом голосования согласно предложенной модели: там можно посмотреть и явку, и долю <<лидера>>. Видно, что примерно тот же результат можно получить, выбрав точку максимума и воспринимая её как <<восстановленные результаты>>. Это даёт близкие оценки к оценкам предыдущего метода (что, однако, не так удивительно, так как они исходят из общих предположений о природе фальсификаций).

Попытка усовершенствовать метод из~\cite{klimek-et-al} была предложена Калининым и Мебане. Они получили принципиально другие результаты~\cite{mebane-2016,kalinin-mebane-2017}; о них рассказано в интервью Калинина~\cite{kalinin-meduza}. Отличие весьма значительно: например, для голосования 2008 года в таблице 3 в~\cite{kalinin-mebane-2017} приведена оценка менее $0.12\%$, а для голосования 2011 года там же даётся оценка менее $1.4\%$ (самая большая, кроме голосования 2016 года, где оценка $3.5\%$). Эти оценки решительно (на порядок, если не на два) расходятся с другими оценками, в том числе с оценками из~\cite{klimek-et-al}.

Вероятную причину такого различия можно найти в работе Мебане 2016 года~\cite{mebane-2016}, в которой описан применяемый метод (раздел 3). Судя по (не очень внятному) описанию в тексте, он состоит в нахождении наилучшего приближения к данным с помощью изменения параметров модели. Эти параметры включают в себя три неотрицательных числа, $f_0, f_\mathrm{i}, f_\mathrm{e}$, в сумме равных единице. Они интерпретируются как вероятности того, что на участке не будет фальсификаций ($f_0$) и того, что будут фальсификации одного или другого вида ($f_1,f_2$).  Величина, которую предлагается максимизировать, указана в формуле $(6)$  на странице 10 как
\[
\sum_{j\in\{0,\mathrm{i},\mathrm{e}\}} f_j \prod_{i=1}^{n} g_{jW}(W_i\mid N_i;\Psi)g_{jA}(A_i\mid N_i;\Psi).
\]
Здесь для нас существенно, что величины $f_j$ не входят в выражение под знаком произведения. Поэтому, если в какой-то точке (для какого-то значения параметров) достигается максимум, то этот максимум, в частности, будет достигаться и в ограниченной задаче, где остальные параметры фиксированы, а меняются лишь параметры $f_0+f_\mathrm{i}+f_\mathrm{e}=1$. Область изменения этих параметров является треугольником, а максимизируемая функция является линейной функцией на треугольнике. Если, как в~\cite{kalinin-mebane-2017}, в найденной точке максимума все три параметра отличны от нуля, то либо максимум найден неправильно (он в невырожденном случае должен достигаться на границе треугольника), либо максимизируемая (линейная) функция постоянна. В последнем случае найденные значения параметров также не имеют смысла, поскольку для постоянной функции максимум достигается в любой точке, и потому значения параметров можно считать произвольными.

Влияние этой ошибки на материале выборов обсуждается в~\cite{kalinin-reply}.

\subsection{Оценки Сулакшина}

Отметим также любопытную работу С.\,С.\,Сулакшина <<О незаконности\footnote{Оценка законности~--- в каком бы то ни было смысле~--- мероприятия электорального типа, состоявшегося 18 марта 2018 года, не относится к статистическому анализу объявленных результатов и далеко выходит за рамки настоящего обзора.} перевыборов Путина 18 марта 2018 года>>~\cite{sulakshin}. Согласно его оценкам, приведённым на с.~2 в~\cite{sulakshin}, <<реальная явка составила $50$--$55\%$, а не официальные $67.5\%$>> и <<реально Путин В.В. получил не $76\%$ голосов от числа проголосовавших, а всего $49\%$>>, из чего автор заключает, что <<от числа зарегистрированных избирателей это составляет $33\%$, от всего населения России $24.5\%$. Уже эти оценки вызывают некоторое недоумение: если явилось чуть больше половины зарегистрированных избирателей (а именно так считается явка), и примерно половина проголосовала <<за>>, то число проголосовавших <<за>> должно составить чуть больше четверти всех зарегистирированных избирателей, а не $33\%$.

\begin{figure}
\begin{center}
\includegraphics[width=0.9\textwidth]{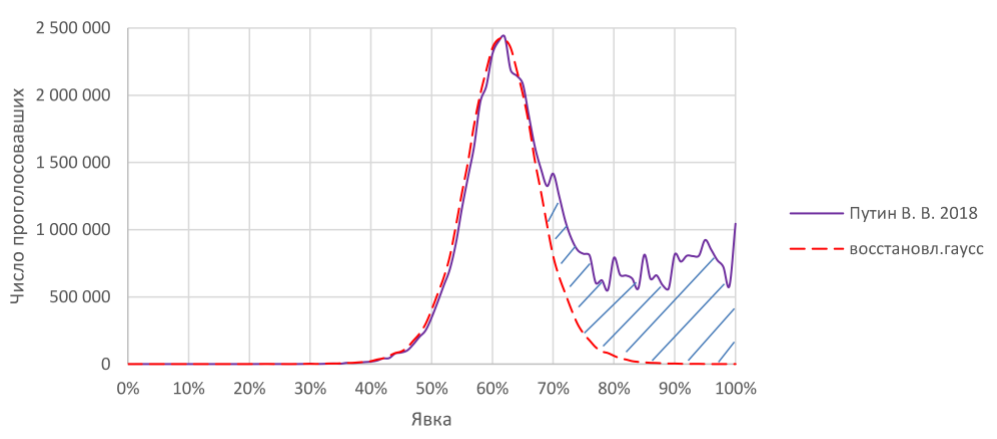}
\end{center}
\caption{Рисунок 14 из~\cite{sulakshin}, где он снабжён подписью: <<Рис. 14. Итоговая официальная кривая и восстановленная по гауссу [здесь и далее строчная буква оригинала] достоверная. Отношение площадей под кривыми ``восстановленная кривая гаусса'' / ``официальный результат Путина'' даёт коэффициент фальсификации в $1{,}57$ раза.>>}\label{sulakshin-page54.pdf}
\end{figure}

Работа состоит из нескольких частей. В первой приводятся (среди прочего) фотографии и свидетельства нарушений на выборах. Далее, на с.~53, идёт раздел <<Факт и масштаб фальсификаций. Основные результаты и их анализ>>. Там говорится (с.~54):
\begin{quote}
Главный вывод проведённой по <<модели гаусса>> [строчная буква в оригинале] реконструкции таков, что если правое крыло распределения голосов от явки восстановить по левой части гауссового вида, и вычесть из общей, в том числе фальсифицированной, площади заштрихованный кусок, то результат Путина на выборах становится $49{,}08\%$. Рис.14. [см. рис.~\ref{sulakshin-page54.pdf}]. Это означает~--- Путин не выиграл даже в первом туре.
\end{quote}
К этой оценке возникают очевидные вопросы (даже если отвлечься от правомерности приближения <<кривой Гаусса>>): на рисунке не указано, отложены ли по вертикали голоса <<за Путина>> или все проголосовавшие. Но в любом случае коэффициент $1{,}57$, полученный, по словам автора, сравнением площадей под графиками, не имеет отношения к делу. Если по оси ординат отложены все голоса, то удаление <<фальсифицированных>> голосов, даже если предположить, что все они были поданы за Путина, приводит к равному уменьшению и числителя, и знаменателя дроби, выражающей результат Путина (долю от проголосовавших), и нельзя уменьшать всю дробь в такое число раз (это бы соответствовало уменьшению только числителя). Если по оси ординат отложены голоса за Путина, то и в этом случае уменьшение площади в $1{,}57$ раз соответствует уменьшению числителя дроби, а надо учитывать также и уменьшение знаменателя.

\begin{figure}
\begin{center}
\includegraphics[width=0.9\textwidth]{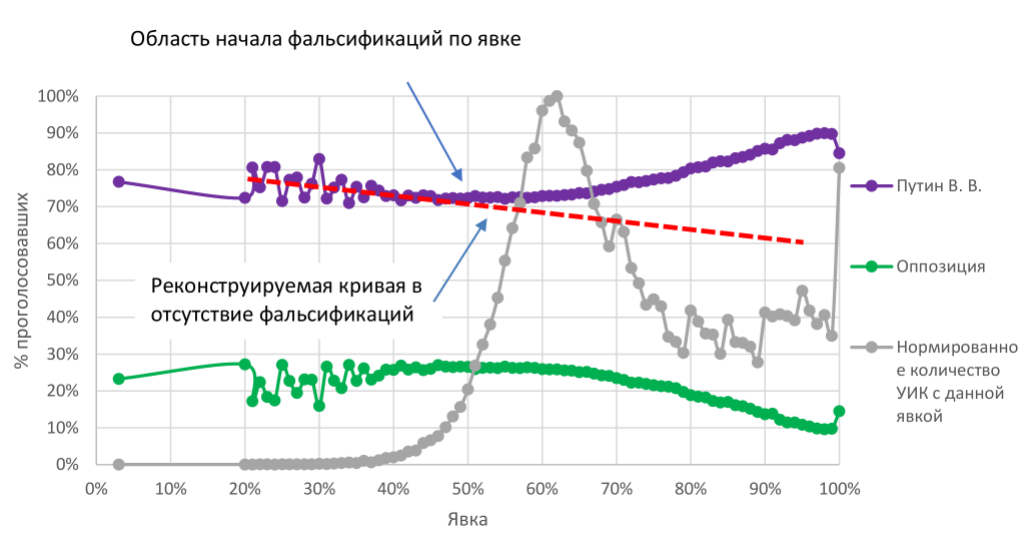}
\end{center}
\caption{Рисунок 29 из~\cite{sulakshin}, где он снабжён подписью: <<Рис.~29. Экстраполируемая сходимость на больших явках свидетельствует, что результат Путина существенно меньше (на двузначные числа), чем официальные $77\%$.>>}\label{sulakshin-page70.pdf}
\end{figure}

Другие аргументы в пользу приводимой оценки также выглядят малоубедительными. Линейная интерполяция (с. 69 и рис. 29, воспроизведённый здесь как рис.~\ref{sulakshin-page70.pdf}), опирается на показанный автором в виде пунктирной прямой <<понижающийся тренд>> зависимости результата Путина от явки, который не вытекает из данных рисунка (основная часть голосовавших приходится на максимум кривой явки). Но даже если принять этот <<тренд>>, то нет никаких причин рассматривать <<экстраполируемую сходимость на больших явках>>.
Оценки явки и доли голосов <<за>> по графикам зависимости от размера участка опираются на странные соображения, которые трудно комментировать: <<Явка не должна зависеть от размера участка. А фальсифицировать тем труднее, чем больше участок. Поэтому для гипотетического бесконечно большого участка явка будет истинной. Из рис.~31 [зависимость явки от размера участка] видно, что истиная явка меньше, чем официальная, данная ЦИК, которая, соответственно, сфальсифицирована.>>

На с.~76 отмечается, что доля проголосовавших <<за>> по данным экзит-поллов ВЦИОМ мало зависит от времени, когда люди приняли участие в голосовании. Из этого автор делает вывод, что <<в естественном случае выборов без фальсификаций предпочтения избирателей при статистически значимой выборке от явки не зависят>>, путая явку как приход избирателя в то или иное время дня при голосовании и явку как долю явившихся в течение всего дня (по отношению ко всем зарегистированным на участке избирателям).

К сожалению, для приводимых графиков автор не указывает ссылок, и потому потенциально ценные материалы пропадают для анализа, поскольку неизвестно, получил ли их автор самостоятельно (и как) или заимствовал из более надёжных источников.

\section{Добавления 2020 года}
\subsection{Голосование в июне--июле 2020 года}

Весной 2020 года был принят закон <<О поправке к Конституции
Российской Фе\-де\-ра\-ции>>~\cite{2020law}, согласно которому в поправляемый
документ были внесены различные изменения, в том числе и такое:

\begin{quote}
Положение части 3 статьи 81 Конституции Российской Федерации,
ограничивающее число сроков, в течение которых одно и то же лицо
может занимать должность Президента Российской Федерации,
применяется к лицу, занимавшему и (или) занимающему должность
Президента Российской Федерации, без учёта числа сроков, в
течение которых оно [лицо] занимало и (или) занимает эту
должность на момент вступления в силу поправки к Конституции
Российской Федерации, вносящей соответствующее ограничение, и не
исключает для него возможность занимать должность Президента
Российской Федерации в течение сроков, допустимых указанным
положением.
\end{quote}

\begin{figure}[!h]
\begin{center}
\includegraphics[width=0.7\textwidth]{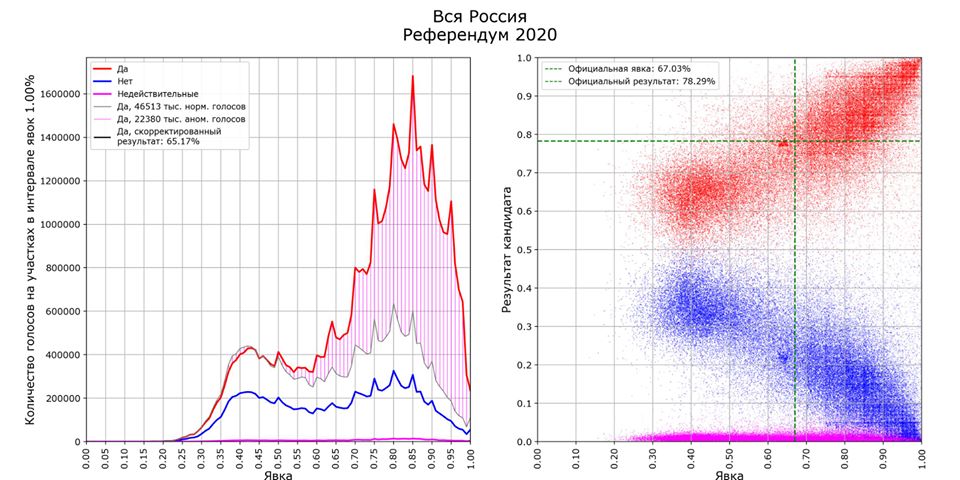}
\end{center}
\caption{Гистограмма и двумерное представление официальных
данных по голосованию в июне--июле 2020 года~\cite{2020allrussia}.}
\label{2020allrussia}
\end{figure}

\begin{figure}[!h]
\begin{center}
\includegraphics[width=0.5\textwidth]{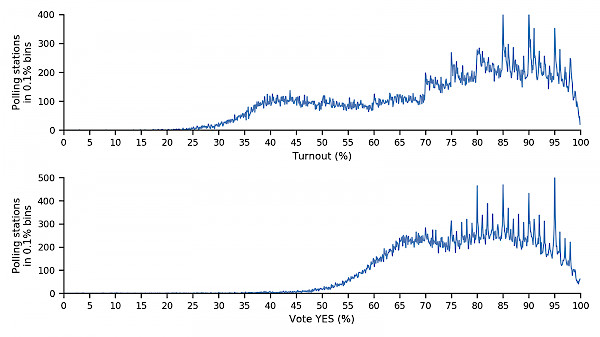}
\end{center}
\caption{Гистограмма по явке и голосам <<за>>~\cite{2020kobak-peaks}.}\label{2020kobak-peaks}
\end{figure}

\begin{figure}[!h]
\begin{center}
\includegraphics[width=0.6\textwidth]{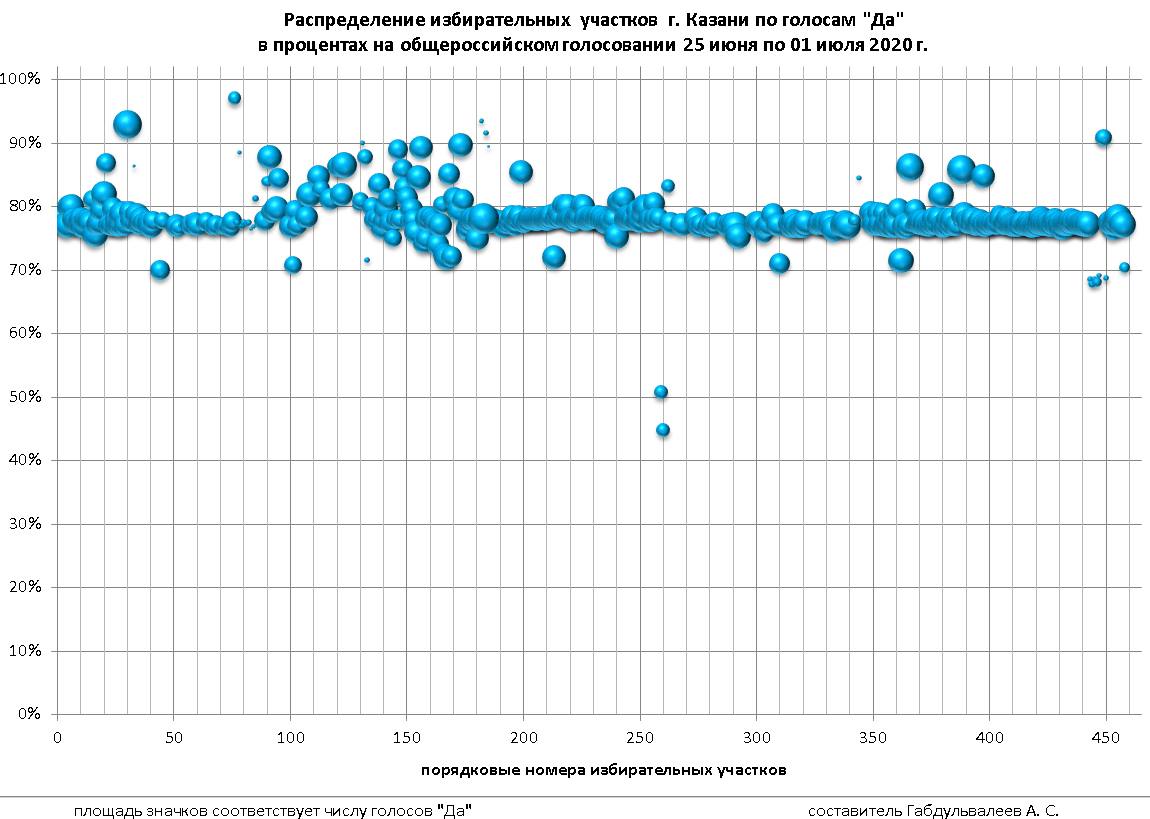}
\end{center}
\caption{Официальные данные по голосованию в Казани~\cite{2020kazan}.}
\label{2020kazan}
\end{figure}

\begin{figure}[!h]
\begin{center}
\includegraphics[width=0.5\textwidth]{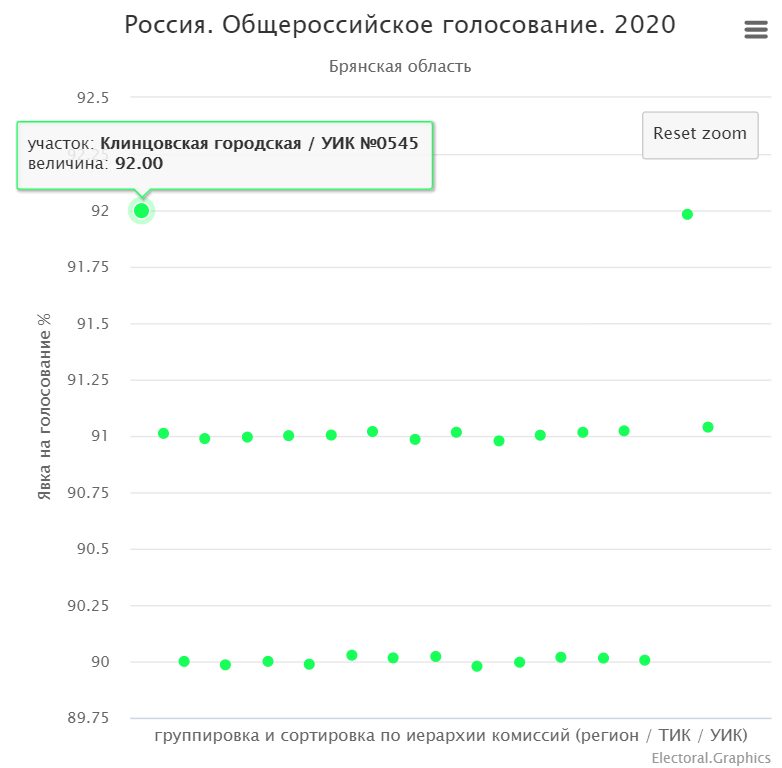}
\end{center}
\caption{Официальные данные по голосованию в Клинцах~\cite{2020klintsy}. По горизонтали отложены участки (в порядке их номеров).}
\label{2020klintsy}
\end{figure}

Эти изменения были внесены статьёй 1 указанного закона.
Одновременно статьёй 2 было назначено <<общероссийское
голосование по вопросу одобрения изменений в Конституцию [так в
тексте закона] Российской Федерации>> (механизм, не
предусмотренный ранее существовавшим избирательным
законодательством, со своими особыми правилами, установленными
тем же законом). Официально опубликованные данные по этому
голосованию (на 2 июля 2020) показаны на
рис.~\ref{2020allrussia}. Отчётливо видны следы сигнала в районе
$40$--$50$\% по явке и $60$--$70$\% по <<да>> среди
явившихся, а также пики на круглых процентах и сетка.
Пики на круглых процентах хорошо видны просто на гистограмме по явке и числу голосов <<за>>, см. рис.~\ref{2020kobak-peaks}~\cite{2020kobak-peaks}. Следует иметь в виду, однако, что эти оценки в принципе не могут
учитывать те регионы, где сигнал был практически полностью
подавлен шумом (см., например, рис.~\ref{2020kazan}
или~\ref{2020klintsy}, где наблюдается забавное чередование явки
в $90\%$ и $91\%$). Но в целом в первом приближении видно, что <<реальных>> голосов <<за>> было, вероятно, чуть больше половины от объявленных. 

\clearpage

\subsection{Голосование в Беларуси}

	Приведём несколько графиков с голосования 2020 года в Беларуси (выборы президента), хотя подробный анализ этих выборов далеко выходит за рамки обзора. В Беларуси результаты по участкам не публикуются централизованно, а только должны вывешиваться на избирательных участках, поэтому статистический анализ сильно затруднён. Однако исследователям выборов удалось собрать данные по фотографиям протоколов, а также по заявлениям голосовавших, из которых видно, что протоколы фальсифицированы (объявленное число голосов за Тихановскую, основного оппозиционного кандидата, меньше, чем число людей, заявивших о голосовании за неё на данном участке). Кроме того, видно, что объявленные суммарные результаты не получены суммированием результатов на участках (см. подробнее в~\cite{2020belarus-otchet,2020belarus-golos}). 
	
\begin{figure}[!h]
\begin{center}
\includegraphics[width=0.5\textwidth]{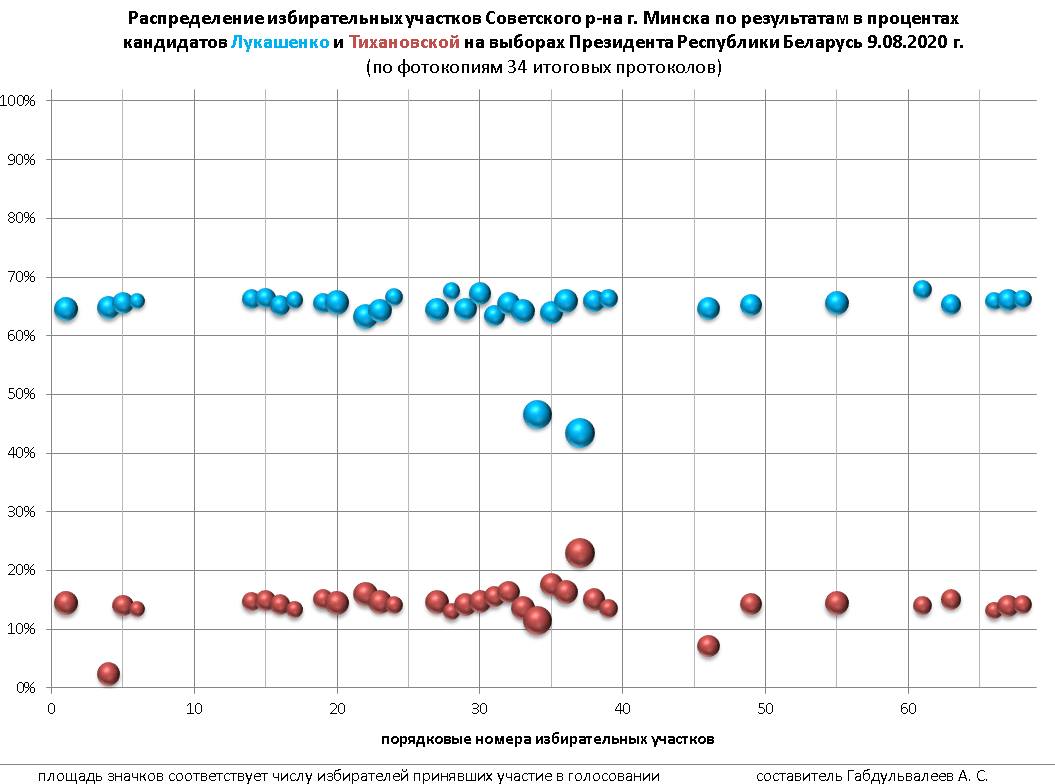}
\end{center}
\begin{center}
\includegraphics[width=0.5\textwidth]{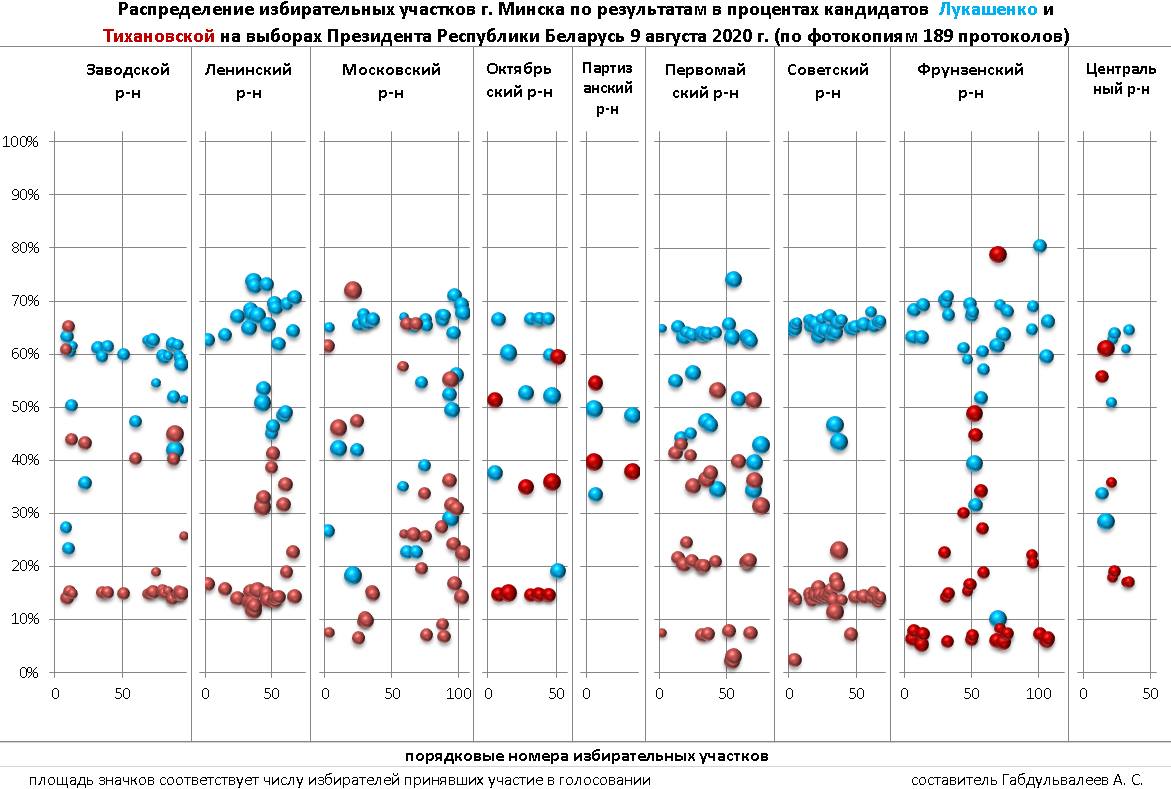}
\end{center}
\caption{Объявленные результаты доли голосов за Лукашенко и Тихановскую по 189 участкам Минска и по участкам Советского района Минска~\cite{2020belarus-azat}.}
\label{2020belarus-azat}
\end{figure}

Фальсификации хорошо видны и на графиках (см. рис.~\ref{2020belarus-azat}).

\newpage
\subsection{Новые попытки критики статистических методов}

Осенью 2020 года был опубликован доклад Борисова, Задорина, Игнатова, Марачевского, Фёдорова\footnote{Именно они указаны в качестве <<авторского коллектива>> на с.~1, хотя в тексте  доклада упоминаются в третьем лице (<<Как отмечает профессор Санкт-Петербургского государственного университета, доктор физико-математических наук В.\,Н.\,Марачевский>>, с. 2, <<Как отмечает социолог И.\,В.\,Задорин>>, с. 10.)} <<Математические инструменты делегитимизации выборов>>~\cite{2020-matinstrumenty}.

Доклад основан на недоразумении: авторы предполагают, что анализ выборов проводится с помощью сравнения распределений с нормальным:

\begin{quote}
Как отмечает профессор Санкт-Петербургского государственного университета, доктор физико-математических наук В.\,Н.\,Ма\-ра\-чевс\-кий, теоретическое описание процесса голосования с помощью одной функции Гаусса возможно только в гипотетическом случае, если вероятности голосования <<за>> во всех городах и регионах примут одно и то же значение, а на каждый избирательный участок в стране придёт одно и то же число избирателей\ldots\

Н.\,Е.\,Шалаев в своём научном исследовании приходит к однозначному выводу: анализ распределения цифр в рамках известных методов неприменим для поиска электоральных аномалий.

В действительности какие-то из эмпирических распределений могут иметь вид близкий к нормальному, но это может происходить не в силу закономерности, а вследствие вариативности распределений, исходя из теории больших чисел (с.3).
\end{quote}

\begin{figure}[!h]
\begin{center}
\includegraphics[width=0.6\textwidth]{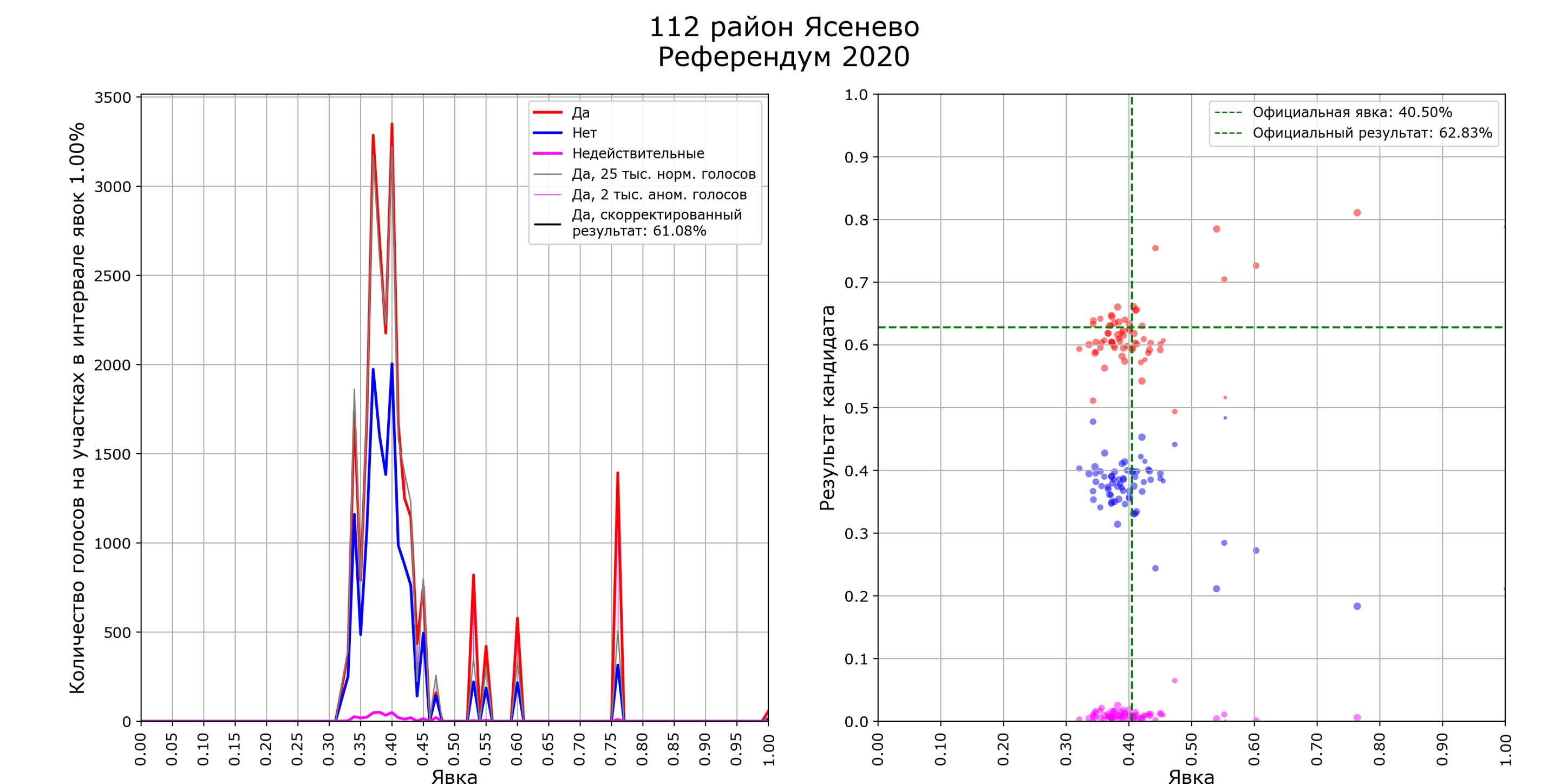}\\
\includegraphics[width=0.6\textwidth]{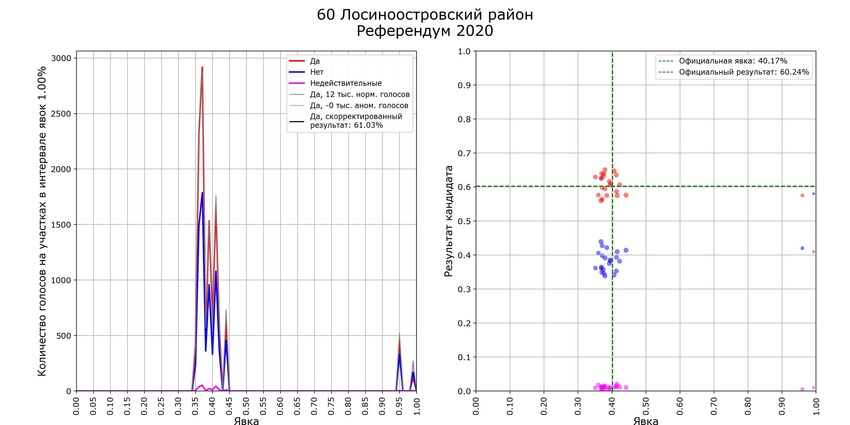}\\
\end{center}
\caption{Рисунки из~\cite{2020-golos}.}
\label{2020-golosinfo}
\end{figure}

И так далее. При этом претензии мало связаны с критикуемыми текстами. Например, на с.~15--16 приводятся два рисунка из статьи на сайте движения <<Голос>>~\cite{2020-golos} (рис.~\ref{2020-golosinfo}) с таким комментарием:
\begin{quote}
Например, на сайте \texttt{golosinfo.org} указана <<аномалия>> в районе Ясенево г. Москвы и в Лосиноостровском районе г.~Москвы. Но если попытаться найти причину различной активности и взглядов участников голосования, то она лежит на поверхности --- в этих районах было большое количество избирателей, поменявших своё место голосования по системе <<мобильный избиратель>>, которая позволяет легально прикрепиться к конкретному избирательному участку, открепившись по месту жительства. И результат с социологической точки зрения понятен и объясним.
\end{quote}
Между тем, если прочесть текст~\cite{2020-golos}, то видно, что эти районы как раз приводятся в качестве примеров корректного голосования (и результаты соответствующие: $60$--$65\%$ <<за>> при явке около $40\%$). Дальше авторы~\cite{2020-matinstrumenty} пишут <<Вызывает ``подозрения'' у авторов-составителей графиков результаты голосования в Черноголовской ТИК Московской области>> --- и приводят на с.~18 рисунок из~\cite{2020-golos}, на котором (на самом рисунке!) указана оценка в <<0 тыс. аномальных голосов>>. (Похоже, что авторы~\cite{2020-matinstrumenty} не читали не только критикуемый текст, но даже надписи на рисунках, которые они воспроизвели в своей статье.)

Далее авторы (справедливо) указывают, что связанные с выборами гистограммы в разных странах бывают далеки от нормальных, приводя многочисленные примеры.

Раздел 5 представляет собой <<копипасту>> фрагментов другого текста (подготовленного в системе \TeX\ и преобразованного в рисунки). Внешне этот текст напоминает математический, но даже при самом благожелательном отношении придать ему смысл нелегко. Автор пишет:
\begin{quote}
С вероятностью $0\le p\le 1$ каждый избиратель отдаёт свой голос `за', с вероятностью $q=1-p$ --- `против', Определим распределение избирателей, проголосовавших `за' в определённом проценте голосов избирателей `за' при большом числе избирательных участков. Вероятность того, что на избирательных участках $m$ человек проголосовало `за' и $n-m$ человек проголосовало `против', определяется величиной
\[
P_n(m)=C_n^m p^m q^{n-m}, \eqno(1)
\]
где 
\[
C_n^m = \frac{n!}{m!(n-m)!} \eqno(2)
\]
является числом независимых способов расположить $m$ избирателей на $n$ местах и одновременно с этим совпадает с числом избирательных участков, на которых $m$ человек голосует `за'. Понятно, что если $m$ избирателей проголосовало `за' и $n-m$ проголосовало `против', то $p^mq^{n-m}$ есть вклад в общую вероятность от одного избирательного участка.
\end{quote}

До сих пор текст можно интерпретировать как (не особенно грамотное) изложение распределения Бернулли, в котором автор смешивает вероятностное пространство с избирательными участками. Но следующий абзац уже не поддаётся никакой разумной интерпретации:
\begin{quote}
Именно такое число избирательных участков $C_m^n$ нужно выбрать для равновероятного учёта всех возможных вкладов голосующих избирателей, отдавших $m$ голосов `за' на каждом из участков. Общее число возможных равновероятных голосований избирателей на участке
\[
Y = \sum_{m=0}^{n} C_n^m \eqno(3)
\]
совпадает с общим числом участков и однозначно накладывает условие на общее число избирателей, равное $N=nY$. При этом строго доказывается, что общее число проголосовавших `за' избирателей есть $Np$.
\end{quote}
Можно только гадать, что с чем перепутал автор и откуда он взял $2^n$ <<равновероятных голосований избирателей>> (число $Y$ равно $2^n$, хотя из текста непонятно, известно ли это автору), и откуда берётся <<однозначно наложенное>> условие равенства числа избирателей числу $n2^n$. Может быть, он как-то смешивает число исходов и их суммарную вероятность в бернуллиевом распределении, потому что дальше пишет:
\begin{quote}
Общее число избирателей, проголосовавших на избирательных участках $N_{[m_1,m_2]}$ в интервале $[m_1,m_2]$, равняется
\[
N_{[m_1,m_2]}=n\sum_{m=m_1}^{m_2} C_n^m. \eqno(6)
\]
\end{quote}
Так или иначе, всё это лишено какого-либо смысла, включая вывод о том, что <<ключевым фактором, затрудняющим прямое использование теоремы Муавра--Лапласа при реальном голосовании, является малость числа избирательных участков по сравнению с величиной $Y=\sum_{m=0}^n C_n^m$>> (то есть по сравнению с $2^n$).

Всё это никак не относится к обсуждаемой теме (поскольку критикуемые в статье~\cite{2020-matinstrumenty} работы не опираются на требование нормальности распределений), и приведено здесь лишь для иллюстрации уровня текста. С этой же целью приведём без комментариев первый абзац раздела 5.5.
\begin{quote}
Может возникнуть вопрос: как может человек голосовать с вероятностью $p$? Проясним данный вопрос. Когда каждый избиратель голосует, он не знает, какова вероятность `за' $p$, он лишь голосует `за' или `против'. Вероятность $p$ возникает только после окончания голосования, и фокус в том, что одновременно возникает и распределение результатов при наличии большого числа избирательных участков. Как это возможно? Ответ на данный вопрос состоит в том, что избиратели, которые голосуют `за' или `против', случайным образом распределены по избирательным участкам по месту проживания. Поэтому на каждый избирательный участок приходит случайное число избирателей, которые хотят проголосовать `за', именно в этот момент случайное распределение избирателей по участкам срабатывает как естественный генератор случайных чисел при избирательном процессе. При достаточно большом числе избирателей $nY$ из-за случайности прикрепления каждого избирателя, голосующего `за' или `против', к своему участку, возникают все возможные равновероятные комбинации избираетелей `за' и `против' на участках $Y$. В однородной системе возниковение всех возможных равновероятных комбинаций `за' и `против' при большом числе участков $Y$ математически ведёт к нормальному (гауссову) распределению числу проголосовавших `за' избирателей с максимумом при проценте проголосовавших `за' $100p$~\%.
\end{quote}

\end{document}